\documentclass[lettersize,onecolumn]{IEEEtran}
\normalsize
\usepackage{graphicx}
\usepackage{float}
\usepackage{subfigure}
\usepackage{lineno}
\usepackage{verbatim}
\usepackage{cite}
\usepackage{latexsym}
\usepackage{amsmath}
\usepackage{amssymb}
\usepackage{mathrsfs}
\usepackage{balance}
\usepackage{setspace}
\usepackage{epstopdf}
\usepackage{graphics}

\usepackage{caption}
\usepackage{graphicx}
\usepackage{float} 
\usepackage{subcaption}

\usepackage{hyperref}

\usepackage[usenames]{color}
\usepackage{mdwmath}
\usepackage{mdwtab}
\usepackage{array}
\usepackage{picins}
\usepackage{eqparbox}
\usepackage{lettrine}
\usepackage{bbm}
\usepackage{dsfont}
\usepackage[noend]{algpseudocode}
\usepackage{algorithmicx,algorithm}

\newtheorem{remark}{Remark}
\newtheorem{theorem}{Theorem}
\newtheorem{lemma}{Lemma}
\newtheorem{definition}{Definition}
\newtheorem{prop}{Proposition}

\newtheorem{example}{Example}
\newtheorem{corollary}{Corollary}

\allowdisplaybreaks[4]
\begin{document}

	\title{On the Analysis of Random Linear Streaming Codes in Stochastic Channels}

    \author{Kai Huang, Wenjie Guan, Xiaoran Wang
    , Jinbei Zhang, Kechao Cai
    \\

School of Electronic and Communication Engineering, Sun Yat-sen University, Shenzhen, China\\

Email: \{huangk56,guanwj9,wangxr77\}@mail2.sysu.edu.cn
, \{zhjinbei,caikch3\}@mail.sysu.edu.cn 
}
    {}
	\maketitle

	\begin{abstract}
		\boldmath
        Random Linear Streaming Codes (RLSCs) can dramatically reduce the queuing delay of block codes in real-time services. 
        In this paper, we aim to explore the fundamental limit of large-field-size RLSCs in stochastic symbol erasure channels (SEC).
        The performance of \textit{Non-systematic} RLSCs (NRLSCs) in i.i.d. SEC has been analyzed in [Pinwen Su et al. 2022].
        In this work, we first characterize the closed-form expression on the exact error probability of NRLSCs in Gilbert-Elliott symbol erasure channels (G-ESEC). 
        Compared to i.i.d SEC, the erasure probability of G-ESEC depends on current channel state, thus the transitions between different states should be considered. 
        To deal with the stochastic state transitions, we introduce two novel techniques.
        (i) To account for the impact of switching states on probability terms, we find and leverage the \textit{recursive structure} of the state transition traces.
        (ii) To obtain the expected number of error timeslots, we derive the \textit{stationary initial distribution of the states}, and formulate \textit{iterative equation} to characterize the expectation terms. 
        Then we analyze the \textit{Systematic} RLSCs (SRLSCs) in a special SEC, i.e., the packet erasure channel (PEC), where the encoded symbols will be either all received or all erased in each slot. 
        In this scenario, SRLSCs could save some source symbols which should have exceeded the decoding delay in NRLSCs, and thus could significantly reduce the error probability.   
        To this point, our contributions are two-folds.
        (i) Through a case study, we first find a counter-intuitive phenomenon that SRLSCs can cause unexpected error events comparing to NRLSCs in some erasure patterns. 
        Then we fully characterize the error event of SRLSCs for any erasure pattern.  
        (ii) For i.i.d. PEC, we derive an analytical expression on the exact error probability of SRLSCs when the length of the memory approaches infinity and the coding rate equals to 1/2. 
        Simulations are conducted to verify the accuracy of our analysis on the exact error probabilities, and compare the performance of NRLSCs, SRLSCs, and existing streaming codes for sliding window erasure channels. 
	\end{abstract}


    \IEEEpeerreviewmaketitle

	\section{Introduction} \label{section:introduction}

The rapid advancement of 5G mobile communication networks has placed unprecedented demands on data transmission, particularly in applications classified under Ultra-Reliable Low-Latency Communications (URLLC). URLLC is essential for enabling critical real-time services, including autonomous driving, remote medical interventions, and industrial automation, where even minimal delay or data loss can lead to significant safety or performance risks. 
To alleviate the impact of inevitable packet loss in practical scenarios such as unstable wireless channel or network congestion in the peak hours, two categories of schemes, i.e., Automatic Repeat reQuest (ARQ) and Forward Error Correction (FEC) are proposed.
Traditional ARQ protocols, though effective in ensuring reliable data delivery, often fall short in meeting URLLC requirements due to their inherent feedback and retransmission delays. 
This limitation has driven the exploration of 
streaming codes protected by FEC that can provide both high reliability and reduced latency.

The prior studies on FEC-protected streaming codes primarily focused on two types of channel model. 
\cite{bursty only,b1,b2,b3,bn1,bn2,bn3,adversarial channel,SLFTCOM,SLFTIT,variable size,simple1,simple2} considered the first model, i.e, \textit{sliding window packet erasure channels (SWPEC)}, where within a fixed-size window, either a bounded number of burst erasures or arbitrary erasures may occur. 
Particularly, \cite{bursty only} first investigated a bursty $B$-erasure channel with decoding delay $T$. 
The follow-up works extended the results from only bursty erasures \cite{b1,b2,b3} to both bursty and isolated erasures \cite{bn1,bn2,bn3,adversarial channel,SLFTCOM,SLFTIT,variable size}.
Specifically, it is a main stream to consider the $(W,B,M)$-SWPEC, which introduces either one burst erasure with length no longer than $B$ or multiple arbitrary erasures with total count no larger than $M$ within any window with length $W$. 
Their aim is to design optimal (achieving the capacity) error-free streaming codes over a predefined deterministic class of channel erasure patterns. 
Moreover, in order to reduce the computational complexity and power consumption in encoding and decoding, \cite{simple1,simple2} proposed new code constructions to reduce the order of the finite field. 
In the aforementioned studies regarding the SWPEC, well-designed but complex deterministic coding construction are proposed for its parity matrix.


On the other hand, the second model addresses \textit{stochastic channels}, where erasures occur according to probabilistic.
In \cite{RLSCs ISIT,RLSCs,asymptotics1,asymptotics2,asymptotics3}, RLSCs under sufficiently large finite size regime were intensively studied in i.i.d. SEC. Particularly, \cite{RLSCs ISIT,RLSCs} generalized the concept of \textit{information debt}, which was first proposed in \cite{phd}, and characterized the error event of large-finite-field RLSCs for any finite memory length $\alpha < \infty$ and any finite decoding deadline $\Delta < \infty$. Then the closed-form expression of the exact error probability was derived with a novel random-walk-based analysis framework. 
In \cite{asymptotics1,asymptotics2,asymptotics3}, asymptotic results (with some parameters being asymptotically large) were developed.
In these studies, the generator matrix of streaming codes are simply assumed to be almost completely random, and their main focus is the theoretical performance analysis in the stochastic channels. 

In this paper, we aim to further explore the fundamental limit of large-field-size RLSCs in stochastic symbol erasure channels.
In practical scenarios, the channel errors could be bursty, which can not be captured by the i.i.d SEC. 
On the other hand, the channels featuring state-switching behavior such as the Gilbert-Elliott Channel \cite{Gilbert,Elliott} and the Fritchman Channel \cite{Fritchman} can model the erasure scenario with both burst and arbitrary errors, and thus are of great practical interest.
Our first aim is to investigate the NRLSCs in G-ESEC, where the channel can be in good state or bad state. 
The erasure probabilities and also the transition matrices of the information debt are therefore state-dependent.
Thus, the \textit{state transition trace}, which consists of all sequential states along with the timeslots should be considered. 
And the expected number of error timeslots should be averaged over all possible transition traces, which is analytically challenging. 
Furthermore, when the state-switching behavior is considered, the stationary initial distribution of the states also has a non-trivial impact on the error probability and thus should be treated properly for the exact characterization.
Our second aim is to stochastically investigate the theoretical performance limit of SRLSCs in PEC.
Different from the NRLSCs, for SRLSCs in PEC, the destination can decode the source symbols $\mathbf{s}(t)$ immediately, upon perfectly receiving the systematic-encoded packet $\mathbf{x}(t)$ sent from the encoder. 
Consequently, this feature can save some source symbols which should have exceeded the decoding delay in NRLSCs, and thus could significantly reduce the error probability.
However, the characterization of the error event in SRLSCs is still an open question.
And the analysis of SRLSCs is far more challenging than that in NRLSCs. 
This is mainly because the transition matrices of the information debt is homogeneous in NRLSCs, while they become heterogeneous in SRLSCs, which makes the closed-form expression intractable to obtain.

Our main contributions can be summarized as follows.
\begin{itemize}
    \item First, we characterize the exact error probability of NRLSCs in G-ESEC under the finite memory length and the decoding delay constraints.
    The derivations are based on two novel techniques:
    \begin{enumerate}
        \item To account for the exact impact of the switching states on the probability terms, we find and leverage the \textit{recursive structure} of the state transition traces and derive the probability terms in the form of multiplies of state transition matrices and information debt transition matrices. 
        \item To obtain the expected number of error timeslots, we derive the \textit{stationary initial distribution of the states}, and formulate \textit{iterative equation} to characterize the expectation terms. 
    \end{enumerate}  
    These results generalize the stochastic analyses of \cite{RLSCs ISIT,RLSCs} into G-ESEC and could be independent of interest.
    Moreover, the results can also be further generalized into any hidden Markov channel with more than two hidden states.
    We also numerically verify the correctness of our theoretical results by comparing to Monte-Carlo simulations.  

    \item Second, we investigate the theoretical performance of SRLSCs in PEC, where the encoded symbols will be either all received or all erased in each slot. Our contributions are two-folds:
    \begin{enumerate}
        \item Through a case study, we find a counter-intuitive phenomenon that comparing to NRLSCs, SRLSCs can actually cause unexpected error events for some erasure patterns. 
        The main cause is that when some symbols are delivered successfully, the process of eliminating their impact on other unknowns will cause de-correlation between the preceding and the following information.  
        With insights obtained from the case study, we fully characterize the error event of SRLSCs for any erasure pattern.  
  
        \item We derive an analytical expression on the exact error probability of SRLSCs when the length of memory $\alpha \rightarrow \infty$ and the coding rate equals to 1/2 in i.i.d. PEC. 
        The derivation of this novel result involves the decomposition of tridiagonal Toeplitz matrix\cite{Toeplitz} and the derivation of distribution on the Catalan Number\cite{Catalan} and could be independent of interest.
    \end{enumerate}
    
    \item  Third, we conduct extensive numerical simulations on comparisons between NRLSCs, SRLSCs and the streaming codes for SWPEC \cite{adversarial channel}.
    With the simulation results, we have the following observation. 
        \begin{enumerate}
        \item SRLSCs display a lower error probability than NRLSCs in most of the scenarios. 
        The gap increases along with the channel erasure rate.
        \item RLSCs present a better resistance to the increasing channel erasure rate than the streaming codes in \cite{adversarial channel}.
        This result implies that the complex and deterministic design on the parity matrix could be counter-productive in the stochastic channels with high and time-varying erasure probability, compared to the simple and all-random generator matrix.
    \end{enumerate}
\end{itemize}

The rest of the paper is organized as follows. In Section \ref{section:model statement}, we describe the system model of NRLSCs and the definitions. In Section \ref{section:Main Results1}, we present our first contribution, the characterization of the exact error probability of NRLSCs in G-ESEC. 
In Section \ref{section:Main Results2}, we present our second contribution, the investigation on the SRLSCs in PEC. 
The numerical comparisons are presented in Section \ref{section:numerical results}. We conclude in Section \ref{section:conclusion}.

\textit{Notations:} In this paper, for some integers $a$ and $b$, $\{a,a+1,\dots,b\}$ is denoted as  $[a,b]$ and $\{1,2,\dots,a\}$ is denoted as $[a]$. $\Phi$ represents the empty set. The probability is denoted by $\text{Pr}(\cdot)$, and the expectation is denoted by $\mathbb{E}\{\cdot\}$. We use $(\cdot)^\top$ to represent the transpose of a matrix or a vector. We use $\mathbf{s}_a^b \triangleq [\mathbf{s}^\top(a),\mathbf{s}^\top(a+1),\dots, \mathbf{s}^\top(b)]^\top$ to represent the \textit{cumulative} column vector. $\mathds{1}\{\cdot\}$ is the indicator function. In the presented partitioned matrices, the omitted entries are all zeros.
$\vec{\mathbf{1}}$ and $\vec{\mathbf{0}}$ are used to represent column vectors of all 1s or 0s, respectively.
$\vec{\delta}_{k}$ is a column vector where the $k$-th entry is one and all other entries are zeros.
Identity matrix of size $n$ is denoted by $\mathbf{I}_n$. 
Denote $\mathbf{T}_{n} = \begin{bmatrix}
    (1-p)\mathbf{I}_n & p\mathbf{I}_n\\
    r\mathbf{I}_n & (1-r)\mathbf{I}_n
\end{bmatrix}_{2n\times 2n}$, where $n\ge 1$ is a positive integer, and $p,r\in(0,1)$ are transition probabilities defined afterwards.
Let $H(X)$ represent the entropy of $X$ and $I(X;Y)$ represent the mutual information between $X$ and $Y$.
Let $c = a|b$ be the modulus operator, which represents that $c$ is the remainder of the Euclidian division of $a$ by $b$.
Let $diag[x(1),\cdots,x(n)]$ be the $n$-by-$n$ diagonal matrix with $x(1),\cdots,x(n)$ listed sequentially in its main diagonal.


\section{System Model and Definitions} \label{section:model statement}
In this section, we first describe the general model of NRLSCs and the G-ESEC, which will be further analyzed in the next section. In Section \ref{section:Main Results2}, the model with SRLSCs will be introduced. 

\textbf{Encoder:} For any timeslot $t\ge 1$, $K$ source symbols $\mathbf{s}(t)=[s_1(t),\dots, s_K(t)]^\top$
will arrive at the encoder. Each symbol $s_k(t), k\in [K]$ is an i.i.d. sample from the finite field $GF(2^q)$. 
The arrived source symbols will be cached into the encoder's memory, which is a shift register with length $\alpha$.
Thus, there are at most $\alpha \cdot K$ symbols from previous $\alpha$ timeslots $\{\mathbf{s}(t):t\in [t-\alpha,t-1]\}$ stored in the memory of the encoder.
Then the $(\alpha+1)K$ symbols (the arrived $K$ symbols and the $\alpha \cdot K$ symbols in the memory) will be jointly encoded into $N$ symbols $\mathbf{x}(t) = [x_1(t),\dots,x_N(t)]^\top \in (GF(2^q))^N$ for transmission. $\mathbf{x}(t)$ is also referred to as the ``\textit{packet}" herein.
Throughout the paper, all the encoding/decoding operations are defined over $GF(2^q)$. 
Let $\mathbf{G}_t$ be the \textit{generator matrix} for timeslot $t$, thus 
\begin{equation}
    \mathbf{x}(t) = \mathbf{G}_t \cdot \mathbf{s}_{\max(t-\alpha,1)}^t.
\end{equation}

\textbf{Symbol erasure channel:}
In every timeslot $t$, the $N$ encoded symbols $\mathbf{x}(t)$ will be transmitted into the channel by the transmitter. A random subset of $\mathbf{x}(t)$, denoted by $\mathcal{C}_t \subseteq [N]$, can be received successfully by the destination, while the rest will be considered as ``erasure". Denote the number of received symbols as $C_t = |\mathcal{C}_t|$. Note that the minimum element of erasure in SEC is a symbol. On the other hand, the PEC could be different and will be otherwise introduced in Section \ref{section:Main Results2}. 

\textbf{Gilbert-Elliott erasure behavior and state transition stochastics:} 
The G-ESEC is a special Hidden Markov Chain (HMC) with only two hidden states, i.e., the good state and the bad state. 
The symbol erasure rates in these two states could be different. 
Denote the channel state at timeslot $t$ as $a_t$.
In each timeslot, the number of received symbols $C_t = |\mathcal{C}_t|$ depends on $a_t$.
When $a_t=G$ ($G$ is the abbreviation for good state), $C_t$ could be very large and close to $N$. When $a_t=B$ ($B$ is the abbreviation for bad state), $C_t$ could be small and close to zero. Define the emission probability at each state as $P_i^G \triangleq \text{Pr}(C_t=i|a_t=G)$ and $P_i^B \triangleq \text{Pr}(C_t=i|a_t=B), i\in [N]$, respectively.
The channel state can transition between good and bad at the end of each timeslot. 
The transition matrix between good and bad states is defined as 
\begin{equation}
    \mathbf{T}_1 = \begin{bmatrix}
    1-p & p\\
    r & 1-r
\end{bmatrix}.
\end{equation}
Specifically, when $a_t=G$, the channel will stay in good state, i.e., $a_{t+1}=G$ with probability $1-p$ or it will switch to bad state, i.e., $a_{t+1}=B$ with probability $p$.
Denote the stationary distribution of the hidden states as $\pi = \begin{bmatrix}
    \pi_G&\pi_B\end{bmatrix}=\begin{bmatrix}
    \frac{r}{p+r}&\frac{p}{p+r}\end{bmatrix}$. 
Without loss of generality, we assume that the G-ESEC is ergodic.

\begin{figure}[!hbtp]\setcounter{subfigure}{0}
    \centering
    \subfigure[$\mathbf{x}_1^t = \mathbf{G}^{(t)} \mathbf{s}_1^t$.]{\includegraphics[width=0.49\textwidth]{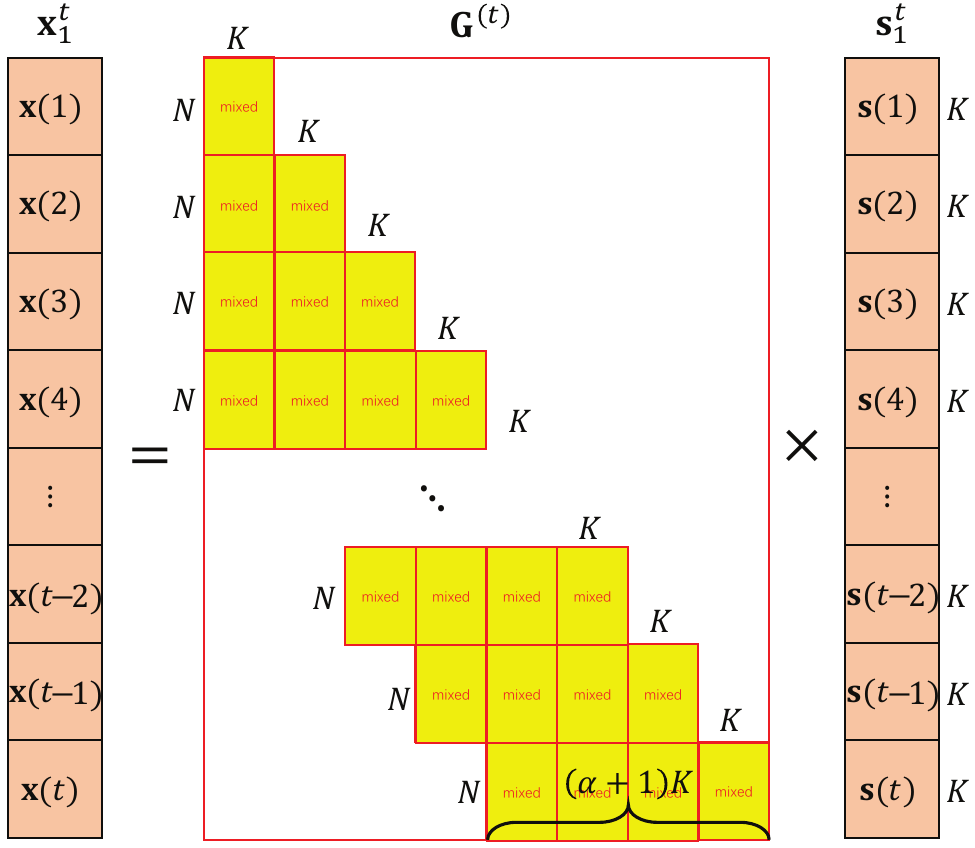}}
    \subfigure[$\mathbf{y}_1^t = \mathbf{H}^{(t)} \mathbf{s}_1^t$.]{\includegraphics[width=0.49\textwidth]{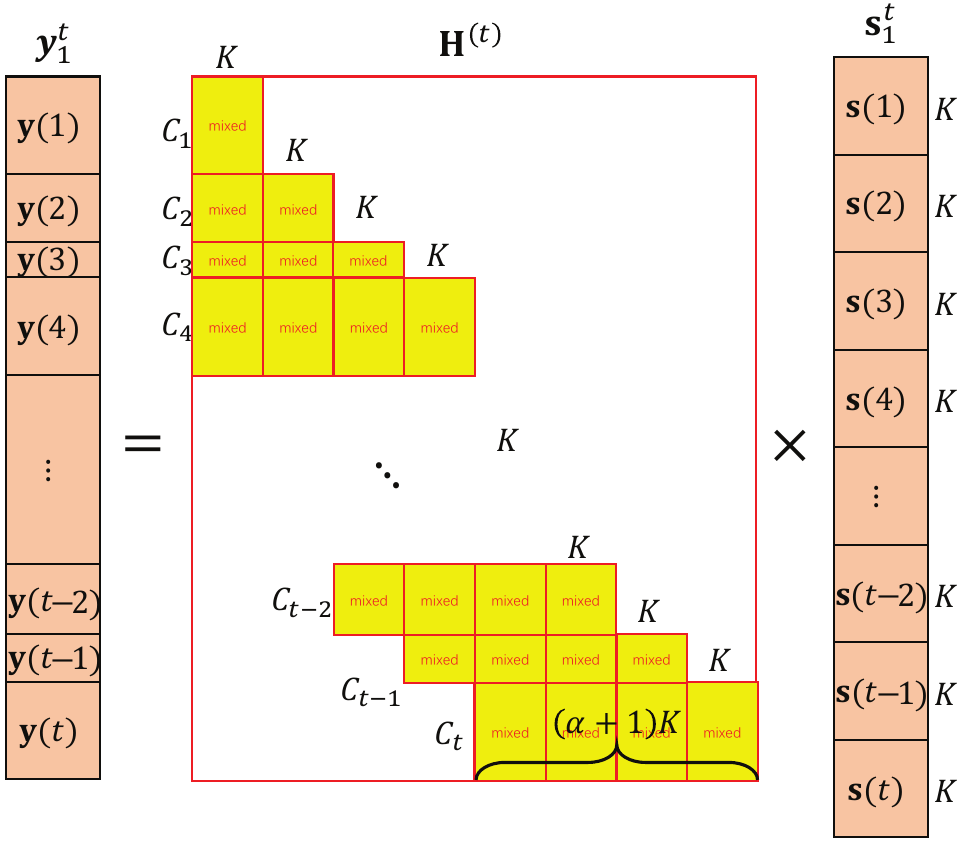}}
    \caption{Illustrations of non-systematic RLSCs with $\alpha=3$.}
    \label{nonsys}
\end{figure}

\textbf{Decoder:} Denote the $C_t$ received symbols at timeslot $t$ as $\mathbf{y}(t) = [y_1(t),\dots,y_{C_t}(t)]^\top$. Then we have $\mathbf{y}(t) = \mathbf{H}_t\cdot \mathbf{s}_{\max(t-\alpha,1)}^t$, where $\mathbf{H}_t$ is the projection of $\mathbf{G}_t$ onto the random set $\mathcal{C}_t$.
After properly shifting and stacking the $\mathbf{G}_t$ and $\mathbf{H}_t$ along with  timeslots, we can obtain the \textit{cumulative generator matrices} $\mathbf{G}^{(t)}$ and \textit{cumulative receiver matrices} $\mathbf{H}^{(t)}$ satisfying that 
\begin{equation}\label{equation:encoding}
    \mathbf{x}_1^t = \mathbf{G}^{(t)} \mathbf{s}_1^t, \quad\mathbf{y}_1^t = \mathbf{H}^{(t)} \mathbf{s}_1^t.
\end{equation}
Illustrations of $\mathbf{x}_1^t$ and $\mathbf{y}_1^t$ with $\alpha =3$ are displayed in Fig. \ref{nonsys} (a) and (b), respectively. All entries in the yellow blocks are non-zero, while all entries in the white space are zeros.

\textbf{Decodability:} At timeslot $t$, the destination can observe $\mathbf{y}(1)$ to $\mathbf{y}(t)$. The decoder should try to decode $\mathbf{s}(t)$ at timeslot $t+\Delta$ with observations $\mathbf{y}(1)$ to $\mathbf{y}(t+\Delta)$, where $\Delta$ is the decoding delay.
The decodabilities are defined as follows:
\begin{definition}
The symbol $s_k(t)$ is $\Delta$-decodable if the vector $\vec{\delta}_{(t-1)K+k}^\top$ is in the row space of $\mathbf{H}^{(t+\Delta)}$, where $\vec{\delta}_{(t-1)K+k}$ is a column vector such that its $((t-1)K+k)$-th element is one and all the other elements are zeros. 
\end{definition}
\begin{definition}
The vector $\mathbf{s}(t)$ is $\Delta$-decodable if all symbols $\{s_k(t):k\in[K]\}$ are $\Delta$-decodable. 
\end{definition}

In this paper, we aim at characterizing the exact value of the slot error probability of RLSCs in G-ESEC, defined as
\begin{equation}
    p_{e,[1,T]}^{\text{RLSC}(q)} \triangleq \frac{1}{T}\!\sum_{t=1}^T \text{Pr}(\mathbf{s}(t) \text{ is not } \Delta\text{-decodable}).
\end{equation}
We are exclusively interested in the \textit{long term} slot error probability under the \textit{sufficiently large finite field size regime}, which is defined by
\begin{equation}
    p_e \triangleq \lim_{T\rightarrow \infty}\lim_{q\rightarrow \infty}  p_{e,[1,T]}^{\text{RLSC}(q)}.
\end{equation}

To simplify the discussion, we impose two assumptions on randomness of the generator matrix.

\noindent \textbf{(I) Non-systematic Random linear streaming codes:}
The entries of $\mathbf{G}_t$, represented by the yellow blocks in Fig. \ref{nonsys} (a), are chosen uniformly and randomly from GF($2^q$), excluding 0. 
This assumption indicates the NRLSCs, where the $N$ symbols sent in each timeslot is a mixture (linear combination) of $(\alpha+1)K$ symbols from the present and previous timeslots. 
In Section \ref{section:Main Results2}, we will discuss and analyze the SRLSCs.

\noindent \textbf{(II) The Generalized MDS Condition (GMDS):} For any $t$ and any finite sequence of pairs $\{(i_l,j_l):l\in[L]\}$ satisfying the following two conditions: \textbf{(a)} $i_{l_1}\neq i_{l_2}$ and $j_{l_1}\neq j_{l_2}$ for any $l_1\neq l_2$ and \textbf{(b)} the $(i_l, j_l)$-th entry of $\mathbf{G}^{(t)}$ is non-zero for all $l \in [L]$, define the corresponding row and column index sets $S_R \triangleq \{i_l:l \in [L]\}$ and $S_C \triangleq \{j_l:l \in [L]\}$.
The GMDS condition requires that the submatrix of the cumulative generator matrix $\mathbf{G}^{(t)}$ induced by $S_R$ and $S_C$ is always invertible.

These two assumptions jointly ensure all successfully delivered symbols can carry as much information as possible for decoding, and thus avoid the discussion on some corner cases. 
In this way, all the randomness is a result of random channel realization, not the random code construction.
It is worthy noting that we do not specify any code construction on the generator matrix $\mathbf{G}_t$ at each timeslot, leaving it almost randomly chosen in the asymptotically large finite field.
And our main focus is the stochastic analysis and theoretical characterization of the exact error probability, different from the explicit designs on the generator matrix\cite{bursty only,b1,b2,b3,bn1,bn2,bn3,adversarial channel,SLFTCOM,SLFTIT,variable size,simple1,simple2}. 

\section{The Analysis of Non-systematic RLSCs in the G-ESEC}\label{section:Main Results1}
In this section, we present our first result, i.e., the characterization of the error probability of NRLSCs in G-ESEC. 

\subsection{Preliminaries}
As in \cite{RLSCs}, we first reuse the definition of \textit{information debt} $I_d(t)$, which was originally introduced in \cite{phd}. The concept $I_d(t)$ is used to describe how many linear equations the destination still needs for successful decoding. 
\begin{definition}\label{definition:informtion debt old}
    Let $\zeta = \alpha K+1$ and $I_d(0)=0$. For any $t\ge 1$, the information debt $I_d(t)$ of NRLSCs in SEC is calculated iteratively by 
    \begin{align}
        \hat{I}_d(t) &\triangleq \left(K-C_t+\min(I_d(t-1),\alpha K)\right)^+\label{equation:7}\\
        I_d(t) &\triangleq \min\left(\zeta,\hat{I}_d(t)\right).\label{equation:8}
    \end{align}
\end{definition}
$\zeta$ is the maximum value of the information debt. $I_d(t)$ hits $\zeta$ indicates the de-correlation between the latest symbols and some previous symbols, which implies that some previous symbols erased by the channel are lost forever, no matter how many linear equations can be received thereafter. This is generally because the symbols transmitted thereafter will not contain any information (linear combinations) of the previous lost symbols due to the limit of the memory length $\alpha$.

The \textit{hitting time sequences} of $I_d(t)$, i.e., $\{t_i:i\in[0,\infty]\}$ and $\{\tau_j:j\in[0,\infty]\}$ are defined as follows.
\begin{definition}\label{definition:hitting time old}
    Initialize that $t_0\triangleq0$ and $\tau_0\triangleq0$ and define iteratively 
    \begin{align}
        t_i &\triangleq \inf \{t':t'>t_{i-1}, I_d(t')=0\}\label{equation:ti_old}\\
        \tau_j &\triangleq \inf \{t':t'>\tau_{j-1}, I_d(t')=\zeta\}
    \end{align}
    as the $i$-th and $j$-th time that $I_d(t)$ hits 0 and $\zeta$, respectively.
\end{definition}

In \cite{RLSCs}, the error event is characterized for NRLSCs as follows:

\begin{prop}\label{Proposition:characterization of the error event}
(\textit{Proposition 3 in \cite{RLSCs}})
     Assume \textbf{GMDS} holds. For any index $i_0\ge 0$,
\textbf{(a)} if $\nexists \tau_j \in (t_{i_0},t_{i_0+1})$, $\mathbf{s}(t)$ is not $\Delta$-decodable $\forall t\in (t_{i_0},t_{i_0+1}-\Delta)$; 
\textbf{(b)} if $\exists \tau_j \in (t_{i_0},t_{i_0+1})$, let $\tau_{j^*}$ be the one with the largest $j$, then $\mathbf{s}(t)$ is not $\Delta$-decodable $\forall t\in (t_{i_0},\max(\tau_{j^*}-\alpha+1,t_{i_0+1}-\Delta))$;
\textbf{(c)} for the rest of $t$, $\mathbf{s}(t)$ is $\Delta$-decodable.
\end{prop}

\textit{Although i.i.d. SEC is assumed in \cite{RLSCs}, we note that this proposition also holds for G-ESEC.} Actually, the correctness of the proposition is independent of the number of channel states. 
This is because the proof of Proposition \ref{Proposition:characterization of the error event} (which can be referred to \cite{RLSCs}) is only based on the analysis on the cumulative receiver matrices $\mathbf{H}^{(t)}$, which is the joint outcome of code structure $\mathbf{G}^{(t)}$ and the channel emission $C_t$.
Therefore, given the realization of $\mathbf{H}^{(t)}$, the error event for NRLSCs will be determined, regardless of which hidden state the channel emissions $C_t$ were generated by.
In other words, for a given erasure pattern, by \cite[Proposition 3]{RLSCs}, one can uniquely derive the error regime, irrespective of the successful deliveries and erasures in the pattern are generated by how many states.
Therefore, \cite[Proposition 3]{RLSCs} also holds for G-ESEC.
With this argument, the following lemma holds directly, with which we can analyze the exact error probability.


\begin{lemma}\label{lemma:1} (\textit{Lemma 4 in \cite{RLSCs}})
    Assume the transmission rate is within the capacity. The error probability of NRLSCs, denoted as $p_e^{ns}$, can be given by 
    \begin{equation}
        p_e^{ns} = \frac{\mathbb{E}\{L_G+L_{B_1}+L_{B_2}\}}{\mathbb{E}\{t_{i_0+1} - t_{i_0}\}},
    \end{equation}
    where $i_0 \ge 0$ is any arbitrary but fixed index,
    \begin{align}
        &L_G \triangleq \mathds{1}\{\nexists \tau_j \in (t_{i_0},t_{i_0+1})\}\cdot(t_{i_0+1} - \Delta - 1 - t_{i_0})^+,\\
        &L_{B_1} \triangleq \mathds{1}\{\exists \tau_j \in (t_{i_0},t_{i_0+1})\}\cdot(\tau_{j^*}-t_{i_0}),\\
        &L_{B_2} \!\triangleq\! \mathds{1}\{\exists \tau_j \!\in \!(t_{i_0},t_{i_0+1})\}\!\cdot\!\max(-\alpha,t_{i_0+1} - \Delta - 1 - \tau_{j^*}),
    \end{align}
    and $\tau_{j^*}$ is the largest $\tau_j$ within the interval $(t_{i_0},t_{i_0+1})$.
\end{lemma}

\textit{Proof:} Lemma \ref{lemma:1} holds from Proposition \ref{Proposition:characterization of the error event} by calculating the ratio of expected error timeslots to the expected interval of the zero-hitting times.
Notice that $t_i$ defined in (\ref{equation:ti_old}) is a Markov renewal process.
By \cite[Theorem 3.3]{Renewal process}, Lemma \ref{lemma:1} holds directly.

\subsection{Characterization of $p_e^{ns}$ in G-ESEC}
This subsection is devoted for characterizing the exact value of $\mathbb{E}\{t_{i_0+1}-t_{i_0}\}, \mathbb{E}\{L_G\}, \mathbb{E}\{L_{B_1}\}$ and $\mathbb{E}\{L_{B_2}\}$ in G-ESEC. Since the change of $I_d(t)$ depends on $a_t$, the analyses of above formulas will be significantly different from that in \cite{RLSCs}.
Recall that G-ESEC is a HMC with two hidden states $a_t \in \{G,B\}$ and $\zeta+1$ emission observations $I_d(t)\in\{0,1,\dots,\zeta\}$. 
At each hidden state, there exists a distinct transition matrix of the emission observations $I_d(t)$.
We denote the $(\zeta+1)\times (\zeta+1)$ transition matrix of $I_d(t)$ in state $G$ and $B$ as $\Gamma^G = [\gamma_{i,j}^G]$ and $\Gamma^B = [\gamma_{i,j}^B]$, respectively, where $\gamma_{i,j}^G = \text{Pr}(I_d(t+1)=j|I_d(t)=i,a_t=G)$ and $\gamma_{i,j}^B = \text{Pr}(I_d(t+1)=j|I_d(t)=i,a_t=B)$, $\forall i,j\in [0,\zeta]$. Matrices $\Gamma^G$ and $\Gamma^B$ are determined by the distribution of $P_i^G$ and $P_i^B, i\in [N]$ respectively. Denote the intermediate states as $\phi = \{1,\dots,\zeta-1\}$. Thus, $\Gamma^G$ and $\Gamma^B$ can be partitioned as follows:
\begin{equation}
    \Gamma^G \!= \!\!\begin{bmatrix}
        \Gamma_{0,0}^G & \Gamma_{0,\phi}^G &\Gamma_{0,\zeta}^G \\
        \Gamma_{\phi,0}^G & \Gamma_{\phi,\phi}^G &\Gamma_{\phi,\zeta}^G \\
        \Gamma_{\zeta,0}^G & \Gamma_{\zeta,\phi}^G &\Gamma_{\zeta,\zeta}^G 
    \end{bmatrix}\!\!, 
    \Gamma^B \!= \!\!\begin{bmatrix}
        \Gamma_{0,0}^B & \Gamma_{0,\phi}^B &\Gamma_{0,\zeta}^B \\
        \Gamma_{\phi,0}^B & \Gamma_{\phi,\phi}^B &\Gamma_{\phi,\zeta}^B \\
        \Gamma_{\zeta,0}^B & \Gamma_{\zeta,\phi}^B &\Gamma_{\zeta,\zeta}^B 
    \end{bmatrix}\!\!,     
\end{equation}
where $\Gamma_{A,B}^{(\cdot)}\triangleq [\gamma_{i,j}^{(\cdot)}], \forall i\in A,\forall j\in B$.

Since the change of $I_d(t)$ depends on the state, all sequential states should be considered.
Define the \textit{state transition trace} as a sequence of states.  
Let $S(k)$ be the set of all transition traces with length $k$. 
Since at each timeslot, the state can be $G$ or $B$, we have $|S(k)| = 2^k$.
Let $S_i^k$ be the $i$-th  transition trace with length $k$, $i \in [2^k]$. 
For example, transition trace $GGGBB$ can be represented by a binary string $11100$, where 1 denotes $G$ and 0 denotes $B$. Since $(11100)_2=(28)_{10}$, we can use $S_{28}^5$ to represent transition trace $GGGBB$.
Then denote the \textit{stationary initial probability distribution of the states when $I_d(t)$ starts from zero} as $\pi^{(0)} = \begin{bmatrix}
    \pi_G^{(0)} & \pi_B^{(0)}
\end{bmatrix}_{1\times2}\triangleq\begin{bmatrix}
    \text{Pr}(a_t=G|I_d(t)=0) & \text{Pr}(a_t=B|I_d(t)=0)
\end{bmatrix}$. Note that $\pi^{(0)}$ is different from the stationary distribution of the states, i.e., $\pi =\begin{bmatrix}\frac{r}{p+r}&\frac{p}{p+r}\end{bmatrix}$ which was defined in Section \ref{section:model statement}. 
This is because $\pi$ accounts for every timeslot $t\ge 0$, while $\pi^{(0)}$ only accounts for the timeslots when the information debt initials from $I_d(t)=0$.

In order to derive $\mathbb{E}\{t_{i_0+1}-t_{i_0}\}$, we should first consider the probability of $t_{i_0+1} - t_{i_0} = k$, i.e., it takes $k$ timeslots for the information debt to start from zero and then come back.
This probability should include the occurrence of all possible state transition traces of length $k$.
Thus, we have
\begin{equation}\label{equation:summation of Pr(A)}
	\text{Pr}(t_{i_0+1} - t_{i_0} = k) =  \sum_{i\in S(k)}  \text{Pr}(t_{i_0+1} - t_{i_0} = k|S_i^k) \cdot \text{Pr}(S_i^k).
\end{equation}
Each term in the summation of (\ref{equation:summation of Pr(A)}) should be calculated. Take the transition trace $GGBBG$ with length $k=5$ as an example. 
\begin{align}
    &\text{Pr}(t_{i_0+1} - t_{i_0} = 5|GGBBG) \cdot \text{Pr}(GGBBG)\nonumber\\
    = &\begin{bmatrix}
    \Gamma_{0,\phi}^G & \Gamma_{0,\zeta}^G
\end{bmatrix} \begin{bmatrix}
    \Gamma_{\phi,\phi}^G & \Gamma_{\phi,\zeta}^G\\
    \Gamma_{\zeta,\phi}^G & \Gamma_{\zeta,\zeta}^G
\end{bmatrix}\begin{bmatrix}
    \Gamma_{\phi,\phi}^B & \Gamma_{\phi,\zeta}^B\\
    \Gamma_{\zeta,\phi}^B & \Gamma_{\zeta,\zeta}^B
\end{bmatrix} 
\begin{bmatrix}
    \Gamma_{\phi,\phi}^B & \Gamma_{\phi,\zeta}^B\\
    \Gamma_{\zeta,\phi}^B & \Gamma_{\zeta,\zeta}^B
\end{bmatrix}\begin{bmatrix}
    \Gamma_{\phi,0}^G \\
    \Gamma_{\zeta,0}^G
\end{bmatrix} \cdot \left(\pi_G^{(0)}(1-p) p (1-r) r\right).\label{equation:GGBBG}
\end{align}
Then sum over all possible lengths of the state transition traces, $\mathbb{E}\{t_{i_0+1} - t_{i_0}\}$ can be given by
\begin{equation}\label{equation:expectation of 0 to 0=k}
	\mathbb{E}\{t_{i_0+1} - t_{i_0}\} = \sum_{k=1}^\infty k \cdot \sum_{i\in S(k)}  \text{Pr}(t_{i_0+1} - t_{i_0} = k|S_i^k) \cdot \text{Pr}(S_i^k).
\end{equation}
Note that the expressions like  (\ref{equation:GGBBG}) are complex and heterogeneous for the summations (\ref{equation:summation of Pr(A)}) and  (\ref{equation:expectation of 0 to 0=k}).
We present the result of above summation (\ref{equation:summation of Pr(A)}) in the following proposition. 

\begin{prop}\label{proposition:1}
Denote $\begin{bmatrix}
    \Gamma_{0,\phi}^G & \Gamma_{0,\zeta}^G
\end{bmatrix}_{1\times\zeta}
= \Gamma_s^G, 
\begin{bmatrix}
    \Gamma_{\phi,0}^G \\
    \Gamma_{\zeta,0}^G
\end{bmatrix}_{\zeta\times1}
= \Gamma_e^G$, 
$\begin{bmatrix}
    \Gamma_{\phi,\phi}^G & \Gamma_{\phi,\zeta}^G\\
    \Gamma_{\zeta,\phi}^G & \Gamma_{\zeta,\zeta}^G
\end{bmatrix}_{\zeta\times\zeta} = Q^G$, and similarly denote $\Gamma_s^B,\Gamma_e^B$ and $Q^B$. 
Assume the stationary initial distribution of the states starting from $I_d(t)=0$, i.e.,  $\pi^{(0)}=\begin{bmatrix}
    \pi_G^{(0)} & \pi_B^{(0)}
\end{bmatrix}$ is given.
For $k=1$, $\text{Pr}(t_{i_0+1}-t_{i_0} = k)$ can be given by
\begin{equation}\label{equation:Pr(T=1)}
    \text{Pr}(t_{i_0+1}-t_{i_0} = 1) = \pi_G^{(0)}\cdot \Gamma_{0,0}^G + \pi_B^{(0)}\cdot \Gamma_{0,0}^B.
\end{equation}
For any $k\ge 2$, $\text{Pr}(t_{i_0+1}-t_{i_0} = k)$ can be characterized by 
\begin{equation}\label{equation:P(T=k)}
    \text{Pr}(t_{i_0+1}-t_{i_0} = k) \!=\! 
        \begin{bmatrix}\pi_G^{(0)} & \pi_B^{(0)}\end{bmatrix}\!\!
        \begin{bmatrix}
            \Gamma_{s}^G & \\
             & \Gamma_{s}^B
        \end{bmatrix}
        \!\!\left\{\!
        \begin{bmatrix}
            (1-p)\mathbf{I}_\zeta & p\mathbf{I}_\zeta\\
            r\mathbf{I}_\zeta & (1-r)\mathbf{I}_\zeta
        \end{bmatrix}\!\!
        \begin{bmatrix}
            Q^G & \\
             & Q^B
        \end{bmatrix}
        \!\right\}^{k-2}
        \begin{bmatrix}
            (1-p)\mathbf{I}_\zeta & p\mathbf{I}_\zeta\\
            r\mathbf{I}_\zeta & (1-r)\mathbf{I}_\zeta
        \end{bmatrix}\!
        \begin{bmatrix}
            \Gamma_{e}^G \\
            \Gamma_{e}^B
        \end{bmatrix}\!.
\end{equation}
\end{prop}

The proof of Proposition \ref{proposition:1} can be found in Appendix \ref{appendix:proposition 1}. 
Proposition \ref{proposition:1} is proved by leveraging the \textit{recursive structure} of the transition trace. 
Here we briefly introduce the idea.
Note that $\text{Pr}(t_{i_0+1}-t_{i_0} = 2)$ is obtained by summing over all possible transition traces with length $k=2$, one of which is the  transition trace $GG$. 
Also note that all transition traces with length $k=3$ can be obtained from the transition traces with length $k=2$ by further proceeding one more timeslot.
For example, transition trace $GG$ can further derive $GGG$ and $GGB$ by proceeding one more timeslot. Due to this recursive structure, the probability terms $\sum_{i\in S(k)}\text{Pr}(t_{i_0+1} - t_{i_0} = k|S_i^k) \cdot \text{Pr}(S_i^k)$ for any $k$ and $k+1$ can be related to each other. Finally we can prove Proposition \ref{proposition:1} by mathematical deduction.

One can notice that equation (\ref{equation:P(T=k)}) is in the form of the multiplies of state transition matrices and information debt transition matrices. 
Between each two terms of the information debt transition matrices, there is a state transition matrix $\mathbf{T}_{\zeta}$.
This structure actually reflects the essence of the HMC: at every timeslot, the channel state first transitions, and then the observation emission is generated depending on current state. 

Similarly, we can obtain $\text{Pr}(t_{i_0+1} - t_{i_0} = k|\nexists \tau_j \in (t_{i_0},t_{i_0+1}))$ and $\text{Pr}(\tau_{j_0} - t_{i_0} = k|t_{i_0} > \tau_{j_0-1}, t_{i_0+1} > \tau_{j_0})$, etc., which are fundamental for deriving $\mathbb{E}\{L_G\}, \mathbb{E}\{L_{B_1}\}$ and $\mathbb{E}\{L_{B_2}\}$.
The results are presented as a corollary of Proposition \ref{proposition:1} as follows.

\begin{corollary}\label{corollary:1}
    Assume that $\pi^{(0)}$ is given, similar to (\ref{equation:P(T=k)}), for $k\ge 2$, the value of $\text{Pr}(t_{i_0+1} - t_{i_0} = k|\nexists \tau_j \in (t_{i_0},t_{i_0+1})), \text{Pr}(\tau_{j_0} - t_{i_0} = k|t_{i_0} > \tau_{j_0-1}, t_{i_0+1} > \tau_{j_0})$ can be given as follows.
    \begin{align}
        &\quad\text{Pr}(t_{i_0+1} - t_{i_0} = k|\nexists \tau_j \in (t_{i_0},t_{i_0+1})) \nonumber\\
        &= \begin{bmatrix}\pi_G^{(0)} & \pi_B^{(0)}\end{bmatrix}\!\!
        \begin{bmatrix}
            \Gamma_{0,\phi}^G & \\
             & \Gamma_{0,\phi}^B
        \end{bmatrix}
        \!\!\left\{\!
        \begin{bmatrix}
            (1-p)\mathbf{I}_{\zeta-1} & p\mathbf{I}_{\zeta-1}\\
            r\mathbf{I}_{\zeta-1} & (1-r)\mathbf{I}_{\zeta-1}
        \end{bmatrix}\!\!
        \begin{bmatrix}
            \Gamma_{\phi,\phi}^G & \\
             & \Gamma_{\phi,\phi}^B
        \end{bmatrix}
        \!\right\}^{k-2}
        \begin{bmatrix}
            (1-p)\mathbf{I}_{\zeta-1} & p\mathbf{I}_{\zeta-1}\\
            r\mathbf{I}_{\zeta-1} & (1-r)\mathbf{I}_{\zeta-1}
        \end{bmatrix}\!
        \begin{bmatrix}
            \Gamma_{\phi,0}^G \\
            \Gamma_{\phi,0}^B
        \end{bmatrix}\!.\\
        &\quad\text{Pr}(\tau_{j_0} - t_{i_0} = k|t_{i_0} > \tau_{j_0-1}, t_{i_0+1} > \tau_{j_0}) \nonumber\\
        &= \begin{bmatrix}\pi_G^{(0)} & \pi_B^{(0)}\end{bmatrix}\!\!
        \begin{bmatrix}
            \Gamma_{0,\phi}^G & \\
             & \Gamma_{0,\phi}^B
        \end{bmatrix}
        \!\!\left\{\!
        \begin{bmatrix}
            (1-p)\mathbf{I}_{\zeta-1} & p\mathbf{I}_{\zeta-1}\\
            r\mathbf{I}_{\zeta-1} & (1-r)\mathbf{I}_{\zeta-1}
        \end{bmatrix}\!\!
        \begin{bmatrix}
            \Gamma_{\phi,\phi}^G & \\
             & \Gamma_{\phi,\phi}^B
        \end{bmatrix}
        \!\right\}^{k-2} \begin{bmatrix}
            (1-p)\mathbf{I}_{\zeta-1} & p\mathbf{I}_{\zeta-1}\\
            r\mathbf{I}_{\zeta-1} & (1-r)\mathbf{I}_{\zeta-1}
        \end{bmatrix}\!
        \begin{bmatrix}
            \Gamma_{\phi,\zeta}^G \\
            \Gamma_{\phi,\zeta}^B
        \end{bmatrix}\!.
    \end{align}
\end{corollary}

On the other hand, $\pi^{(0)}$, assumed to be given in Proposition \ref{proposition:1}, could also significantly influence the outcome of (\ref{equation:Pr(T=1)}) and (\ref{equation:P(T=k)}).
To derive $\pi^{(0)}$, we first denote the \textit{transition matrix of the initial probability distribution of the states between any two adjacent times that $I_d(t)$ hits zero} as $T_{0\rightarrow 0}$.
Specifically, $\forall l\ge 1$, let $\pi^{(l)}$ be the probability distribution of the states at timeslot $t_l$, where $t_l$ is the $l$-th time $I_d(t)$ hits zero, then the probability distribution of the states at timeslot $t_{l+1}$, i.e., $\pi^{(l+1)}$ equals to $\pi^{(l)} \cdot T_{0\rightarrow 0}$.
Moreover, when the Markov chain becomes stationary, we have $\pi^{(0)} = \pi^{(0)} \cdot T_{0\rightarrow 0}$. 
Then we derive $T_{0\rightarrow 0}$ and the value of $\pi^{(0)}$ in the following proposition. 

\begin{prop}\label{proposition:2}
Denote $\begin{bmatrix}
    \Gamma_{0,0}^G & \\
     & \Gamma_{0,0}^B
\end{bmatrix}_{2\times2} = \mathbf{\Gamma}_{0,0}, 
\begin{bmatrix}
    \Gamma_{s}^G & \\
     & \Gamma_{s}^B
\end{bmatrix}_{2\times2\zeta} = \mathbf{\Gamma}_{s}, 
\begin{bmatrix}
    \Gamma_{e}^G & \\
     & \Gamma_{e}^B
\end{bmatrix}_{2\zeta\times2} = \mathbf{\Gamma}_{e},
\begin{bmatrix}
    Q^G & \\
     & Q^B 
\end{bmatrix}_{2\zeta\times2\zeta} = \mathbf{Q}$ and recall that $\mathbf{T}_{n} = \begin{bmatrix}
    (1-p)\mathbf{I}_n & p\mathbf{I}_n\\
    r\mathbf{I}_n & (1-r)\mathbf{I}_n
\end{bmatrix}$. 
Then $\pi^{(0)}=\begin{bmatrix}
    \pi_G^{(0)} & \pi_B^{(0)}
\end{bmatrix}$ is the solution of the following equations:
\begin{equation}\label{equation:solution of initial}
    \begin{bmatrix}
    (T_{0\rightarrow 0} - \mathbf{I}_2)^\top \\
     \begin{array}{cc}
        1  & 1 
     \end{array}
\end{bmatrix}_{3\times2} \cdot 
\begin{bmatrix}
    \pi_G^{(0)} \\
    \pi_B^{(0)}
\end{bmatrix}_{2\times1}
=\begin{bmatrix}
    0 \\
    0 \\
    1
\end{bmatrix}_{3\times1},
\end{equation}
where 
\begin{align}
    T_{0\rightarrow 0} = \big[\mathbf{\Gamma}_{0,0} + \mathbf{\Gamma}_{s}(\mathbf{I}_{2\zeta}-\mathbf{T}_{\zeta}\mathbf{Q})^{-1}\mathbf{T}_{\zeta}\mathbf{\Gamma}_{e}\big]\mathbf{T}_1.
\end{align}
\end{prop}

\begin{remark}
    Note that although there are three equations in (\ref{equation:solution of initial}), only two out of them are actually effective for the solution.
    This is because the first two equations are linearly dependent.
    Specifically, $T_{0\rightarrow 0}$ is a stochastic matrix in form of $\begin{bmatrix}
        1-a & a \\
        b & 1-b 
    \end{bmatrix}$, where $a,b\in(0,1)$. 
    Therefore, $(T_{0\rightarrow 0} - \mathbf{I}_2)^\top$ should be in form of $\begin{bmatrix}
        -a & b \\
        a & -b 
    \end{bmatrix}$, where the second equation is redundant.
\end{remark}   

The proof of Proposition \ref{proposition:2} can be found in Appendix \ref{appendix:proposition 2}. 
Proposition \ref{proposition:2} can be  proved utilizing Proposition \ref{proposition:1}.
According to Proposition \ref{proposition:2}, the value of $\pi^{(0)}$ can be calculated by solving the linear equations (\ref{equation:solution of initial}).
With Proposition \ref{proposition:1} and \ref{proposition:2}, we are ready to characterize the exact expressions of the terms defined in Lemma \ref{lemma:1}.

\begin{lemma}\label{lemma:2}
    Denote $\begin{bmatrix}
    \Gamma_{0,\phi}^G & \\
     & \Gamma_{0,\phi}^B
\end{bmatrix}_{2\times(2\zeta-2)} = \mathbf{\Gamma}_{0,\phi},\begin{bmatrix}
    \Gamma_{0,\zeta}^G & \\
     & \Gamma_{0,\zeta}^B
\end{bmatrix}_{2\times2} = \mathbf{\Gamma}_{0,\zeta}$, $\begin{bmatrix}
    \Gamma_{\phi,\phi}^G & \\
     & \Gamma_{\phi,\phi}^B 
\end{bmatrix}_{(2\zeta-2)\times(2\zeta-2)} = \mathbf{\Gamma}_{\phi,\phi}$ and similarly denote $\mathbf{\Gamma}_{\phi,\zeta},\mathbf{\Gamma}_{\zeta,\zeta},\mathbf{\Gamma}_{\zeta,\phi}$. 
Denote $(\mathbf{I}_{2\zeta-2} - \mathbf{T}_{\zeta-1}\mathbf{\Gamma}_{\phi,\phi})^{-1} = \mathbf{M}$. Let $\psi = (\Delta - \alpha - 1)^+$.
Then the terms defined in Lemma \ref{lemma:1} can be characterized by equation (\ref{equation:E(A)}) to (\ref{equation:E(LB2)}) as follows:
    \begin{equation}\label{equation:E(A)}
		\mathbb{E}\{t_{i_0+1} - t_{i_0}\} = 1 + \pi^{(0)}\mathbf{\Gamma}_s(\mathbf{I}_{2\zeta}-\mathbf{T}_{\zeta}\mathbf{Q})^{-1}\cdot\vec{\mathbf{1}}_{2\zeta},
    \end{equation}  
    \begin{align}\label{equation:E(LG)}
            \mathbb{E}\{L_G\} = 
            \pi^{(0)}
            \mathbf{\Gamma}_{0,\phi}
        \mathbf{M}\cdot(\mathbf{T}_{\zeta-1}\mathbf{\Gamma}_{\phi,\phi})^{\Delta}\cdot \mathbf{T}_{\zeta-1}
            \begin{bmatrix}
                \Gamma_{\phi,0}^G \\
                \Gamma_{\phi,0}^B
            \end{bmatrix},
    \end{align}
    \begin{align}\label{equation:E(LB1)}
        \mathbb{E}\{L_{B_{1}}\} &= \pi^{(0)} \cdot \left[T_{0\rightarrow \zeta} \cdot (\mathbf{I}_2-T_{\zeta\rightarrow\zeta})^{-1}\cdot \Vec{m}+\Vec{n}\right],
    \end{align}
    \begin{align}\label{equation:E(LB2)}
        \mathbb{E}\{L_{B_{2}}\} = \pi^{(0)}\cdot T_{0\rightarrow \zeta} \cdot (\mathbf{I}_2-T_{\zeta\rightarrow\zeta})^{-1}\cdot \Vec{b},
    \end{align}
where $\vec{\mathbf{1}}_{2\zeta}$ is a column vector with $2\zeta$ 1s, 
\begin{align}
       T_{0\rightarrow \zeta} = \big[\mathbf{\Gamma}_{0,\zeta} 
        + \mathbf{\Gamma}_{0,\phi} \mathbf{M}\mathbf{T}_{\zeta-1}\cdot\mathbf{\Gamma}_{\phi,\zeta} \big]
        \mathbf{T}_{1},
\end{align}
\begin{align}
       T_{\zeta\rightarrow \zeta} = \big[\mathbf{\Gamma}_{\zeta,\zeta} 
        + \mathbf{\Gamma}_{\zeta,\phi} \mathbf{M}\mathbf{T}_{\zeta-1}\cdot\mathbf{\Gamma}_{\phi,\zeta}\big]
        \mathbf{T}_{1}
\end{align}
are the transition matrices of the initial probability distribution, and vectors $\Vec{m},\Vec{n},\Vec{b}$ are defined as follows:
\begin{align}
        \Vec{m} = \begin{bmatrix}
            \Gamma_{\zeta,\zeta}^G \\
            \Gamma_{\zeta,\zeta}^B\end{bmatrix} 
        +\mathbf{\Gamma}_{\zeta,\phi} 
  (\mathbf{I}_{2\zeta-2}+\mathbf{M})\mathbf{M} \mathbf{T}_{\zeta-1} \begin{bmatrix}
            \Gamma_{\phi,\zeta}^G \\
            \Gamma_{\phi,\zeta}^B
        \end{bmatrix},
\end{align}
\begin{align}        
        \Vec{n} = \begin{bmatrix}
            \Gamma_{0,\zeta}^G \\
            \Gamma_{0,\zeta}^B\end{bmatrix} 
        +\mathbf{\Gamma}_{0,\phi}
  (\mathbf{I}_{2\zeta-2}+\mathbf{M})\mathbf{M}\mathbf{T}_{\zeta-1} \begin{bmatrix}
            \Gamma_{\phi,\zeta}^G \\
            \Gamma_{\phi,\zeta}^B
        \end{bmatrix},
\end{align}
\begin{align}\label{equation:b}
        \Vec{b} = -\min\{\Delta, \alpha\}\begin{bmatrix}
            \Gamma_{\zeta,0}^G \\
            \Gamma_{\zeta,0}^B\end{bmatrix} 
        - \alpha \mathbf{\Gamma}_{\zeta,\phi} \mathbf{M} \mathbf{T}_{\zeta-1} \begin{bmatrix}
            \Gamma_{\phi,0}^G \\
            \Gamma_{\phi,0}^B
        \end{bmatrix}+\mathbf{\Gamma}_{\zeta,\phi}  
        \cdot\left[\mathbf{M}^2+(\alpha+\psi-\Delta)\mathbf{M}\right]\!(\mathbf{T}_{\zeta-1}\mathbf{\Gamma}_{\phi,\phi})^{\psi}\mathbf{T}_{\zeta-1}\begin{bmatrix}
            \Gamma_{\phi,0}^G \\
            \Gamma_{\phi,0}^B
        \end{bmatrix}\!.
    \end{align}
\end{lemma}

The proof of Lemma \ref{lemma:2} is much more involved and technical.
Given Proposition \ref{proposition:1} and \ref{proposition:2}, the expressions of the terms $\mathbb{E}\{t_{i_0+1} - t_{i_0}\}$ and $\mathbb{E}\{L_G\}$ can be directly derived by averaging over all possible state transition traces with all possible lengths as in (\ref{equation:expectation of 0 to 0=k}), which could be similar to the procedure in \cite{RLSCs}.
However, the characterization of the terms $\mathbb{E}\{L_{B_1}\}$ and $\mathbb{E}\{L_{B_2}\}$ are significantly different. 
This is primarily because $L_{B_1}$ and $L_{B_2}$ involve $\tau_{j^*}$, the last time $I_d(t)$ hits $\zeta$ before hitting zero, which is not a stopping time. In \cite{RLSCs}, the \textit{recursive equation} is leveraged to deal with this problem in the i.i.d channel. Nonetheless, the recursive equation no more holds in G-ESEC, due to the varying probability distribution of the states when each time $I_d(t)$ hits $\zeta$ before hitting zero.
In this work, we derive \textit{iterative equation} to accommodate to the varying initial distribution of the states.
The proof of Lemma \ref{lemma:2} can be found in Appendix \ref{appendix:lemma 2}.

With Lemma \ref{lemma:1} and Lemma \ref{lemma:2}, the derivation of the exact error probability of NRLSCs in G-ESEC is straightforward and can be given by the following theorem.

\begin{theorem}\label{theorem:1}
    In Gilbert-Elliott channel with state transition matrix $\mathbf{T}_1 = \begin{bmatrix}
    1-p & p\\
    r & 1-r
\end{bmatrix}$ and information debt transition matrix $\Gamma^G,\Gamma^B$, for any finite memory length $\alpha$ and decoding delay $\Delta$, the exact slot error probability of large-field-size NRLSCs $p_e^{ns}$ can be computed by assembling Lemma \ref{lemma:1} and \ref{lemma:2}.
\end{theorem}

\begin{remark}
    When $\Gamma^G=\Gamma^B$ or $p=r=0$, implying that only one hidden state is effective, the $p_e$ of Theorem \ref{theorem:1} for G-ESEC will degenerated into the $p_e$ for i.i.d. SEC in \cite[Theorem 1]{RLSCs}.
\end{remark}

\begin{corollary}
    All the results regarding the error probability of NRLSCs in G-ESEC (including Proposition \ref{proposition:1},\ref{proposition:2}, Corollary \ref{corollary:1} and Lemma \ref{lemma:2}, Theorem \ref{theorem:1}) can be further generalized into any HMC with $L\ge 2$ hidden states.
\end{corollary}

When the hidden Markov channel with $L$ hidden states is considered, there will be $L$ distinct observation emission matrices of the information debt, i.e., $\Gamma^1$ to $\Gamma^L$. Note that the recursive structure of transition trace still holds. 
Thus the error probability of NRLSCs can be also characterized into a closed-form expression in the HMC. 
Since this generalization is quite straightforward, the corresponding proof is omitted. 
To provide a basic concept for the readers, the generalized form of equation (\ref{equation:P(T=k)}) in HMC is given as follows: Let $\begin{bmatrix}
    \Gamma_{0,\phi}^l & \Gamma_{0,\zeta}^l
\end{bmatrix}
= \Gamma_s^l, 
\begin{bmatrix}
    \Gamma_{\phi,0}^l \\
    \Gamma_{\zeta,0}^l
\end{bmatrix}
= \Gamma_e^l$, and 
$\begin{bmatrix}
    \Gamma_{\phi,\phi}^l & \Gamma_{\phi,\zeta}^l\\
    \Gamma_{\zeta,\phi}^l & \Gamma_{\zeta,\zeta}^l
\end{bmatrix} = Q^l$, $\forall l \in [1,L]$.
Assume that the transition matrix of the hidden states is $\begin{bmatrix}
    t_{11} & \cdots  & t_{1L}\\
    \vdots & \ddots &\vdots \\
    t_{L1} & \cdots & t_{LL}
\end{bmatrix}_{L\times L}$.
For $k\ge 2$, we have
\begin{equation}
    \text{Pr}(t_{i_0+1}-t_{i_0} = k) \!=\! 
\begin{bmatrix}\pi_1^{(0)} & \cdots &\pi_L^{(0)} \end{bmatrix}\!\!
\begin{bmatrix}
    \Gamma_{s}^1\! & \!& \!\\
    \! & \ddots \!& \!\\
    \!& \!&\Gamma_{s}^L\!
\end{bmatrix}\!\!
\left\{\!\!
\begin{bmatrix}
    t_{11}\mathbf{I}_\zeta & \cdots  & t_{1L}\mathbf{I}_\zeta\\
    \vdots & \ddots &\vdots \\
    t_{L1}\mathbf{I}_\zeta & \cdots & t_{LL}\mathbf{I}_\zeta
\end{bmatrix}\!\!\!
\begin{bmatrix}
    Q^1\! & \!& \!\\
    \! & \ddots\! & \!\\
   \! & \!&Q^L\!
\end{bmatrix}\!\!
\right\}^{k-2}\!\!
\begin{bmatrix}
    t_{11}\mathbf{I}_\zeta\! & \cdots\!  & t_{1L}\mathbf{I}_\zeta\!\\
    \vdots\! & \ddots\! &\vdots\! \\
    t_{L1}\mathbf{I}_\zeta\! & \cdots\! & t_{LL}\mathbf{I}_\zeta\!
\end{bmatrix}\!\!
\begin{bmatrix}
    \Gamma_{e}^1 \\
    \vdots\\
    \Gamma_{e}^L
\end{bmatrix}\!\!.
\end{equation}

\section{The Analysis of Systematic RLSCs in Packet Erasure Channel}\label{section:Main Results2}

In this section, we present our second result, i.e., the analysis of SRLSCs in PEC. 

\subsection{Introduction on SRLSCs and the Packet Erasure Channel}

First, we narrow our discussion into the PEC. 

\textbf{Packet erasure channel:} 
In every timeslot $t$, the $N$ encoded symbols $\mathbf{x}(t)$ will be transmitted into the channel and is assumed to be either all received perfectly or all erased. 
Denote the indicator function of erasure as $e(t)$. $e(t) = 1$ if the erasure occurs at timeslot $t$ and $e(t) = 0$ if the packet is delivered successfully.
Denote the received symbols at timeslot $t$ as $\mathbf{y}(t)$, where
\begin{equation}
    \mathbf{y}(t) = \left\{
    \begin{aligned}
        [x_1(t),&\dots,x_N(t)]^\top \quad \text{if} \  e(t) = 0\\
        &* \qquad\qquad\quad\  \text{if} \ e(t) = 1,
    \end{aligned}
    \right
    .
\end{equation}
and $*$ stands for the erased symbol.

We refer to this model as the ``\textit{packet erasure channel} (PEC)" thereafter, since the $N$ symbols will have the same erasure behavior as they are in a packet.
The difference between SEC and PEC is that, the basic element of erasure in SEC is one encoded symbol, while that in PEC is a packet, which consists of $N$ encoded symbols.
Therefore, PEC can be regarded as a special SEC with bigger granularity. 
This scenario is practical and has been widely considered in many works on the streaming codes in erasure channels \cite{bursty only,b1,b2,b3,bn1,bn2,bn3,adversarial channel,SLFTCOM,SLFTIT,variable size,simple1,simple2}. 

\begin{figure}[!hbtp]
    \centering
    \includegraphics[width=0.5\linewidth]{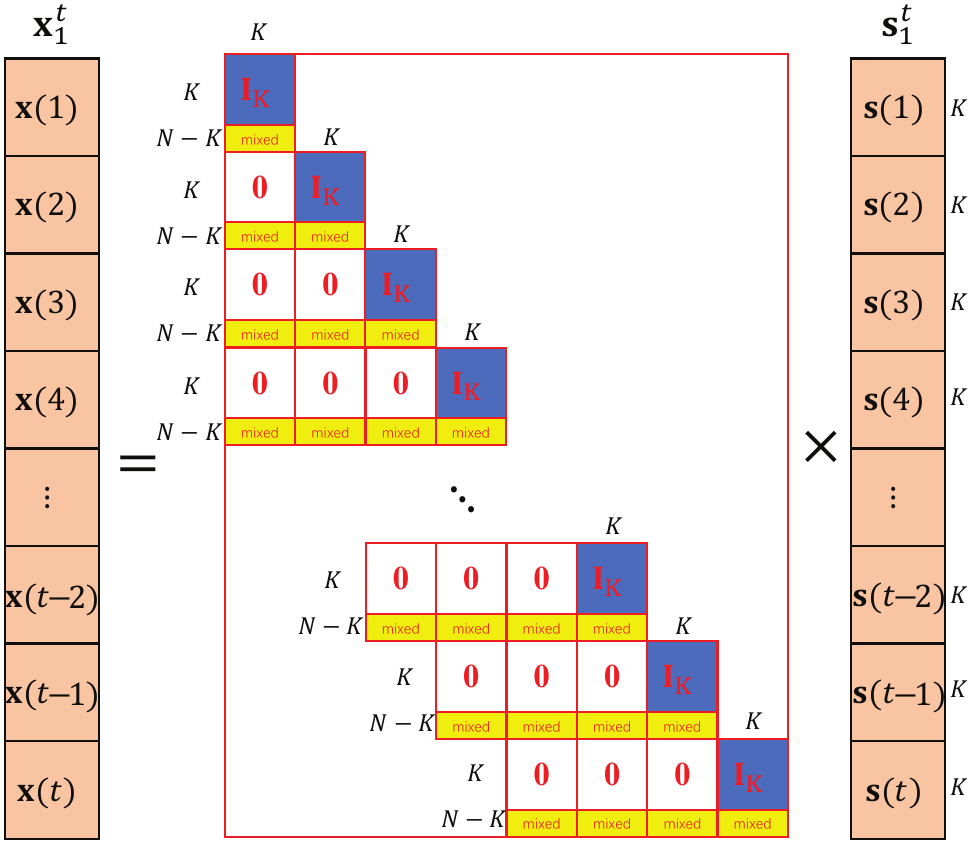}
    \caption{Illustration of $\mathbf{x}_1^t = \mathbf{G}^{(t)} \mathbf{s}_1^t$ in systematic RLSCs with $\alpha=3$.}
    \label{sys_1}
\end{figure}

Then we briefly introduce the SRLSCs.
In SRLSCs, the first $K$ symbols of $\mathbf{x}(t)$ are uncoded source symbols $\mathbf{s}(t)$, while the rest $N-K$ parity symbols are linear combinations of $(\alpha+1)K$ symbols from the present and previous timeslots. 
An illustration of SRLSCs can be found in Fig. \ref{sys_1}.
In other words, the generator matrix $\mathbf{G}_t$ admits the following three features:
(1) The upper left corner of $\mathbf{G}_t$ is a $K$-by-$\alpha K$ zero matrix (the white blocks in Fig. \ref{sys_1}); (2) the upper right corner of $\mathbf{G}_t$ is a $K$-by-$K$ identity matrix $\mathbf{I}_K$ (the blue blocks in Fig. \ref{sys_1});
(3) The entries of the lower $(N-K)$-by-$(\alpha+1) K$ parity check matrix of $\mathbf{G}_t$ (the yellow blocks in Fig. \ref{sys_1}), are chosen uniformly and randomly from GF($2^q$), excluding 0. 

Recall the error event for NRLSCs in SEC in Proposition \ref{Proposition:characterization of the error event}. 
The characterization indicates that NRLSCs should wait $I_d(t)$ to hit zero for any decode.
However, for SRLSCs in PEC, for any timeslot $t$ if the $N$ transmitted encoded symbols $\mathbf{x}(t)$ are all received perfectly, the destination can instantly decode $\mathbf{s}(t)$ from the first $K$ uncoded symbols of $\mathbf{x}(t)$, without waiting for the $I_d(t)$ to hit zero. 
Compared to the NRLSCs, this feature could save some source symbols which should have exceeded the decoding delay $\Delta$, and thus could significantly reduce the error probability in some delay-sensitive settings. 
This feature is referred to as the \textit{instant decodability} of SRLSCs herein.
Moreover, Gaussian elimination will be executed to eliminate the impact of the successfully delivered symbols on the erased symbols (remaining unknowns). 
Due to the instant decodability, compared to NRLSCs, it is used to be thought that SRLSCs can save the symbols beyond the decoding delay $\Delta$ and thus can decrease $p_e$ without any cost. 
However, the example presented in the next subsection shows that sometimes SRLSCs can actually lead to unexpected error events.

\subsection{Characterization on the Error Event of SRLSCs in PEC}

We start with a counter-intuitive phenomenon that SRLSCs can actually lead to unexpected errors comparing to NRLSCs for some erasure patterns.

\begin{example}\label{example:1}
    \begin{figure}
        \centering
        \subfigure[Illustration of $\mathbf{G}^{(8)}$.]{
            \begin{minipage}[b]{0.318\textwidth}
            \includegraphics[width=1\textwidth]{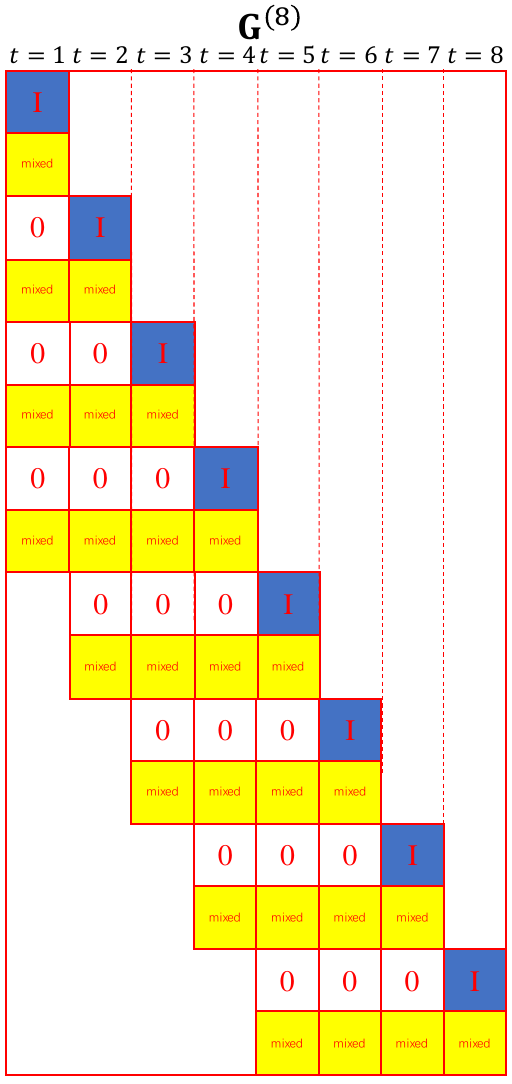} 
            \end{minipage}
        }
        \subfigure[Illustration of $\mathbf{H}^{(8)}$ and the linear equations after eliminating the delivered symbols.]{
            \begin{minipage}[b]{0.4\textwidth}
            \includegraphics[width=1\textwidth]{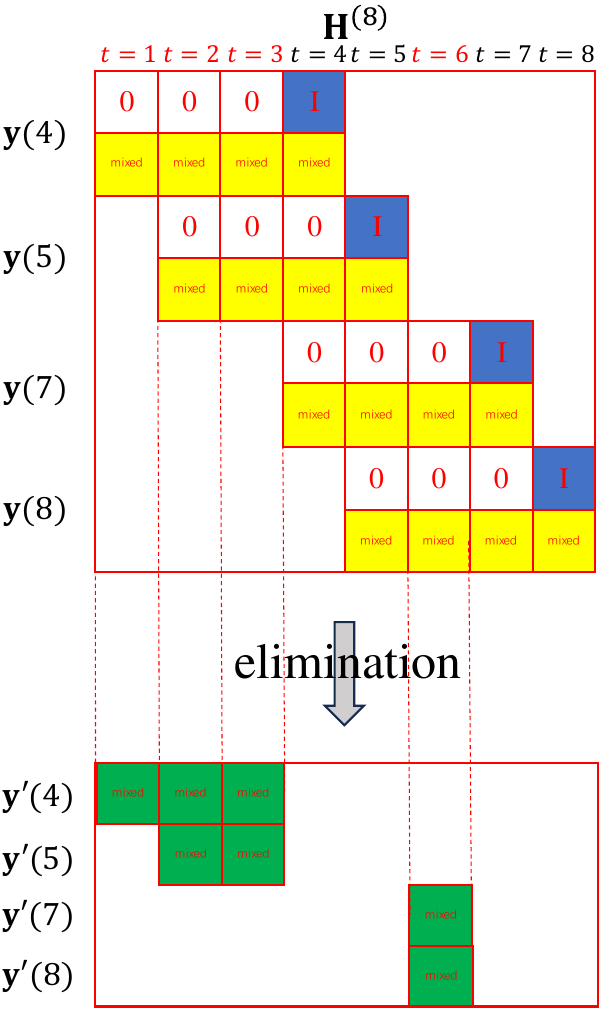} 
            \end{minipage}
        }
        \caption{An example for the decodability of systematic RLSCs with $\alpha=3$.}
        \label{Figure:sys_decodability}
    \end{figure}
    Assume that $K=N-K$
    , $\alpha = 3$ and $\Delta = 6$.
    Consider 8 consecutive timeslots of transmissions.
    The cumulative generator matrix $\mathbf{G}^{(8)}$ is illustrated in Fig. \ref{Figure:sys_decodability} (a).
    Construct the following erasure pattern that 
    $\mathbf{x}(1),\mathbf{x}(2),\mathbf{x}(3)$ and $\mathbf{x}(6)$ are erased, while $\mathbf{x}(4),\mathbf{x}(5),\mathbf{x}(7),\mathbf{x}(8)$ are received successfully. 
    The cumulative receiver matrix $\mathbf{H}^{(8)}$ is illustrated in the upper of Fig. \ref{Figure:sys_decodability} (b).
    Since that $\mathbf{y}(4),\mathbf{y}(5),\mathbf{y}(7),\mathbf{y}(8)$ are successfully received, the decoder can directly obtain $\mathbf{x}(4),\mathbf{x}(5),\mathbf{x}(7),\mathbf{x}(8)$ from the first $K$ symbols of the corresponding packets. 
    Then, the decoder will execute Gaussian elimination to eliminate the impact of the delivered symbols $\mathbf{x}(4),\mathbf{x}(5),\mathbf{x}(7),\mathbf{x}(8)$ on the received combinations $\mathbf{y}(4),\mathbf{y}(5),\mathbf{y}(7),\mathbf{y}(8)$, which contain the information of the remaining unknowns $\mathbf{s}(1),\mathbf{s}(2),\mathbf{s}(3)$ and $\mathbf{s}(6)$.
    We use $\mathbf{y}'(t)$ to represent the linear equations after elimination, which are illustrated by the green blocks in the lower of Fig. \ref{Figure:sys_decodability} (b).
    It seems that we have $4K$ unknowns $\mathbf{s}(1),\mathbf{s}(2),\mathbf{s}(3)$ and $\mathbf{s}(6)$ with $4K$ linear equations $\mathbf{y}'(4),\mathbf{y}'(5),\mathbf{y}'(7)$ and $\mathbf{y'}(8)$.
    However, one can also notice that $\mathbf{y}'(7)$ and $\mathbf{y'}(8)$ are both only related to the same unknown $\mathbf{s}(6)$ and thus the latter one $\mathbf{y'}(8)$ is redundant. 
    This redundancy also indirectly leads to the decoding failure of $\mathbf{s}(1),\mathbf{s}(2),\mathbf{s}(3)$, since there are only $2K$ equations $\mathbf{y}'(4),\mathbf{y}'(5)$ for $3K$ unknowns $\mathbf{s}(1),\mathbf{s}(2),\mathbf{s}(3)$. 
    Therefore, there appears $3K$ errors out of $8K$ symbols for SRLSCs in this erasure pattern.

    On the contrary, if for NRLSCs, there will be $8K$ unknowns with $8K$ equations, thus all $8K$ source symbols will be decodable. 
    This can be easily verified by calculating the information debt, which shows that $I_d(t)$ is always smaller than $\zeta = (\alpha+1)K=4K$ for $t\in[1,8]$. 
    Further take the decoding delay $\Delta=6$ into account, one can notice that there are only $2K$ errors (exceed the decoding delay) out of $8K$ symbols for NRLSCs in this erasure pattern.

    Therefore, in this example, SRLSCs will cause unexpected decoding failure, and thus lead to a larger error rate than NRLSCs.
    It also reflects that the characterization of the error events for NRLSCs in \cite{RLSCs} is no longer available for SRLSCs.
\end{example}

\textit{Discussion on Example \ref{example:1}}
\begin{itemize}
    \item \textbf{The underlying cause of the phenomenon.} The underlying cause is the de-correlation between the preceding and the following unknowns, given the information of intermediate deliveries.
In Example \ref{example:1}, $\mathbf{s}(1)$ to $\mathbf{s}(8)$ are correlated with each other according to the linear combinations $\mathbf{y}(4),\mathbf{y}(5),\mathbf{y}(7)$ and $\mathbf{y}(8)$.
Note that $\mathbf{s}(4),\mathbf{s}(5),\mathbf{s}(7)$ can be instantly decoded from the first $K$ symbols of $\mathbf{y}(4),\mathbf{y}(5),\mathbf{y}(7)$, respectively.
However, when $\mathbf{s}(4),\mathbf{s}(5),\mathbf{s}(7)$ are given, we have 
\begin{equation}
    H\Big(\mathbf{s}(6)|\mathbf{s}(4),\mathbf{s}(5),\mathbf{s}(7),\mathbf{y}(7)\Big)=0,
\end{equation}
which means $\mathbf{s}(6)$ can be directly decoded from linear combination $\mathbf{y}(7)$, given the values of unknowns $\mathbf{s}(4),\mathbf{s}(5),\mathbf{s}(7)$. 
Therefore, we have 
\begin{equation}
    I\Big(\mathbf{s}(1),\mathbf{s}(2),\mathbf{s}(3);\mathbf{s}(6)|\mathbf{s}(4),\mathbf{s}(5),\mathbf{s}(7),\mathbf{y}(7)\Big)=0,
\end{equation}
which means $\mathbf{s}(1),\mathbf{s}(2),\mathbf{s}(3)$ are actually de-correlated with $\mathbf{s}(6)$ (and also all the symbols thereafter), given $\mathbf{s}(4),\mathbf{s}(5),\mathbf{s}(7)$. 

    \item \textbf{The direct cause of the phenomenon.} The direct cause is that $\zeta$ will be no more constantly equal to $\alpha K + 1$ in SRLSCs. Recall that $\zeta-1=\alpha K$ represents the maximum allowable number of unknowns in NRLSCs.  However, in SRLSCs, due to the instant decodability, \textit{the symbols have been decoded instantly from the successfully deliveries should be deducted from the number}. This thus leads to a smaller value of $\zeta$, which will even change over time. In SRLSCs, we denote the time-varying maximum value of information debt as $\zeta(t)$. Since the instantly decoded symbols are deducted from the number, we have $\zeta(t) \le \alpha K + 1$.
\end{itemize}

To formulate $\zeta(t)$ and characterize the error event of SRLSCs in PEC, we modify Definition \ref{definition:informtion debt old} and \ref{definition:hitting time old} as follows.

\begin{definition}\label{definition:informtion debt new}
    Let $\zeta(0) = 1,I_d(0)=0,t_c = 0$ and $e(0)=0$. For any $t\ge 1$, the information debt $I_d(t)$ of SRLSCs in PEC is calculated iteratively by 
    \begin{align}
        \hat{I}_d(t) &\triangleq \left[K-N\big(1-e(t)\big) +\min(I_d(t-1),\zeta(t-1)-1\right]^+\label{equation:11}\\
        \zeta(t) &\triangleq \left[K\cdot e(t) - K\cdot e(\max(t-\alpha,t_c)) + \zeta(t-1)\right] \cdot \mathds{1}\{\hat{I}_d(t)\neq0\} + 1\cdot \mathds{1}\{\hat{I}_d(t)=0\}\label{equation:12}\\
        I_d(t) &\triangleq \min\left(\zeta(t),\hat{I}_d(t)\right) \label{equation:13}\\
        t_c &\triangleq t \cdot \mathds{1}\{I_d(t)=0\} + t_c \cdot \mathds{1}\{I_d(t)\neq0\}.\label{equation:14}
    \end{align}
\end{definition}
\begin{definition}\label{definition:hitting time new}
    Let $t_0\triangleq0$ and $\tau_0\triangleq0$ and define iteratively 
    \begin{align}
        t_i &\triangleq \inf \{t':t'>t_{i-1}, I_d(t')=0\}\label{equation:ti_new}\\
        \tau_j &\triangleq \inf \{t':t'>\tau_{j-1}, I_d(t')=\zeta(t')\}\label{equation:tau_j new}
    \end{align}
    as the $i$-th and $j$-th time that $I_d(t)$ hits 0 and $\zeta(t)$, respectively.
\end{definition}

For simplicity, we use the same notations $I_d(t),t_i,\tau_j$, etc. for both SRLSCs and NRLSCs, which slightly abuses the notation. 
We note that they are actually different. 
Equations (\ref{equation:11}) and (\ref{equation:13}) have the same physical meaning as (\ref{equation:7}) and (\ref{equation:8}) respectively, while (\ref{equation:12}) and (\ref{equation:14}) are a bit different.
Roughly speaking, (\ref{equation:11}) represents the transition of the temporary information debt $\hat{I}_d(t)$ due to the erasure $e(t)$, while (\ref{equation:13}) is to see whether the temporary information debt $\hat{I}_d(t)$ has exceeded the maximum allowable value $\zeta(t)-1$ or not.
$t_c$ represents the latest timeslot that $I_d(t)$ hits zero, and equation (\ref{equation:14}) stands for the update of $t_c$.
$\zeta(t)-1$, which accounts for the maximum allowable number of unknowns at timeslot $t$, is calculated as the number of erasures in the latest (at most) $\alpha$ timeslots after $t_c$. 
The iterative form of this argument gives the first term of (\ref{equation:12}).
The second term of (\ref{equation:12}) indicates that when each time the temporary information debt $\hat{I}_d(t)$ hits zero, $\zeta(t)-1$ will be reset to 0 accordingly, since it is the time for the decoding process of the previous decodable symbols.

With the new definitions, the error event of SRLSCs in PEC can be characterized as follows.

\begin{prop}\label{proposition:error event in SRLSCs}
Assume \textbf{GMDS} holds\footnote{Recall that we have assumed \textbf{GMDS} for NRLSCs. In SRLSCs, for each $\mathbf{x}(t)$, since the first $K$ uncoded symbols are only effective for decoding $\mathbf{s}(t)$ instantly, we assume \textbf{GMDS} only for the cumulative parity check matrices (the yellow blocks in Fig. \ref{sys_1}).}. 
For SRLSCs in PEC, for any fixed index $i_0 \ge 0$, \textbf{(a)} if $\nexists \tau_j \in (t_{i_0}, t_{i_0+1})$, then $\mathbf{s}(t)$ is not $\Delta$-decodable 
\begin{equation}\label{equation:e(t)1}
    \forall t \in \left\{t'|t'\in(t_{i_0}, t_{i_0+1} - \Delta),e(t')=1\right\},
\end{equation}
\textbf{(b)} if $\exists \tau_j \in (t_{i_0}, t_{i_0+1})$, let $\tau_{j^*}$ be the one with the largest $j$, then $\mathbf{s}(t)$ is not $\Delta$-decodable 
\begin{equation}\label{equation:e(t)2}
\forall t \in \left\{t'|t'\!\in\!\left(t_{i_0}, \max(\tau_{j^*} \!-\! \alpha + 1, t_{i_0+1} \!-\! \Delta)\right),e(t')\!=\!1\right\},
\end{equation}
\textbf{(c)} for the rest of $t$, $\mathbf{s}(t)$ is $\Delta$-decodable.
\end{prop}

Proposition \ref{proposition:error event in SRLSCs} directly holds with the proof of \textbf{Proposition 3} in \cite{RLSCs} and the Definitions \ref{definition:informtion debt new} and \ref{definition:hitting time new} stated above. Since the proof of Proposition \ref{proposition:error event in SRLSCs} is straightforward and highly similar to the proof of \textbf{Proposition 3} in \cite{RLSCs}, it is omitted in this paper. Note that the form of the error regime for SRLSCs in Proposition \ref{proposition:error event in SRLSCs} is similar to that for NRLSCs in Proposition \ref{Proposition:characterization of the error event}. The differences on the error event and their impact on the error probabilities are discussed as follows. 

\textit{Discussion on differences of the error event and error probability between SRLSCs and NRLSCs}
\begin{itemize}
    \item \textbf{Instant decodability.} Note from (\ref{equation:e(t)1}), (\ref{equation:e(t)2}), the error regime of SRLSCs is only comprised of the timeslots when the erasure appears, i.e., $e(t) = 1$. This is because the instant decodability excludes the timeslots with perfect delivery. The gain of instant decodability primarily depends on the channel erasure rate. When the erasure rate is higher, the gain becomes more prominent. 
    Also note that the gain of instant decodability will not necessarily reduce the error probability compared to NRLSCs. To account for the impact on error probability, both the instant decodability and also the different hitting times, which will be discussed in the next bullet point, should be jointly considered.
    
    \item \textbf{Different hitting times.} Since the calculation of $I_d(t)$ and $\zeta(t)$ in SRLSCs are different from that in NRLSCs, the hitting time sequences $t_i$ and $\tau_j$ are also different, which will lead to distinct error regimes. It is worthy noting that the impact of the different hitting times may either increase or decrease the error probability compared to NRLSCs, and the impact primarily depends on the parameters $\Delta,\alpha$ and the channel stochastics. Although SRLSCs have a smaller $\zeta(t)$ than NRLSCs, which may intuitively lead to a higher probability for $I_d(t)$ to hit $\zeta(t)$, the error probability of SRLSCs is however not necessarily higher than NRLSCs, even regardless of the gain produced by the instant decodability. This can be interpreted as follows. A higher probability for $I_d(t)$ to hit $\zeta(t)$ can indirectly indicate a smaller value of $\mathbb{E}\{t_{i_0+1} - t_{i_0}\}$. This is because at each time $\hat{I}_d(t)$ exceeds $\zeta(t)$, some previous symbols will be abandoned due to their undecodability and the information debt will be thereby reduced to $\zeta(t)-1$, which can make $I_d(t)$ even closer to 0. Therefore, in an average sense, it could take a shorter time for $I_d(t)$ to hit zero again, which implies a smaller $\mathbb{E}\{t_{i_0+1} - t_{i_0}\}$ in SRLSCs. With this argument, symbols sent in the timeslots that are close to $t_{i_0}$ are less likely to exceed the decoding delay $\Delta$. Therefore, in delay-sensitive scenarios where $\Delta$ is small, the different hitting times of SRLSCs can individually lead to a lower error probability than NRLSCs, even regardless of the instant decodability. In other cases, the different hitting times can have negative effect on the error probability.     
\end{itemize}

    In summary, the comparison between $p_e^{ns}$ and $p_e^{sys}$ depends on the parameters $\Delta,\alpha$ and the channel stochastics. The exact boundary of their advantageous regions is still an open question and can only be derived numerically. We have conducted extensive Monte-Carlo simulations on the comparisons. A general conclusion is that when $\Delta < \alpha$, we have $p_e^{sys} < p_e^{ns}$, and when $\Delta > \alpha$, we have $p_e^{sys} > p_e^{ns}$. The conclusion basing on simulations can be sometimes inaccurate. However, it can provide a general instruction to choose between SRLSCs and NRLSCs. 



In the following, we study the analytical performance of SRLSCs in stochastic channels. To take a first step, we consider the simplest case, the i.i.d. PEC. Recall that the $p_e$ of NRLSCs in i.i.d. SEC and G-ESEC are analyzed in \cite{RLSCs} and Section \ref{section:Main Results1} of this paper, respectively. Both of them are characterized as a closed-form expression in the form of multiplies of state transition matrices. However, the analysis of SRLSCs is much more challenging.
The main reason is that since $\zeta(t)$ is time-varying, the size of state transition matrices can also vary over time (the state transition matrix at timeslot $t$ has the size of $(\zeta(t-1)+1)\times (\zeta(t)+1)$), which makes the terms heterogeneous.
And this also makes the characterization of the hitting time more challenging.
Therefore, having $p_e$ in a closed-form expression is intractable for SRLSCs to the best of our knowledge.
Due to this consideration, we first consider a special asymptotical case.

\subsection{The Analytical Expression of Exact $p_e$ for SRLSCs when $\alpha \rightarrow \infty$ and $\frac{K}{N}=\frac{1}{2}$ in the i.i.d. PEC}

In an i.i.d. PEC, assume that the probability of perfect delivery is $p$, and the corresponding erasure probability equals to $1-p$. The erasure probability is identical for all the timeslots.
To ensure $I_d(t)=0$ is positive recurrent, we assume $p>\frac{1}{2}$.
The reason why we choose the asymptotical case that $\alpha\rightarrow \infty$ for stochastical analysis is as follows. When $\alpha \rightarrow \infty$, the encoder has nearly infinite memory and thus can cache all source symbols from the previous timeslots\footnote{This asymptotical case is also practical in some scenarios. The first scenario is where the prize of memory is not the main concern of the problem. The second scenario is when $\alpha$ is sufficiently larger than $\mathbb{E}\{t_{i_0+1} - t_{i_0}\}$. In this case, the information debt can only hit the maximum value $\zeta(t)$ with a stochastically-ignorable probability.}. According to Definition \ref{definition:informtion debt new} and Proposition \ref{proposition:error event in SRLSCs}, the information debt $I_d(t)$ will never hit the maximum value $\zeta(t)$. This is mainly because when $\alpha\rightarrow \infty$, the information connection between the earliest and the latest arrived source symbol always exists. Therefore, the intractable hitting time $\tau_{j^*}$ and the time-varying $\zeta(t)$ can be circumvented in this case. And Proposition \ref{proposition:error event in SRLSCs} can be also simplified into Corollary \ref{corollary:error event in SRLSCs when r=1/2} below.
Note that the $p_e$ of NRLSCs has been analyzed in \cite{asymptotics2} for asymptotical case $\alpha\rightarrow \infty$ with random source arrival and $K=1$ simultaneously.
As a by-product, our result also extends the asymptotical results in \cite{asymptotics2} into the case that $K>1$.

Let us focus on a round that the information debt starts from zero and hits back to zero, i.e., $t\in (t_{i_0},t_{i_0+1}]$, for any index $i_0$. According to the decoding delay $\Delta$, $(t_{i_0},t_{i_0+1}]$ can be divided into two non-overlapping segments, i.e., $\mathbb{L}_1=\left(t_{i_0},\max( t_{i_0+1}-\Delta,t_{i_0})\right)$ and $\mathbb{L}_2=\left[\max( t_{i_0+1}-\Delta,t_{i_0}),t_{i_0+1}\right]$. In other words, if $t_{i_0+1}-t_{i_0} \le \Delta + 1$, then $\mathbb{L}_1=\Phi$ and $\mathbb{L}_2=(t_{i_0},t_{i_0+1}]$; otherwise, $\mathbb{L}_1=\left(t_{i_0},t_{i_0+1}-\Delta \right)$ and $\mathbb{L}_2=\left[t_{i_0+1}-\Delta,t_{i_0+1}\right]$. Roughly speaking, $\mathbb{L}_1$ represents the timeslots beyond the $\Delta$-decodable regime (also referred to $\Delta$-undecodable regime), while $\mathbb{L}_2$ represents the timeslots within the $\Delta$-decodable regime.

However, not all of the source symbols arrived in $\mathbb{L}_1$ are considered as errors. Due to the instant decodability, the symbols in the timeslots when $e(t) = 0$ can be decoded without any delay. Thus, for any timeslot $t\in \mathbb{L}_1$ and $e(t) = 0$, $\mathbf{s}(t)$ is also $\Delta$-decodable and can be excluded from the error regime. Therefore, only the source symbols in $\mathbb{L}_1$ and with $e(t) = 1$ are considered as errors in SRLSCs when $\alpha \rightarrow \infty$.
The error event in this case is characterized as the following corollary.

\begin{corollary}\label{corollary:error event in SRLSCs when r=1/2}
Assume \textbf{GMDS} holds. 
For SRLSCs in PEC with $\alpha \rightarrow \infty$, for any fixed index $i_0 \ge 0$, \textbf{(a)} $\mathbf{s}(t)$ is not $\Delta$-decodable 
\begin{equation}
\forall t \in \left\{t'|t'\in(t_{i_0}, t_{i_0+1} - \Delta),e(t')=1\right\},
\end{equation}
\textbf{(b)} $\mathbf{s}(t)$ is $\Delta$-decodable for the rest of $t$.
\end{corollary}

Denote $N_e$ as the number of erasures in $\mathbb{L}_1$. Specifically, $N_e = \sum_{t\in \mathbb{L}_1} \mathds{1}\{e(t)=1\}$. Similar to the \textbf{Lemma 2} in \cite{asymptotics2},  with Corollary \ref{corollary:error event in SRLSCs when r=1/2}, the $p_e$ of SRLSCs in i.i.d. PEC when $\alpha\rightarrow \infty$ can be directly given by the following lemma.

\begin{lemma}\label{lemma:pe when r=1/2}
When $\alpha \rightarrow \infty$, the error probability of SRLSCs in i.i.d. PEC can be given by 
    \begin{align}
        p_e^{sys} = \frac{\mathbb{E}\{N_e\}}{\mathbb{E}\{t_{i_0+1} - t_{i_0}\}}.
    \end{align}
\end{lemma}

\textit{Proof:} Lemma \ref{lemma:pe when r=1/2} holds from Corollary \ref{corollary:error event in SRLSCs when r=1/2} by calculating the ratio of expected error timeslots to the expected interval of the zero-hitting times.
Similar to the proof of Lemma \ref{lemma:1}, $t_i$ defined in (\ref{equation:ti_new}) is a Markov renewal process.
By \cite[Theorem 3.3]{Renewal process}, Lemma \ref{lemma:pe when r=1/2} holds directly.

The exact characterization of $\mathbb{E}\{t_{i_0+1} - t_{i_0}\}$ and $\mathbb{E}\{N_e\}$ in Lemma \ref{lemma:pe when r=1/2} can be given in the following two lemmas.
\begin{lemma}\label{lemma:E(ti0+1-ti0) when r=1/2}
    When $\alpha \rightarrow \infty$ and $\frac{K}{N}=\frac{1}{2}$, in i.i.d. PEC where the probability of perfect delivery is $p$, the exact value of $\mathbb{E}\{t_{i_0+1} - t_{i_0}\}$ can be given by
    \begin{align}
        \mathbb{E}\{t_{i_0+1} - t_{i_0}\} 
        =p + 2p(1-p)\cdot \int_0^1 \left(1+\frac{1}{1-2\sqrt{p(1-p)}\cos{\pi x}}\right)\frac{\sin^2{\pi x}}{1-2\sqrt{p(1-p)}\cos{\pi x}} dx.
    \end{align}
\end{lemma}
\begin{lemma}\label{lemma:E(N_e) when r=1/2}
    When $\alpha \rightarrow \infty$ and $\frac{K}{N}=\frac{1}{2}$, in i.i.d. PEC where the probability of perfect delivery is $p$, the exact value of $\mathbb{E}\{N_e\}$ can be given by
    \begin{align}
        \mathbb{E}\{N_e\}
        =\sum_{l=\lceil\frac{\Delta}{2}\rceil+1}^\infty  p^l(1-p)^l  \left[\Big(2l-\Delta-1\Big)\cdot C_{l-1} - \sum_{i=0}^{l-\lfloor\frac{\Delta}{2}\rfloor - 3} C_i \cdot C_{l-2-i} \cdot \Big(2l - \Delta - 3 - 2i\Big)\right],
    \end{align}
    where $C_n = \frac{1}{n+1} \binom{2n}{n}, \forall n\ge 0$ are the Catalan Numbers \cite{Catalan}.
\end{lemma}

The proof of Lemma \ref{lemma:E(ti0+1-ti0) when r=1/2} is presented in Appendix \ref{appendix:lemma E(ti0+1-ti0) when r=1/2}, while the proof of Lemma \ref{lemma:E(N_e) when r=1/2} is presented in Appendix \ref{appendix:E(N_e) when r=1/2}.
Then, the exact error probability of SRLSCs when $\alpha \rightarrow \infty$ and $\frac{K}{N}=\frac{1}{2}$ can be given by the following theorem.
\begin{theorem}\label{theorem:pe when r=1/2}
When $\alpha \rightarrow \infty$ and $\frac{K}{N}=\frac{1}{2}$, in i.i.d. PEC, where the probability of perfect delivery is $p$, the exact error probability of SRLSCs can be given by 
    \begin{equation}
        p_e^{sys} = \frac{\sum_{l=\lceil\frac{\Delta}{2}\rceil+1}^\infty  p^l(1-p)^l  \left[\Big(2l-\Delta-1\Big)\cdot C_{l-1} - \sum_{i=0}^{l-\lfloor\frac{\Delta}{2}\rfloor - 3} C_i \cdot C_{l-2-i} \cdot \Big(2l - \Delta - 3 - 2i\Big)\right]}{p + 2p(1-p)\cdot \int_0^1 \left(1+\frac{1}{1-2\sqrt{p(1-p)}\cos{\pi x}}\right)\frac{\sin^2{\pi x}}{1-2\sqrt{p(1-p)}\cos{\pi x}} dx},\label{equation:p_e^sys}
    \end{equation}
where $C_n = \frac{1}{n+1} \binom{2n}{n}, \forall n\ge 0$.
\end{theorem}

Theorem \ref{theorem:pe when r=1/2} holds directly from Lemma \ref{lemma:pe when r=1/2}, Lemma \ref{lemma:E(ti0+1-ti0) when r=1/2} and Lemma \ref{lemma:E(N_e) when r=1/2}.

\begin{remark}
    \textit{(The convergence analysis on the numerator)} It can be easily proved that the summation in the numerator will converge. 
    From Appendix \ref{appendix:E(N_e) when r=1/2}, one can observe that the numerator $\mathbb{E}\{N_e\} = \sum_{l = \lceil\frac{\Delta}{2}\rceil + 1}^\infty p^l(1-p)^l  \sum_{i=1}^{C_{l-1}}  N_e(l,i)$.
    Let $a_l = (1-p)^l\cdot p^l \cdot \sum_{i=1}^{C_{l-1}}N_e(l,i)$. 
    When $l\rightarrow \infty$, we can obtain $\sum_{i=1}^{C_{l-1}}N_e(l,i) \overset{l \gg \Delta}{\longrightarrow}  l\cdot C_{l-1} = \binom{2l-2}{l-1}$.
    With the Stirling's formula, we obtain $\binom{2l-2}{l-1} \sim  \frac{4^{l-1}}{\sqrt{\pi (l-1)}}$.
    Therefore, $a_l \sim \frac{[4p(1-p)]^l}{4\sqrt{\pi (l-1)}}$. Let $p' = 4p(1-p)$. Since $\frac{1}{2}<p<1$, we have $|p'|<1$. Thus, $\lim_{l\rightarrow\infty}|\frac{a_{l+1}}{a_l}|=\lim_{l\rightarrow\infty}|\frac{p'^{l+1}}{\sqrt{l}}\cdot\frac{\sqrt{l-1}}{p'^l}| = |p'|<1$. By the D'Alembert's test, $\sum_{l=1}^\infty a_l$ converges. One can also notice that the speed of convergence is highly dependent on parameters the $\Delta$ and $p$, especially $p$. A fast convergence can be expected with a larger value of $p$ and a smaller value of $\Delta$.
\end{remark}

\begin{remark}
    \textit{(Discussion on numerical approximation of Theorem \ref{theorem:pe when r=1/2})}
    Note that in Theorem \ref{theorem:pe when r=1/2}, only an analytical expression is derived on the exact $p_e^{sys}$ in this special asymptotical case. The summation to infinity in the numerator and the integral in the denominator are still intractable to have a closed-form expression and this can be an interesting future work to solve. To evaluate $p_e^{sys}$ numerically with a stable convergence value of the summation, an appropriate upper threshold, denoted by $l_{max}$, should be chosen according to the value of $\Delta$ and $p$. $l_{max}$ can have a small value with a large $p$ and a small $\Delta$.
    For example, when $p=0.8$ and $\Delta=20$, $l_{max}=20$ can be chosen.
    However, when $p=0.55$ and $\Delta = 100$, $l_{max}=500$ can be chosen.
\end{remark}

\begin{remark}
    \textit{(Technical differences to the related work \cite{asymptotics2})} In \cite{asymptotics2}, an analytical expression of error probability is also derived in asymptotic case $\alpha\rightarrow\infty$, but for NRLSCs and random arrivals with $K=1$. In \cite{asymptotics2}, the $p_e$ is given as $p_e = \frac{\mathbb{E}\{(t_{i_0+1} - \Delta - 1 - t_{i_0})^+\}}{\mathbb{E}\{t_{i_0+1} - t_{i_0}\}}$. The derivation of the numerator involves the eigendecomposition of a tridiagonal Toeplitz matrice \textit{where its two first off-diagonals are with constants}.
    On the other hand, in our work, $p_e$ is derived as $p_e^{sys} = \frac{\mathbb{E}\{N_e\}}{\mathbb{E}\{t_{i_0+1} - t_{i_0}\}}$. The derivation of $\mathbb{E}\{t_{i_0+1} - t_{i_0}\}$ in Lemma \ref{lemma:E(ti0+1-ti0) when r=1/2} involves the eigendecomposition of a tridiagonal Toeplitz matrice where \textit{the two symetrical off-diagonals are positioned $K$ steps from the main diagonal instead of only one}. Therefore, regarding of this term, our result generalizes the analysis framework of \cite{asymptotics2} into $K>1$.
    Furthermore, the derivation of $\mathbb{E}\{N_e\}$ is fundamentally different. Rather than the expected length of the $\Delta$-undecodable regime $(t_{i_0}, t_{i_0+1} - \Delta)$, the number of erasures in it should be characterized. 
    However, one can notice that $\mathbb{E}\{N_e\} \neq \mathbb{E}\{(t_{i_0+1} - \Delta - 1 - t_{i_0})^+\} \cdot (1-p)$, which means the $\mathbb{E}\{N_e\}$ can not be directly obtained by multiplying the length of undecodable regime and the erasure rate.
    This is because the erasures are not distributed evenly in $(t_{i_0},t_{i_0+1})$. Intuitively, the erasures are distributed intensively at the beginning of the round (close to $t_{i_0}$), and distributed sparsely at the end of the round (close to $t_{i_0+1}$).
    Therefore, in our work, we innovatively leverage the feature of Catalan Numbers \cite{Catalan} to characterize the exact distribution the erasures within $(t_{i_0},t_{i_0+1})$, and then further derive $\mathbb{E}\{N_e\}$.
\end{remark}

\section{Numerical Results}\label{section:numerical results}

\begin{figure}[!hbtp]\setcounter{subfigure}{0}
    \centering
    \subfigure[$p_e^{ns}$.]{\includegraphics[width=0.49\textwidth]{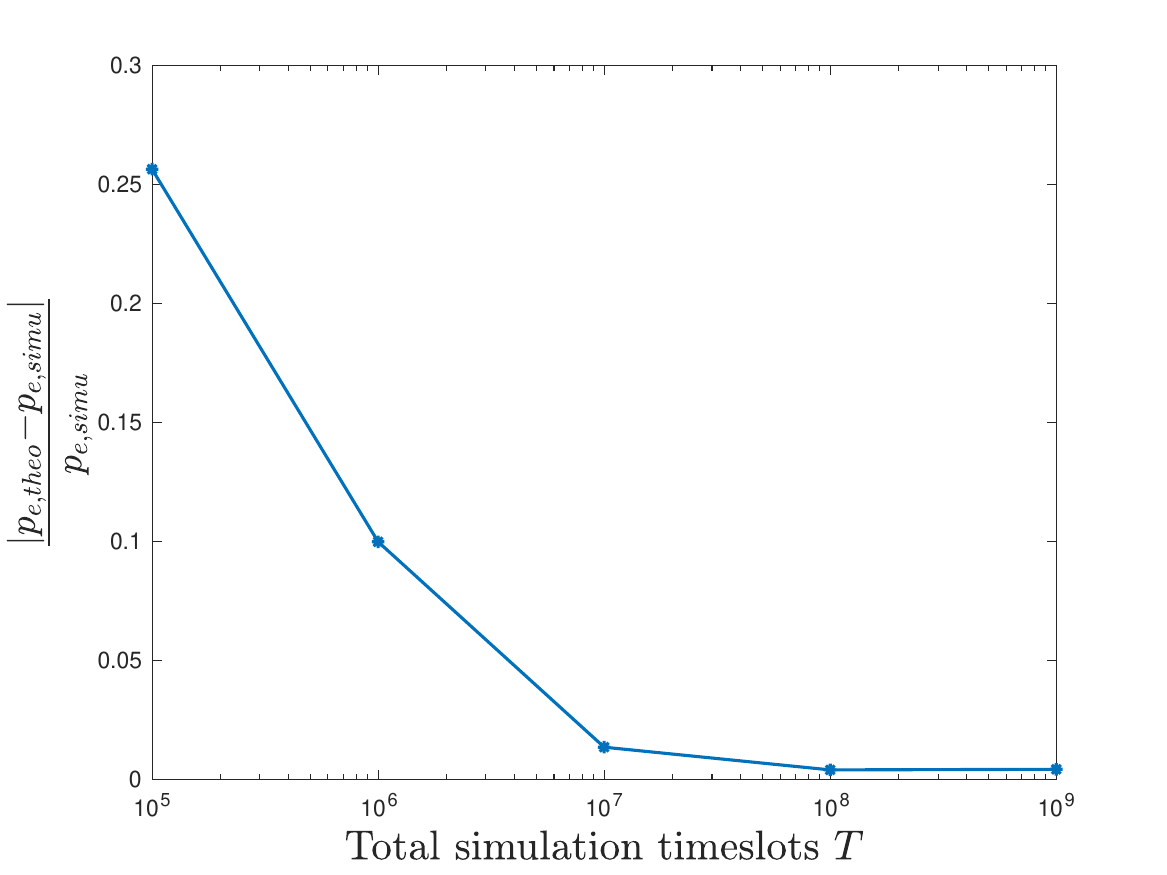}}
    \subfigure[$p_e^{sys}$.]{\includegraphics[width=0.49\textwidth]{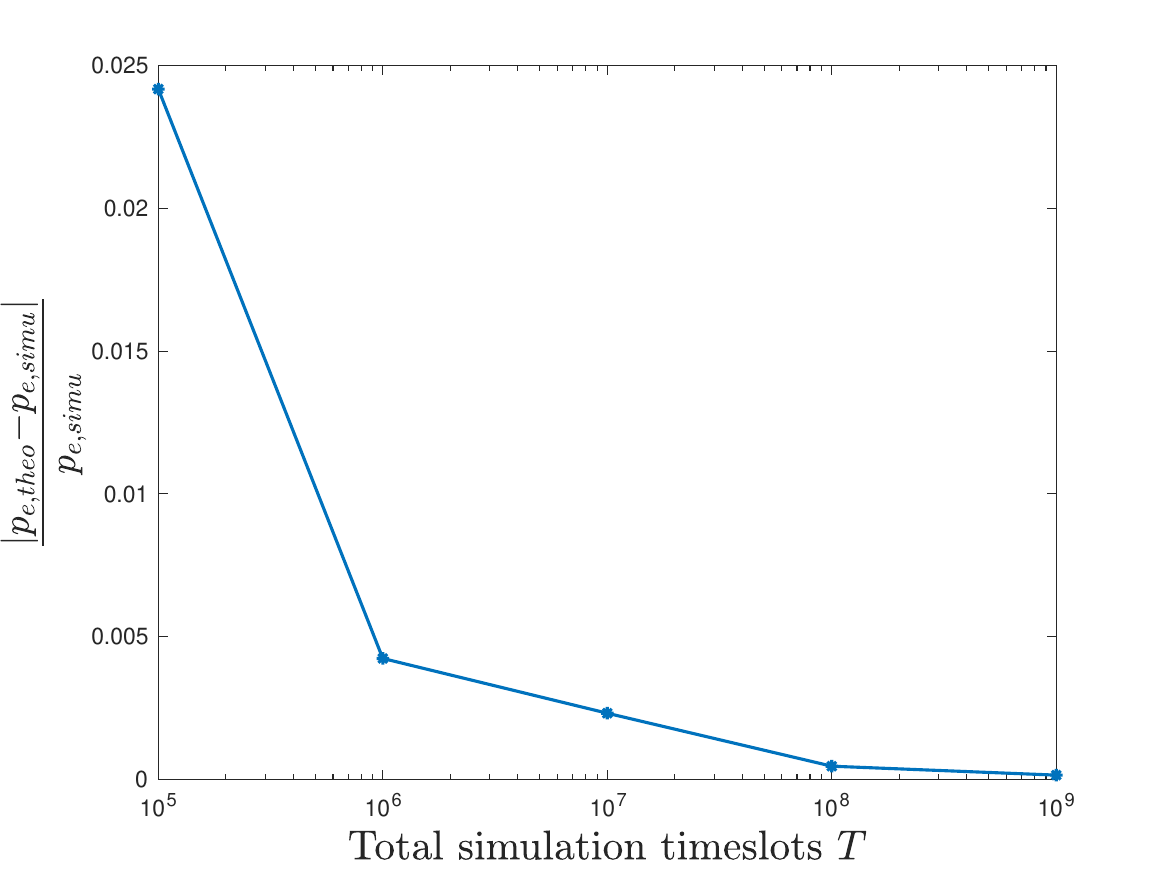}}
    \caption{Relative deviation of theoretical expression from its simulation value.}
    \label{deviation_vs_T}
\end{figure}

We first numerically compare the theoretical $p_e^{ns}$ derived by Theorem \ref{theorem:1} and the corresponding $p_e^{ns}$ derived by Monte-Carlo simulation. 
The system parameters are set to $K=5,N=10,\Delta=5,\alpha = 4$. 
We consider the G-ESEC with state transition probabilities $(p,r) = (10^{-4},0.5)$.
Assume that $C_t$ follows binomial distributions with success probability $p_G=0.7$ in good state, and with $p_B=0$ in bad state, i.e., $C_t \sim B(10,0.7)$ when in $G$ and $C_t=0$ when in $B$.
In the Monte-Carlo simulation, in each round we sample $T$ timeslots of channel realizations and determine the error events accordingly. 
In Fig. \ref{deviation_vs_T} (a), we plot the relative deviation of the theoretical error probability from its simulation value, i.e., $\frac{|p_{e,theo}-p_{e,simu}|}{p_{e,simu}}$ versus, the sampling timeslots $T$. 
At each value of $T$, the deviation is averaged over 10 rounds of experiments.
One can notice that the relative deviation decreases rapidly and asymptotically approaches zero when $T$ increases and approaches infinity. 
When $T = 10^9$, the relative deviation is only $0.4\%$.
This could show the correctness of Theorem \ref{theorem:1}.  

Then we compare the $p_e^{sys}$ derived by Theorem \ref{theorem:pe when r=1/2} and the corresponding $p_e^{sys}$ derived by Monte-Carlo simulation. 
In this setting, $K=5,N=10$, satisfying that $\frac{K}{N}=0.5$. We also assume $\alpha\rightarrow \infty$ and $\Delta=5$. We consider the i.i.d. PEC with probability of perfect delivery $p=0.7$.
The summation to infinity in the numerator of Theorem \ref{theorem:pe when r=1/2} is set to 150.
In Fig. \ref{deviation_vs_T} (b), we plot again the relative deviation of the theoretical value from its simulation value versus $T$. 
At each value of $T$, the deviation is also averaged over 10 rounds of experiments.
One can also notice that the deviation decreases and asymptotically approaches zero when $T$ increases and approaches infinity. 
When $T = 10^9$, the relative deviation is surprisingly only $0.014\%$.
This shows the correctness of Theorem \ref{theorem:pe when r=1/2}.  

\begin{figure}[thbp!]
    \centering
    \begin{minipage}[t]{0.49\linewidth}
        \centering
        \includegraphics[width=0.9\linewidth]{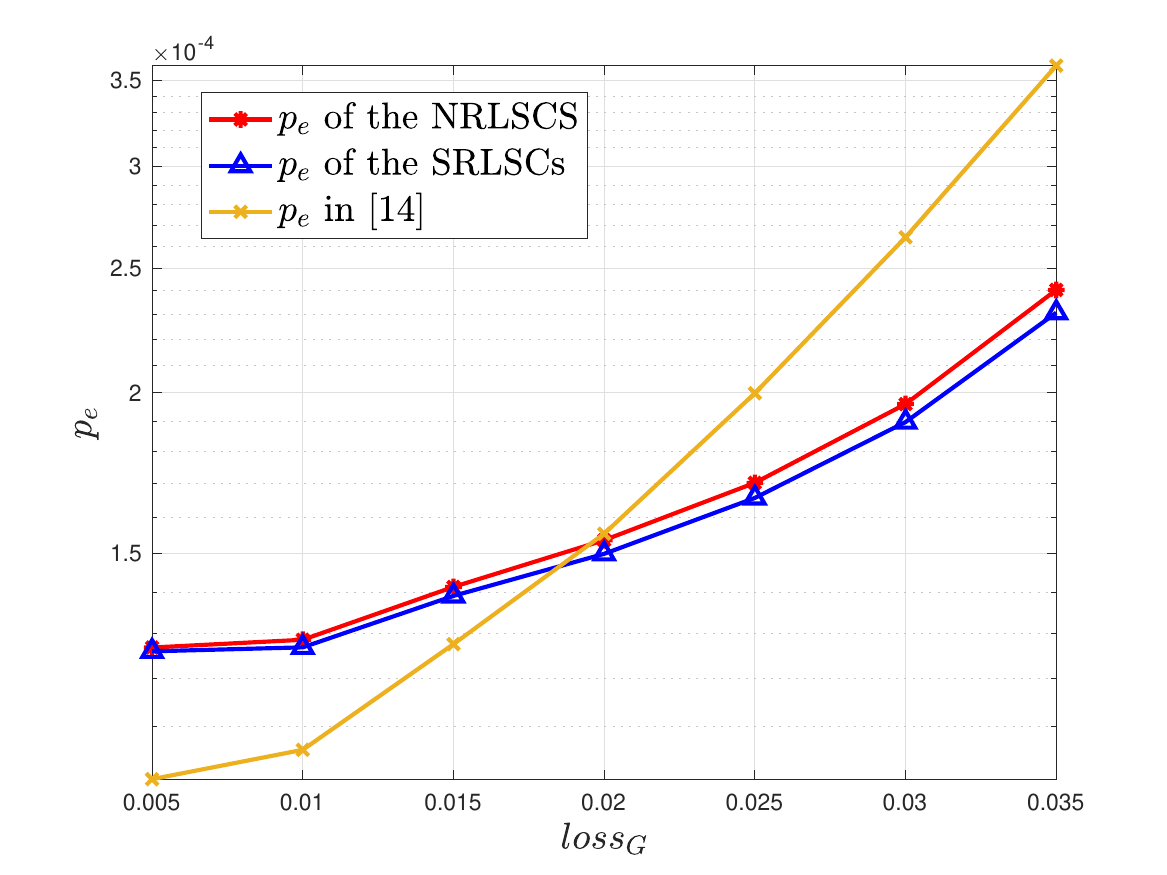}
        \caption{$K=3,N=6,\Delta=4,\alpha=5,(10^{-4},0.4)$.}
        \label{figure:comparison_between23_1}
    \end{minipage}
    \begin{minipage}[t]{0.49\linewidth}
        \centering
        \includegraphics[width=0.9\linewidth]{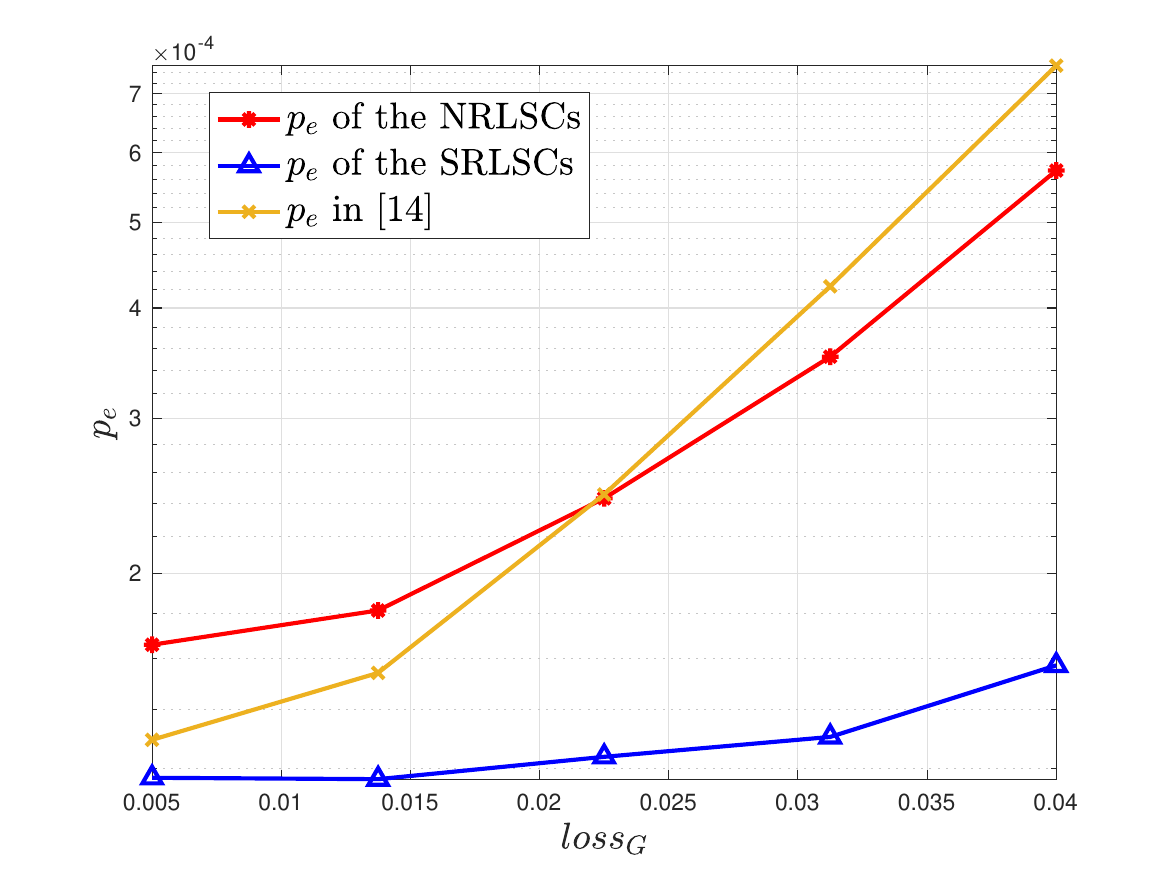}
        \caption{$K=4,N=7,\Delta=5,\alpha=6,(10^{-4},0.4)$.}
        \label{figure:a6d5k4n7r04500}
    \end{minipage}
    \begin{minipage}[t]{0.49\linewidth}
        \centering
        \includegraphics[width=0.9\linewidth]{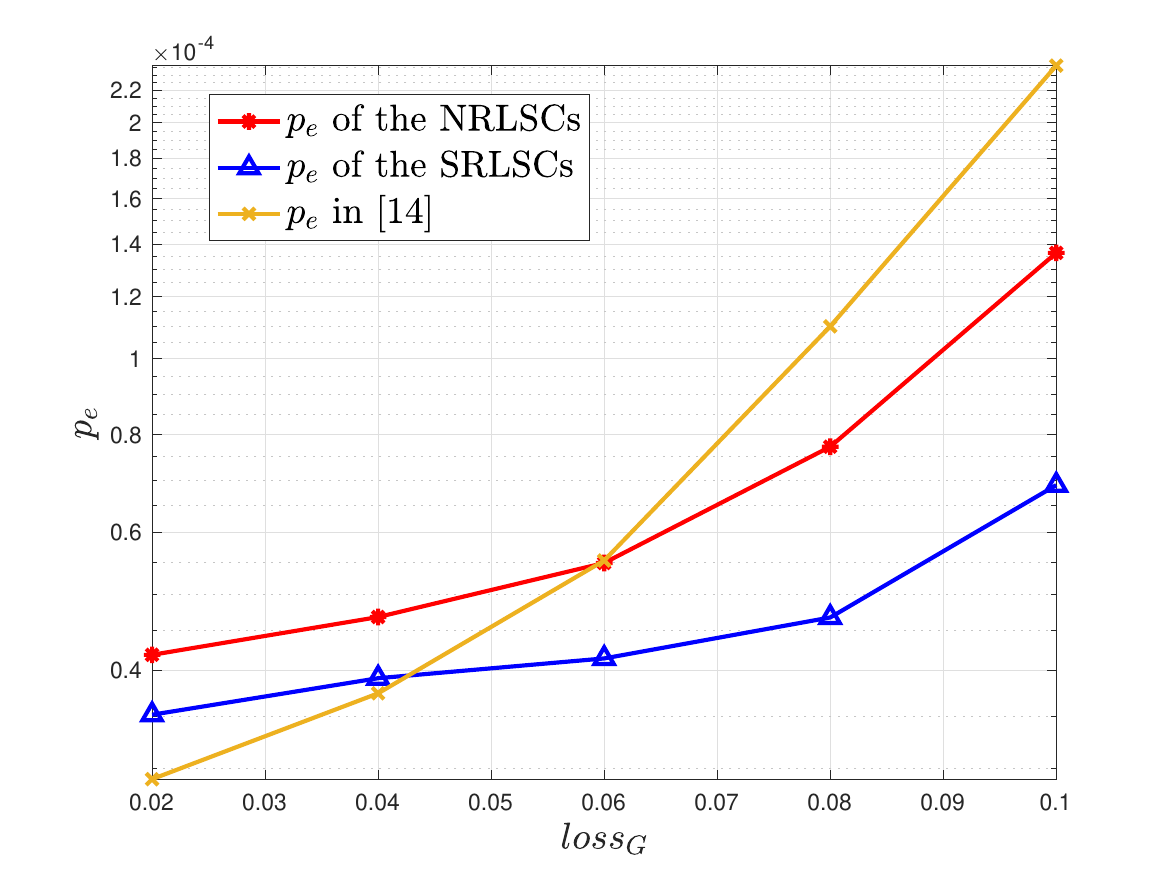}
        \caption{$K=4,N=10,\Delta=7,\alpha=9,(10^{-4},0.4)$.}
        \label{figure:a9d7k4n10r04500}
    \end{minipage}
    \begin{minipage}[t]{0.49\linewidth}
        \centering
        \includegraphics[width=0.9\linewidth]{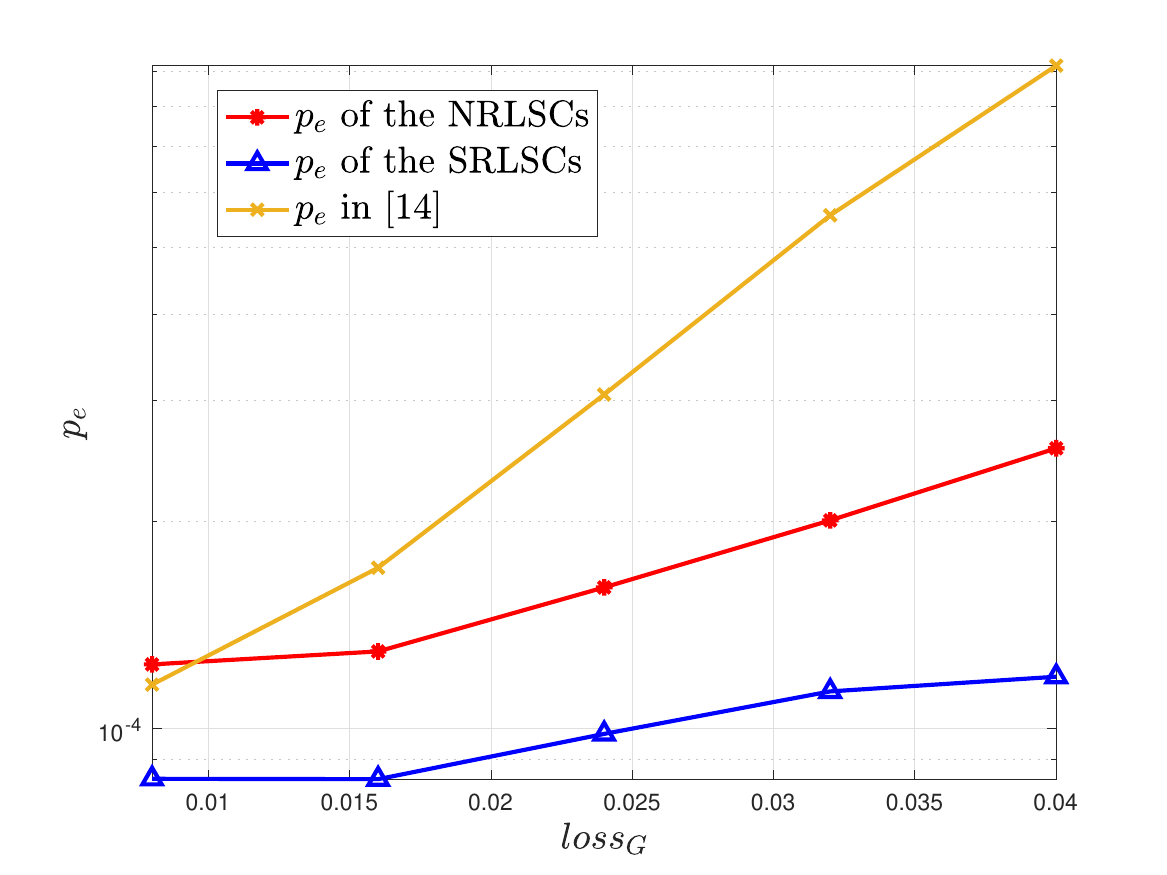}
        \caption{$K=6,N=10,\Delta=7,\alpha=9,(10^{-4},0.4)$.}
        \label{figure:a9d7k6n10r04100}
    \end{minipage}
 \end{figure}

We then numerically compare the RLSCs in this paper and the streaming codes proposed in \cite{adversarial channel} which are optimal for the $(W,B,M)$-sliding window packet erasure channels (SWPEC). The $(W,B,M)$-SWPEC introduces either one burst erasure with length no longer than $B$ or multiple arbitrary erasures with total count no larger than $M$ within any window with length $W$. 
In \cite{adversarial channel}, optimal streaming codes with parameters $K=\Delta-M+1,N=K+B$ and $\alpha=N-1$ is proposed for both $\frac{K}{N}\ge\frac{1}{2}$ and $\frac{K}{N}<\frac{1}{2}$.
We consider the Gilbert-Elliott packet erasure channel (G-EPEC).
The erasure probability in the bad state is 1, while the erasure probability in the good state, denoted as $loss_G$, is plotted against the error probability $p_e$. 
At each value of $loss_G$, $p_e$ is averaged over 100 rounds of simulations, each contains $10^6$ timeslots of channel realizations.
In Fig. \ref{figure:comparison_between23_1} to Fig. \ref{figure:a9d7k6n10r04100}, the $p_e$ of both SRLSCs and NRLSCs are compared to the $p_e$ in \cite{adversarial channel}. 
In Fig. \ref{figure:comparison_between23_1}, the parameters are set to $K=3,N=6,(p,r) = (10^{-4},0.4),\Delta = 4, \alpha = 5$, the generator matrix of \cite{adversarial channel} is chosen as $\begin{bmatrix}
    1& 0 &0 &1 &1 &0\\
    0& 1 &0 &0 &1 &1\\
    0& 0 &1 &0 &1 &2
\end{bmatrix}$ with rate $\frac{K}{N}=\frac{1}{2}$.
In Fig. \ref{figure:a6d5k4n7r04500}, the parameters are set to $K=4,N=7,\Delta=5,\alpha=6,(10^{-4},0.4)$, the generator matrix in \cite{adversarial channel} is changed to 
$\begin{bmatrix}
    1& 0 &0 &0 &1 &2 &0\\
    0& 1 &0 &0 &0 &1 &3\\
    0& 0 &1 &0 &0 &2 &1\\
    0& 0 &0 &1 &1 &1 &1
\end{bmatrix}$ with rate $\frac{K}{N}=\frac{4}{7}$. 
In Fig. \ref{figure:a9d7k4n10r04500}, the parameters are set to $K=4,N=10,\Delta=7,\alpha=9,(10^{-4},0.4)$, the generator matrix in \cite{adversarial channel} is changed to  
$\begin{bmatrix}
    1& 0 &0 &0 &1 &4 &16 &64 &0 &0\\
    0& 1 &0 &0 &1 &3 &0 &27 &81 &0\\
    0& 0 &1 &0 &1 &2 &0 &0 &16 &32\\
    0& 0 &0 &1 &1 &1 &0 &0 &1 &1
\end{bmatrix}$ with rate $\frac{K}{N}=\frac{4}{10}$.
In Fig. \ref{figure:a9d7k6n10r04100}, the parameters are set to $K=6,N=10,\Delta=7,\alpha=9,(10^{-4},0.4)$, the generator matrix in \cite{adversarial channel} is changed to  
$\begin{bmatrix}
    1& 0 &0 &0 &0 &0 &1 &6 &0 &0\\
    0& 1 &0 &0 &0 &0 &0 &5 &25 &0\\
    0& 0 &1 &0 &0 &0 &0 &0 &16 &64\\
    0& 0 &0 &1 &0 &0 &0 &0 &9 &27\\
    0& 0 &0 &0 &1 &0 &1 &2 &4 &8\\
    0& 0 &0 &0 &0 &1 &1 &1 &1 &1
\end{bmatrix}$ with rate $\frac{K}{N}=\frac{6}{10}$.

Fig. \ref{figure:comparison_between23_1} to Fig. \ref{figure:a9d7k6n10r04100} show that the SRLSCs can outperform the NRLSCs in these settings\footnote{Note that the above simulations only include the regime $\Delta<\alpha$. This is mainly because \cite{adversarial channel} assumed $1\le M\le B\le \Delta <W$ for the SWPEC. Together with $K=\Delta-M+1,N=K+B$ and $\alpha=N-1$, this condition implicitly indicates the constraint that $\Delta\le \alpha$ for the proposed streaming codes. However, for RLSCs in this paper, we have not imposed such constraint. Therefore, for the comparisons to \cite{adversarial channel}, we only consider $\Delta<\alpha$ as in these settings.}. Also note that the gap between SRLSCs and NRLSCs generally increases along with $loss_G$.
This is mainly because when the channel erasure rate is higher, the effect of instant decodability could be more prominent. 
It is also shown that the RLSCs can outperform the streaming codes in \cite{adversarial channel} when $loss_G$ is relatively high, while the streaming codes in \cite{adversarial channel} have a smaller $p_e$ when $loss_G$ is relatively low.
In the figures, the yellow curve representing \cite{adversarial channel} manifests a steeper slope than the curves representing RLSCs. 
This phenomenon could be interpreted as a potential advantage of the almost totally random generator matrix in RLSCs over the well-designed but complex generator matrix in the optimal streaming codes for SWPEC. 
When $loss_G$ is relatively low, all the erasure patterns generated by G-EPEC could be virtually included in the predetermined erasure patterns of SWPEC. Thus, well-designed code construction of \cite{adversarial channel} can handle the erasures almost perfectly and largely outperforms the RLSCs without any specific designs. 
However, when $loss_G$ is relatively high, the stochastically generated erasure patterns could frequently exceed the predetermined collection of erasure patterns, causing severe performance degradation. 
Surprisingly, when $loss_G$ is increasing, the degradation of \cite{adversarial channel} is more prominent than the RLSCs. 
This observation implies that the complex and deterministic design for SWPEC on the parity matrix could be counter-productive in the stochastic channels with high erasure probability (especially when $loss_G$ can even vary largely over time, where its value could be very large in the worst case), comparing to the simple and all-random generator matrix.
On the other hand, the simple structure of RLSCs with almost totally random entries could however provide a better resistance to the increasing of $loss_G$.
In practical scenarios, the erasure probability of channel could be very high and even time-varying. Thus, the simulation results can shed light on the choice of codes in reality.
A lesson for practical implementation is to choose the all-random RLSCs of this paper in unstable or uncertain channels with probably high and time-varying erasure probability.
If the streaming codes for SWPEC have to be chosen, one should leave a margin when designing the code parameters, in order to avoid the potential rapid performance degradation.

\section{Conclusion}\label{section:conclusion}
In this paper, we mainly investigate the fundamental performance limit of RLSCs under sufficiently large finite size regime in stochastic channels. 
We first characterize the closed-form expression of the error probability for NRLSCs with finite memory length and decoding delay in G-ESEC.  
Then we analyze the theoretical performance of SRLSCs in i.i.d. PEC. 
It is found that SRLSCs can actually cause performance degradation in some cases, due to the de-correlation between the preceding and the following information. 
Then we characterize the error event of SRLSCs and derive the exact analytical expression on the error probability when the length of the memory $\alpha \rightarrow \infty$ and the coding rate equals to 1/2. 
Numerical simulations show that SRLSCs can have a lower error probability compared to NRLSCs in most of the scenarios. It can even outperform some delicately designed optimal streaming codes in some cases.
Future works could include the exact characterization of the SRLSCs and some code constructions with improved and stochastically analyzable performance in the i.i.d. PEC or G-EPEC.

\begin{appendices}
\section{Proof of Proposition \ref{proposition:1}}\label{appendix:proposition 1}

Recall that we assume the stationary initial distribution of the states starting from $I_d(t)=0$, i.e.,  $\pi^{(0)}=\begin{bmatrix}
    \pi_G^{(0)} & \pi_B^{(0)}
\end{bmatrix}$ is given.
When $k=1$, there are only two possible state transition traces, i.e., $G$ and $B$. Thus, we have 
$$\text{Pr}(t_{i_0+1} - t_{i_0} = 1) = \pi_G^{(0)} \cdot \Gamma_{0,0}^G + \pi_B^{(0)} \cdot \Gamma_{0,0}^B.$$ 
Then we prove that equation (\ref{equation:P(T=k)}) holds for all integers $k\ge 2$ by mathematical deduction.  

First we verify that (\ref{equation:P(T=k)}) holds for $k=2$.
When $k=2$, there are only four possible state transition traces, i.e., $GG$, $GB$, $BG$ and $BB$. Thus, we have 
\begin{align}
	&\text{Pr}(t_{i_0+1} - t_{i_0} = 2) = \text{Pr}(GG)\cdot \text{Pr}(t_{i_0+1} - t_{i_0} = 2|GG) + \text{Pr}(GB)\cdot \text{Pr}(t_{i_0+1} - t_{i_0} = 2|GB)\nonumber\\
	 &\qquad\qquad\qquad\qquad+ \text{Pr}(BG)\cdot \text{Pr}(t_{i_0+1} - t_{i_0} = 2|BG)+ \text{Pr}(BB)\cdot \text{Pr}(t_{i_0+1} - t_{i_0} = 2|BB)\\
=& \pi_G^{(0)}(1-p) \begin{bmatrix}
		\Gamma_{0,\phi}^G & \Gamma_{0,\zeta}^G
	\end{bmatrix} \!\!\begin{bmatrix}
	\Gamma_{\phi,0}^G \\
	\Gamma_{\zeta,0}^G
\end{bmatrix}  + \pi_G^{(0)} p \begin{bmatrix}
	\Gamma_{0,\phi}^G & \Gamma_{0,\zeta}^G
\end{bmatrix} \!\!\begin{bmatrix}
	\Gamma_{\phi,0}^B \\
	\Gamma_{\zeta,0}^B
\end{bmatrix}  + \pi_B^{(0)} r \begin{bmatrix}
	\Gamma_{0,\phi}^B & \Gamma_{0,\zeta}^B
\end{bmatrix} \!\!\begin{bmatrix}
	\Gamma_{\phi,0}^G \\
	\Gamma_{\zeta,0}^G
\end{bmatrix}  + \pi_B^{(0)} (1-r) \begin{bmatrix}
	\Gamma_{0,\phi}^B & \Gamma_{0,\zeta}^B
\end{bmatrix} \!\!\begin{bmatrix}
	\Gamma_{\phi,0}^B \\
	\Gamma_{\zeta,0}^B
\end{bmatrix}  \\
=& \begin{bmatrix}\pi_G^{(0)} & \pi_B^{(0)} \end{bmatrix}
        \begin{bmatrix}
            \Gamma_{s}^G & 0\\
            0 & \Gamma_{s}^B
        \end{bmatrix}
        \begin{bmatrix}
            (1-p)\mathbf{I}_\zeta & p\mathbf{I}_\zeta\\
            r\mathbf{I}_\zeta & (1-r)\mathbf{I}_\zeta
        \end{bmatrix}
        \begin{bmatrix}
            \Gamma_{e}^G \\
            \Gamma_{e}^B
        \end{bmatrix}.\label{equation:k=2}
\end{align}

Then we prove that (\ref{equation:P(T=k)}) holds $\forall k>2$. For any $k>2$, we assume that (\ref{equation:P(T=k)}) holds, and then prove (\ref{equation:P(T=k)}) also holds for $k+1$.

Let $S_{i,j}^k$ be the $j$-th state in sequential order of transition trace $S_i^k$. 
Therefore, $S_{i,1}^k$ and $S_{i,k}^k$ are the first and the last state of transition trace $S_i^k$, respectively.
Then equation (\ref{equation:summation of Pr(A)}) can be written as
\begin{align}
	\text{Pr}(t_{i_0+1} - t_{i_0} = k) &= \sum_{i\in S(k)} \text{Pr}(S_i^k) \cdot \Gamma_s^{S_{i,1}^k} \cdot \prod_{j=2}^{k-1}Q^{S_{i,j}^k}\Gamma_e^{S_{i,k}^k}\\
	&=  \sum_{\substack{i\in S(k)\\i|2=1}} \text{Pr}(S_i^k) \cdot \Gamma_s^{S_{i,1}^k} \cdot \prod_{j=2}^{k-1}Q^{S_{i,j}^k}\Gamma_e^{G} + \sum_{\substack{i\in S(k)\\i|2=0}} \text{Pr}(S_i^k) \cdot \Gamma_s^{S_{i,1}^k} \cdot \prod_{j=2}^{k-1}Q^{S_{i,j}^k}\Gamma_e^{B}\label{equation:subset}\\
       &= \begin{bmatrix}
		\displaystyle\sum_{\substack{i\in S(k)\\i|2=1}} \text{Pr}(S_i^k) \cdot \Gamma_s^{S_{i,1}^k} \cdot \prod_{j=2}^{k-1}Q^{S_{i,j}^k} & \displaystyle\sum_{\substack{i\in S(k)\\i|2=0}} \text{Pr}(S_i^k) \cdot \Gamma_s^{S_{i,1}^k} \cdot \prod_{j=2}^{k-1}Q^{S_{i,j}^k}
	\end{bmatrix} \begin{bmatrix}
		\Gamma_{e}^G \\
		\Gamma_{e}^B
	\end{bmatrix},\label{equation:align}
\end{align}
where equation (\ref{equation:subset}) is derived by dividing the set $S(k)$ into two disjoint subsets $\{i:i\in S(k), i|2=1\}$ and $\{i:i\in S(k), i|2=0\}$, which is according to \textit{the last state} of the transition trace. 
Recall that $S_i^k$, the $i$-th  transition trace with length $k$, $i \in [2^k]$, can be represented by a binary stream, where 1 denotes $G$ and 0 denotes $B$. 
Therefore, $S_{i,k}^k = G$ is equivalent to $i|2=1$ and $S_{i,k}^k = B$ is equivalent to $i|2=0$.

For $k+1$, we have
\begin{align}
	&\text{Pr}(t_{i_0+1} - t_{i_0} = k+1) = \sum_{i\in S(k+1)} \text{Pr}(S_i^{k+1}) \cdot \Gamma_s^{S_{i,1}^{k+1}} \cdot \prod_{j=2}^{k}Q^{S_{i,j}^{k+1}}\Gamma_e^{S_{i,k+1}^{k+1}}\\
	&=  \sum_{\substack{i\in S(k)\\i|2=1}} \text{Pr}(S_i^k) \cdot \Gamma_s^{S_{i,1}^k} \cdot \prod_{j=2}^{k-1}Q^{S_{i,j}^k}\cdot Q^G \cdot \left[(1-p)\Gamma_e^{G}+p\Gamma_e^{B}\right] + \sum_{\substack{i\in S(k)\\i|2=0}} \text{Pr}(S_i^k) \cdot \Gamma_s^{S_{i,1}^k} \cdot \prod_{j=2}^{k-1}Q^{S_{i,j}^k}\cdot Q^B \cdot \left[r\Gamma_e^{G}+(1-r)\Gamma_e^{B}\right]\label{equation:subset2}\\
            &= \begin{bmatrix}
		\displaystyle\sum_{\substack{i\in S(k)\\i|2=1}} \text{Pr}(S_i^k) \cdot \Gamma_s^{S_{i,1}^k} \cdot \prod_{j=2}^{k-1}Q^{S_{i,j}^k} & 
        \displaystyle\sum_{\substack{i\in S(k)\\i|2=0}} \text{Pr}(S_i^k) \cdot \Gamma_s^{S_{i,1}^k} \cdot \prod_{j=2}^{k-1}Q^{S_{i,j}^k}
	\end{bmatrix} 
\begin{bmatrix}
	Q^G & \\
	 &Q^B
\end{bmatrix}
\begin{bmatrix}
	(1-p)\mathbf{I}_\zeta & p\mathbf{I}_\zeta\\
	r\mathbf{I}_\zeta & (1-r)\mathbf{I}_\zeta
\end{bmatrix}
\begin{bmatrix}
		\Gamma_{e}^G \\
		\Gamma_{e}^B
	\end{bmatrix}.\label{equation:align2}
\end{align}

\begin{figure}[!hbtp]
    \centering
    \includegraphics[width=0.5\linewidth]{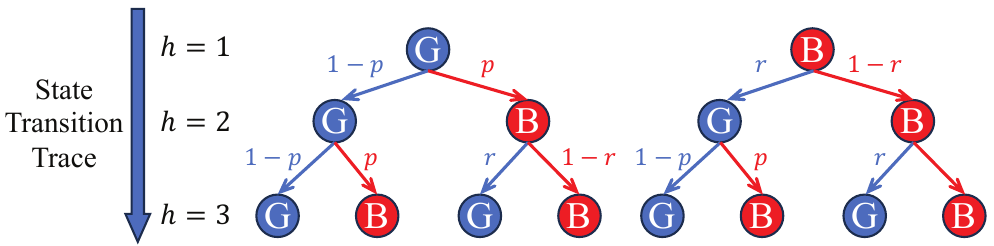}
    \caption{Illustration of the recursive structure of the state transition trace.}
    \label{fig:transistion trace}
\end{figure}

Equation (\ref{equation:subset2}) is more subtle. Different from (\ref{equation:subset}) slightly, (\ref{equation:subset2}) is derived by dividing the set $S(k+1)$ into two subsets, according to \textit{the second-to-last state} of the transition trace.
Consider the last step of the state transition.
When the second-to-last state is $G$, it will stay in $G$ with probability $(1-p)$ or will transition to $B$ with probability $p$. This step yields the term $Q^G \cdot \left[(1-p)\Gamma_e^{G}+p\Gamma_e^{B}\right]$ for the transition of information debt. 
When the second-to-last state is $B$, it will transition to $G$ with probability $r$ or will stay in $B$ with probability $(1-r)$. This step yields the term $Q^B \cdot \left[r\Gamma_e^{G}+(1-r)\Gamma_e^{B}\right]$ for the transition of information debt. 
Except for the last step discussed above, all previous transition steps of any transition trace $S_i^{k+1} \in S(k+1)$ can be found in $S(k)$, and thus the corresponding terms of the transition of information debt are the exactly same.
We refer to this argument as the \textit{recursive structure} of the state transition trace, which is illustrated in Fig. \ref{fig:transistion trace}.
Therefore, equation (\ref{equation:subset2}) holds.

Recall that we assume (\ref{equation:P(T=k)}) holds for $k>2$.
Compare the form of (\ref{equation:P(T=k)}) and (\ref{equation:align}), one can notice that equation (\ref{equation:deduction}) holds. Then substitute (\ref{equation:deduction}) into (\ref{equation:align2}), we can obtain (\ref{equation:align3}).
With equation (\ref{equation:k=2}) and (\ref{equation:align3}), by mathematical induction, we complete the proof.

\begin{align}\label{equation:deduction}
    &\begin{bmatrix}
        \displaystyle\sum_{\substack{i\in S(k)\\i|2=1}} \text{Pr}(S_i^k) \cdot \Gamma_s^{S_{i,1}^k} \cdot \prod_{j=2}^{k-1}Q^{S_{i,j}^k} & \displaystyle\sum_{\substack{i\in S(k)\\i|2=0}} \text{Pr}(S_i^k) \cdot \Gamma_s^{S_{i,1}^k} \cdot \prod_{j=2}^{k-1}Q^{S_{i,j}^k}
	\end{bmatrix} \\
    =&
    \begin{bmatrix}\pi_G^{(0)} & \pi_B^{(0)}\end{bmatrix}
        \begin{bmatrix}
            \Gamma_{s}^G & \\
             & \Gamma_{s}^B
        \end{bmatrix}
        \left\{
        \begin{bmatrix}
            (1-p)\mathbf{I}_\zeta & p\mathbf{I}_\zeta\\
            r\mathbf{I}_\zeta & (1-r)\mathbf{I}_\zeta
        \end{bmatrix}
        \begin{bmatrix}
            Q^G & \\
             & Q^B
        \end{bmatrix}
        \right\}^{k-2}
        \begin{bmatrix}
            (1-p)\mathbf{I}_\zeta & p\mathbf{I}_\zeta\\
            r\mathbf{I}_\zeta & (1-r)\mathbf{I}_\zeta
        \end{bmatrix}, \forall k>2.
\end{align}

\begin{align}\label{equation:align3}
    &\text{Pr}(t_{i_0+1} \!-\! t_{i_0} \!=\! k+1) \!=\!\! \begin{bmatrix}
		\displaystyle\sum_{\substack{i\in S(k)\\i|2=1}} \text{Pr}(S_i^k) \cdot \Gamma_s^{S_{i,1}^k} \cdot \prod_{j=2}^{k-1}Q^{S_{i,j}^k} & \displaystyle\sum_{\substack{i\in S(k)\\i|2=0}} \text{Pr}(S_i^k) \cdot \Gamma_s^{S_{i,1}^k} \cdot \prod_{j=2}^{k-1}Q^{S_{i,j}^k}
	\end{bmatrix} 
 \!\!\!\begin{bmatrix}
		Q^G & \\
		 &Q^B
	\end{bmatrix}\!\!\!\begin{bmatrix}
		(1-p)\mathbf{I}_\zeta & p\mathbf{I}_\zeta\\
		r\mathbf{I}_\zeta & (1-r)\mathbf{I}_\zeta
	\end{bmatrix}\!\!\!
	\begin{bmatrix}
		\Gamma_{e}^G \\
		\Gamma_{e}^B
	\end{bmatrix}  \\
 &=\!\!\begin{bmatrix}\pi_G^{(0)} & \pi_B^{(0)}\end{bmatrix}\!\!
\begin{bmatrix}
	\Gamma_{s}^G & \\
	 & \Gamma_{s}^B
\end{bmatrix}\!\!
\left\{\!
\begin{bmatrix}
	(1-p)\mathbf{I}_\zeta & p\mathbf{I}_\zeta\\
	r\mathbf{I}_\zeta & (1-r)\mathbf{I}_\zeta
\end{bmatrix}\!\!\!
\begin{bmatrix}
	Q^G & \\
	 & Q^B
\end{bmatrix}\!
\right\}^{k-2}\!\!
\begin{bmatrix}
	(1-p)\mathbf{I}_\zeta & p\mathbf{I}_\zeta\\
	r\mathbf{I}_\zeta & (1-r)\mathbf{I}_\zeta
\end{bmatrix}\!\!\!
	\begin{bmatrix}
	Q^G & \\
	 &Q^B
\end{bmatrix}\!\!\!
\begin{bmatrix}
	(1-p)\mathbf{I}_\zeta & p\mathbf{I}_\zeta\\
	r\mathbf{I}_\zeta & (1-r)\mathbf{I}_\zeta
\end{bmatrix}\!\!\!
\begin{bmatrix}
	\Gamma_{e}^G \\
	\Gamma_{e}^B
\end{bmatrix}  \\
&=\!\!
\begin{bmatrix}\pi_G^{(0)} & \pi_B^{(0)}\end{bmatrix}\!\!
\begin{bmatrix}
	\Gamma_{s}^G & \\
	 & \Gamma_{s}^B
\end{bmatrix}\!\!
\left\{\!
\begin{bmatrix}
	(1-p)\mathbf{I}_\zeta & p\mathbf{I}_\zeta\\
	r\mathbf{I}_\zeta & (1-r)\mathbf{I}_\zeta
\end{bmatrix}\!\!\!
\begin{bmatrix}
	Q^G & \\
	 & Q^B
\end{bmatrix}\!
\right\}^{k-1}\!
\begin{bmatrix}
	(1-p)\mathbf{I}_\zeta & p\mathbf{I}_\zeta\\
	r\mathbf{I}_\zeta & (1-r)\mathbf{I}_\zeta
\end{bmatrix}\!\!
\begin{bmatrix}
	\Gamma_{e}^G \\
	\Gamma_{e}^B
\end{bmatrix}\!.
\end{align}

\section{Proof of Proposition \ref{proposition:2}}\label{appendix:proposition 2}

We first derive $T_{0\rightarrow 0}$, the transition matrix of the probability distribution of the states between any two adjacent times that $I_d(t)$ hits zero.

Denote $\pi^{(l)} = \begin{bmatrix}
    \pi_G^{(l)} & \pi_B^{(l)}
\end{bmatrix}, l\ge 1$ the probability distribution of the states at timeslot $t_l$, where $t_l$ is the $l$-th time $I_d(t)$ hits zero.
Formally, $\pi_G^{(l)} \triangleq \text{Pr}(a_{t_l} = G)$ and $\pi_B^{(l)} \triangleq \text{Pr}(a_{t_l} = B)$. 
By the definition, we have $\pi^{(l+1)} = \pi^{(l)} \cdot T_{0\rightarrow 0}$.
Further denote $\pi^{(l),k} = \begin{bmatrix}
    \pi_{G}^{(l),k} & \pi_{B}^{(l),k}
\end{bmatrix}$ as the joint probability distribution of the states at timeslot $t_l$ and the event that $t_l - t_{l-1} = k$.
Formally, $\pi_{G}^{(l),k} \triangleq \text{Pr}(a_{t_l} = G,t_l - t_{l-1} = k)$ and $\pi_{B}^{(l),k} \triangleq \text{Pr}(a_{t_l} = B, t_l - t_{l-1} = k)$.
Thus, by the
law of total probability, $\pi^{(l)}$ can be derived by $\pi^{(l)} = \sum_{k=1}^\infty \pi^{(l),k}$.

Then we focus on the state transition trace from timeslot $t_{l}$ to $t_{l+1} -1$ and derive $\pi^{(l+1),k}$ from $\pi^{(l)}$.
For simplicity, denote the first and the last state of this transition trace as $F_s^{(l)}$ and $L_s^{(l)}$, respectively. Equivalently, $F_s^{(l)} = a_{t_l},L_s^{(l)} = a_{t_{l+1}-1}$.

For $k=1$, there is only one state in the transition trace, thus $L_s^{(l)} = F_s^{(l)}$. Then we have 
\begin{align}
    \pi_{G}^{(l+1),1}& = \pi_G^{(l)}\cdot \text{Pr}(t_{l+1} - t_l = 1|F_s^{(l)} = G) \cdot (1-p) + \pi_B^{(l)}\cdot \text{Pr}(t_{l+1} - t_l = 1|F_s^{(l)} = B) \cdot r,\\
    \pi_{B}^{(l+1),1} &= \pi_G^{(l)}\cdot \text{Pr}(t_{l+1} - t_l = 1|F_s^{(l)} = G) \cdot p + \pi_B^{(l)}\cdot \text{Pr}(t_{l+1} - t_l = 1|F_s^{(l)} = B) \cdot (1-r).
\end{align}
Note that 
\begin{align}
    \text{Pr}(t_{l+1} - t_l = 1|F_s^{(l)} = G) = \Gamma_{0,0}^G, \\
    \text{Pr}(t_{l+1} - t_l = 1|F_s^{(l)} = B) = \Gamma_{0,0}^B.    
\end{align}
Recall the denotations in Proposition \ref{proposition:2}. Thus, for $k=1$, we have 
\begin{equation}
\pi^{(l+1),k} = \pi^{(l)} \cdot\mathbf{\Gamma}_{0,0} \cdot\mathbf{T}_{1}.
\end{equation}

For $k\ge 2$, we can derive equations (\ref{equation:piGkl+1}), (\ref{equation:piBkl+1}), and further derive equation (\ref{equation:equation:pikl+1}) as below.

\begin{align}
    \pi_{G}^{(l+1),k}& = \pi_G^{(l)}\cdot \text{Pr}(t_{l+1} - t_l = k,L_s^{(l)} = G|F_s^{(l)} = G) \cdot (1-p) 
    + \pi_G^{(l)}\cdot \text{Pr}(t_{l+1} - t_l = k,L_s^{(l)} = B|F_s^{(l)} = G) \cdot r \nonumber\\
    &\qquad\quad + \pi_B^{(l)}\cdot \text{Pr}(t_{l+1} - t_l = k,L_s^{(l)} = G|F_s^{(l)} = B) \cdot (1-p) + \pi_B^{(l)}\cdot \text{Pr}(t_{l+1} - t_l = k,L_s^{(l)} = B|F_s^{(l)} = B) \cdot r,\label{equation:piGkl+1}\\
    \pi_{B}^{(l+1),k}& = \pi_G^{(l)}\cdot \text{Pr}(t_{l+1} - t_l = k,L_s^{(l)} = G|F_s^{(l)} = G) \cdot p 
    + \pi_G^{(l)}\cdot \text{Pr}(t_{l+1} - t_l = k,L_s^{(l)} = B|F_s^{(l)} = G) \cdot (1-r)\nonumber\\
    &\qquad\quad + \pi_B^{(l)}\cdot \text{Pr}(t_{l+1} - t_l = k,L_s^{(l)} = G|F_s^{(l)} = B) \cdot p + \pi_B^{(l)}\cdot \text{Pr}(t_{l+1} - t_l = k,L_s^{(l)} = B|F_s^{(l)} = B) \cdot (1-r).\label{equation:piBkl+1}
\end{align}
\begin{equation}\label{equation:equation:pikl+1}
    \begin{bmatrix}
            \pi_{G}^{(l+1),k} & \pi_{B}^{(l+1),k}
    \end{bmatrix}\!\!=\!\!\begin{bmatrix}
        \pi_G^{(l)} & \pi_B^{(l)}
    \end{bmatrix}\!\!
    \begin{bmatrix}
        \text{Pr}(t_{l+1} - t_l = k,L_s^{(l)} = G|F_s^{(l)} = G) & \text{Pr}(t_{l+1} - t_l = k,L_s^{(l)} = B|F_s^{(l)} = G)\\
        \text{Pr}(t_{l+1} - t_l = k,L_s^{(l)} = G|F_s^{(l)} = B) & \text{Pr}(t_{l+1} - t_l = k,L_s^{(l)} = B|F_s^{(l)} = B)
    \end{bmatrix}\!\!
    \begin{bmatrix}
        1-p & p\\
        r & 1-r
    \end{bmatrix}\!\!.
\end{equation} 

Then we derive the intermediate probability matrix in  (\ref{equation:equation:pikl+1}).
Recall the denotations in Proposition \ref{proposition:2}, then (\ref{equation:P(T=k)}) can be re-written as 
\begin{equation}
    \text{Pr}(t_{l+1}-t_l = k) = \begin{bmatrix}
    \pi_G^{(0)} & \pi_B^{(0)}
\end{bmatrix}
\mathbf{\Gamma}_{s}(\mathbf{T}_{\zeta}\mathbf{Q})^{k-2}\mathbf{T}_{\zeta}
    \begin{bmatrix}
            \Gamma_{e}^G \\
            \Gamma_{e}^B
    \end{bmatrix}.
\end{equation}
Denote the intermediate matrices multiplications $\mathbf{\Gamma}_{s}(\mathbf{T}_{\zeta}\mathbf{Q})^{k-2}\mathbf{T}_{\zeta} = \begin{bmatrix}
    \vec{\omega}_{GG} & \vec{\omega}_{GB}\\
    \vec{\omega}_{BG} & \vec{\omega}_{BB}
\end{bmatrix}_{2 \times 2\zeta}$, where $[\vec{\omega}_{(\cdot)(\cdot)}]_{1\times \zeta}$ represents the $(k-1)$-steps transition vector of the both the information debt and the channel state.
For example, $[\vec{\omega}_{GB}]_{1\times \zeta}$ contains $\zeta$ probabilities, i.e., $[\vec{\omega}_{GB}]_{1\times \zeta} = \begin{bmatrix}
    \omega_{GB}(1) & \cdots & \omega_{GB}(\zeta)
\end{bmatrix}$.
The $i$-th term $\omega_{GB}(i)$ is joint the transition probability of $I_d(t_l)=0 \rightarrow I_d(t_l+(k-1))=i$ and $a_{t_l}=G \rightarrow a_{t_l+(k-1)} = B$, i.e., $\omega_{GB}(i) = \text{Pr}\left(I_d(t_l+k-1) = i,L_s^{(l)} = B|I_d(t_l) = 0,F_s^{(l)} = G\right)$.
Notice the physical meaning of $\begin{bmatrix}
    \vec{\omega}_{GG} & \vec{\omega}_{GB}\\
    \vec{\omega}_{BG} & \vec{\omega}_{BB}
\end{bmatrix}$, the intermediate probability matrix in  (\ref{equation:equation:pikl+1}) can be derived by 
\begin{align}
        \begin{bmatrix}
    \text{Pr}(t_{l+1} - t_l = k,L_s^{(l)} = G|F_s^{(l)} = G) & \text{Pr}(t_{l+1} - t_l = k,L_s^{(l)} = B|F_s^{(l)} = G)\\
    \text{Pr}(t_{l+1} - t_l = k,L_s^{(l)} = G|F_s^{(l)} = B) & \text{Pr}(t_{l+1} - t_l = k,L_s^{(l)} = B|F_s^{(l)} = B)
\end{bmatrix} =& \begin{bmatrix}
    \vec{\omega}_{GG} & \vec{\omega}_{GB}\\
    \vec{\omega}_{BG} & \vec{\omega}_{BB}
\end{bmatrix}
\begin{bmatrix}
            \Gamma_{e}^G &  \\
             & \Gamma_{e}^B
    \end{bmatrix}.\\
    =&\mathbf{\Gamma}_{s}(\mathbf{T}_{\zeta}\mathbf{Q})^{k-2}\mathbf{T}_{\zeta}
    \cdot\mathbf{\Gamma}_{e}.\label{equation:transition Pr}
\end{align}

Therefore, (\ref{equation:equation:pikl+1}) can be further derived as 
\begin{equation}
    \pi^{(l+1),k} = \pi^{(l)} \mathbf{\Gamma}_{s}(\mathbf{T}_{\zeta}\mathbf{Q})^{k-2}\mathbf{T}_{\zeta}\mathbf{\Gamma}_{e}\mathbf{T}_{1}.
\end{equation}
Sum over all possible lengths $k\in[1,\infty]$ of the state transition trace, we obtain 
\begin{align}
        \pi^{(l+1)} &= \sum_{k=1}^\infty \pi^{(l+1),k} \\
        &= 
    \pi^{(l)} \cdot\mathbf{\Gamma}_{0,0} \cdot\mathbf{T}_{1} + \pi^{(l)}\mathbf{\Gamma}_{s}\left[\sum_{k=2}^\infty (\mathbf{T}_{\zeta}\mathbf{Q})^{k-2}\right]\mathbf{T}_{\zeta}\mathbf{\Gamma}_{e}\mathbf{T}_{1}\\
    &\overset{(a)}{=} \pi^{(l)}\left[\mathbf{\Gamma}_{0,0} + \mathbf{\Gamma}_{s}(\mathbf{I}_{2\zeta}-\mathbf{T}_{\zeta}\mathbf{Q})^{-1}\mathbf{T}_{\zeta}\mathbf{\Gamma}_{e}\right]\mathbf{T}_{1}.\label{equation:pil+1topil}
\end{align}
Note that $\mathbf{T}_{\zeta}$ is a stochastic matrix with the values in each row summing up to 1, and $\mathbf{Q}$ is a matrix with the values in each row summing no larger than 1.
Therefore, the spectral radius of 
$\mathbf{T}_{\zeta}\mathbf{Q}$ satisfies $\rho(\mathbf{T}_{\zeta}\mathbf{Q})<1$, and thus we have $\lim_{k\rightarrow\infty}\left(\mathbf{T}_{\zeta}\mathbf{Q}\right)^k = \mathbf{0}$.
Therefore, equality (a) holds. 
From (\ref{equation:pil+1topil}) we can notice that 
\begin{equation}
    T_{0\rightarrow 0} = \big[\mathbf{\Gamma}_{0,0} + \mathbf{\Gamma}_{s}(\mathbf{I}_{2\zeta}-\mathbf{T}_{\zeta}\mathbf{Q})^{-1}\mathbf{T}_{\zeta}\mathbf{\Gamma}_{e}\big]\mathbf{T}_1.
\end{equation} 

When the Markov chain becomes stationary, or equivalently, when $l\rightarrow \infty$, we have $\pi^{(0)} = \pi^{(0)} \cdot T_{0\rightarrow 0}$. 
Notice that $\pi_G^{(0)} + \pi_B^{(0)} = 1$.
Thus, one can derive $\pi^{(0)}$ by solving linear equations 
\begin{equation}\label{equation:solution of initial_rewrite}
    \begin{bmatrix}
    (T_{0\rightarrow 0} - \mathbf{I}_2)^\top \\
     \begin{array}{cc}
        1  & 1 
     \end{array}
\end{bmatrix}_{3\times2} \cdot 
\begin{bmatrix}
    \pi_G^{(0)} \\
    \pi_B^{(0)}
\end{bmatrix}_{2\times1}
=\begin{bmatrix}
    0 \\
    0 \\
    1
\end{bmatrix}_{3\times1}.
\end{equation}
Therefore, Proposition \ref{proposition:2} is proved.

\section{Proof of Lemma \ref{lemma:2}}\label{appendix:lemma 2}

We characterize the exact values of $\mathbb{E}\{t_{i_0+1}-t_{i_0}\}, \mathbb{E}\{L_G\}, \mathbb{E}\{L_{B_1}\}$ and $\mathbb{E}\{L_{B_2}\}$ in the following four subsections separately. 

\subsection{$\mathbb{E}\{t_{i_0+1}-t_{i_0}\}$}
By definition, we have 
\begin{align}\label{equation: EA=k_proof}
    \mathbb{E}\{t_{i_0+1} - t_{i_0} = k\} &= \sum_{k=1}^\infty k \cdot\text{Pr}(t_{i_0+1} - t_{i_0} = k)\\
    &=\pi^{(0)}\left(\begin{bmatrix}
        \Gamma_{0,0}^G\\
        \Gamma_{0,0}^B
    \end{bmatrix}+\mathbf{\Gamma}_s\sum_{k=2}^\infty k \cdot (\mathbf{T}_{\zeta}\mathbf{Q})^{k-2}\mathbf{T}_{\zeta}\begin{bmatrix}
        \Gamma_e^G\\
        \Gamma_e^B
    \end{bmatrix}\right).
\end{align}

Since
\begin{align}
    \mathbf{Q} \cdot\vec{\mathbf{1}}_{2\zeta} &= 
    \begin{bmatrix}
        \Gamma_{\phi,\phi}^G & \Gamma_{\phi,\zeta}^G & & \\
        \Gamma_{\zeta,\phi}^G & \Gamma_{\zeta,\zeta}^G & & \\
         & & \Gamma_{\phi,\phi}^B & \Gamma_{\phi,\zeta}^B\\
         & & \Gamma_{\zeta,\phi}^B & \Gamma_{\zeta,\zeta}^B
    \end{bmatrix}\vec{\mathbf{1}}_{2\zeta}\\
    &=
    \begin{bmatrix}
        1-\Gamma_{\phi,0}^G\\
        1-\Gamma_{\zeta,0}^G\\
        1-\Gamma_{\phi,0}^B\\
        1-\Gamma_{\zeta,0}^B
    \end{bmatrix} \\
    &=\vec{\mathbf{1}}_{2\zeta} - \begin{bmatrix}
        \Gamma_e^G\\
        \Gamma_e^B
    \end{bmatrix},
\end{align}
we have
\begin{align}
    \begin{bmatrix}
        \Gamma_e^G\\
        \Gamma_e^B
    \end{bmatrix} = \left(\mathbf{I}_{2\zeta}-\mathbf{Q}\right)\cdot\vec{\mathbf{1}}_{2\zeta}.
\end{align}
Thus, (\ref{equation: EA=k_proof}) can be further written as 
\begin{align}
    &\mathbb{E}\{t_{i_0+1} - t_{i_0} = k\} \\
    &= \pi^{(0)}\left(\begin{bmatrix}
        \Gamma_{0,0}^G\\
        \Gamma_{0,0}^B
    \end{bmatrix}+\mathbf{\Gamma}_s\sum_{k=2}^\infty k \cdot (\mathbf{T}_{\zeta}\mathbf{Q})^{k-2}\mathbf{T}_{\zeta}\left(\mathbf{I}_{2\zeta}-\mathbf{Q}\right)\cdot\vec{\mathbf{1}}_{2\zeta}\right)\\
    &=\pi^{(0)}\left(\begin{bmatrix}\Gamma_{0,0}^G\\\Gamma_{0,0}^B\end{bmatrix} + \mathbf{\Gamma}_s\sum_{k=2}^\infty k \cdot (\mathbf{T}_{\zeta}\mathbf{Q})^{k-2}\mathbf{T}_{\zeta}\cdot\vec{\mathbf{1}}_{2\zeta} - \mathbf{\Gamma}_s\sum_{k=2}^\infty k \cdot (\mathbf{T}_{\zeta}\mathbf{Q})^{k-2}\mathbf{T}_{\zeta}\mathbf{Q}\cdot\vec{\mathbf{1}}_{2\zeta}\right)\\
    &\overset{(b)}{=}\pi^{(0)}\left(\begin{bmatrix}
        \Gamma_{0,0}^G\\
        \Gamma_{0,0}^B
    \end{bmatrix} + \mathbf{\Gamma}_s
    \left[\sum_{k=2}^\infty k \cdot (\mathbf{T}_{\zeta}\mathbf{Q})^{k-2} - \sum_{k=2}^\infty k \cdot (\mathbf{T}_{\zeta}\mathbf{Q})^{k-1}\right]\cdot\vec{\mathbf{1}}_{2\zeta}\right)\\
    &\overset{(c)}{=}\pi^{(0)}\left(\begin{bmatrix}
        \Gamma_{0,0}^G\\
        \Gamma_{0,0}^B
    \end{bmatrix} + \mathbf{\Gamma}_s \left[\mathbf{I}_{2\zeta}+(\mathbf{I}_{2\zeta}-\mathbf{T}_{\zeta}\mathbf{Q})^{-1}\right]\cdot\vec{\mathbf{1}}_{2\zeta}\right)\\
    &=\pi^{(0)}\left(\begin{bmatrix}\Gamma_{0,0}^G\\\Gamma_{0,0}^B\end{bmatrix} + \mathbf{\Gamma}_s\cdot\vec{\mathbf{1}}_{2\zeta} + \mathbf{\Gamma}_s(\mathbf{I}_{2\zeta}-\mathbf{T}_{\zeta}\mathbf{Q})^{-1}\cdot\vec{\mathbf{1}}_{2\zeta}\right)\\
    &\overset{(d)}{=}\pi^{(0)}\left(\begin{bmatrix}
        1\\
        1
    \end{bmatrix} + 
    \mathbf{\Gamma}_s(\mathbf{I}_{2\zeta}-\mathbf{T}_{\zeta}\mathbf{Q})^{-1}\cdot\vec{\mathbf{1}}_{2\zeta}\right)\\
    &= 1 + \pi^{(0)}\mathbf{\Gamma}_s(\mathbf{I}_{2\zeta}-\mathbf{T}_{\zeta}\mathbf{Q})^{-1}\cdot\vec{\mathbf{1}}_{2\zeta},
\end{align}
which is equation (\ref{equation:E(A)}).
Equality (b) is due to the fact that 
\begin{equation}
\mathbf{T}_{\zeta}\cdot\vec{\mathbf{1}}_{2\zeta} = \begin{bmatrix}
    (1-p)\mathbf{I}_{\zeta} & p\mathbf{I}_{\zeta}\\
    r\mathbf{I}_{\zeta} & (1-r)\mathbf{I}_{\zeta}
\end{bmatrix}\cdot\vec{\mathbf{1}}_{2\zeta}= \vec{\mathbf{1}}_{2\zeta}.
\end{equation}
Equality (c) is derived by dislocation subtraction. Equality (d) is due to the fact that 
\begin{equation}
\mathbf{\Gamma}_s \cdot\vec{\mathbf{1}}_{2\zeta} = \begin{bmatrix}
    \Gamma_{0,\phi}^G & \Gamma_{0,\zeta}^G & &\\
     & & \Gamma_{0,\phi}^B & \Gamma_{0,\zeta}^B\\
\end{bmatrix}\cdot\vec{\mathbf{1}}_{2\zeta}
=\begin{bmatrix}
    1-\Gamma_{0,0}^G\\
    1-\Gamma_{0,0}^B
\end{bmatrix}.
\end{equation}

\subsection{$\mathbb{E}\{L_G\}$}
Denote $\text{Pr}(t_{i_0+1} - t_{i_0} = k|\nexists \tau_j \in (t_{i_0},t_{i_0+1}))$ as the probability of $t_{i_0+1} - t_{i_0} = k$, given that there does not exist any index $j$ satisfying that $\tau_j \in (t_{i_0},t_{i_0+1})$.
In the same method of Proposition \ref{proposition:1}, $\text{Pr}(t_{i_0+1} - t_{i_0} = k|\nexists \tau_j \in (t_{i_0},t_{i_0+1}))$ can be given as follows:
\begin{equation}
    \text{Pr}(t_{i_0+1} - t_{i_0} = k|\nexists \tau_j \in (t_{i_0},t_{i_0+1})) \!=\! \pi^{(0)}
\mathbf{\Gamma}_{0,\phi}(\mathbf{T}_{\zeta-1}\mathbf{\Gamma}_{\phi,\phi})^{k-2}\mathbf{T}_{\zeta-1}\!
    \begin{bmatrix}
                \Gamma_{\phi,0}^G \\
                \Gamma_{\phi,0}^B
    \end{bmatrix}\!\!.
\end{equation}
By definition, we have
\begin{align}
    \mathbb{E}\{L_G\} \!&=\! \sum_{k=1}^\infty (k\!-\!\Delta\!-\!1)^+ \pi^{(0)}
\mathbf{\Gamma}_{0,\phi}(\mathbf{T}_{\zeta-1}\mathbf{\Gamma}_{\phi,\phi})^{k-2}\mathbf{T}_{\zeta-1}\!\!
    \begin{bmatrix}
                \Gamma_{\phi,0}^G \\
                \Gamma_{\phi,0}^B
    \end{bmatrix}\\
    &= \pi^{(0)}
\mathbf{\Gamma}_{0,\phi}\!\left[\!\sum_{k=\Delta+2}^\infty (k-\Delta-1) (\mathbf{T}_{\zeta-1}\mathbf{\Gamma}_{\phi,\phi})^{k-2}\right]\!\mathbf{T}_{\zeta-1}\!
    \begin{bmatrix}
                \Gamma_{\phi,0}^G \\
                \Gamma_{\phi,0}^B
    \end{bmatrix}\\
    &= \pi^{(0)}
            \mathbf{\Gamma}_{0,\phi}
        (\mathbf{I}_{2\zeta-2} \!-\! \mathbf{T}_{\zeta-1}\mathbf{\Gamma}_{\phi,\phi})^{-1}(\mathbf{T}_{\zeta-1}\mathbf{\Gamma}_{\phi,\phi})^{\Delta}\mathbf{T}_{\zeta-1}\!
            \begin{bmatrix}
                \Gamma_{\phi,0}^G \\
                \Gamma_{\phi,0}^B
            \end{bmatrix}\!\!,
\end{align}
which is equation (\ref{equation:E(LG)}).

\subsection{$\mathbb{E}\{L_{B_1}\}$}
The derivations of $\mathbb{E}\{L_{B_1}\}$ and $\mathbb{E}\{L_{B_2}\}$ are more complex, mainly because they involve the not stopping time $\tau_{j}^*$. 
When the information debt starts from zero and before it hits back to zero again, it may hit $\zeta$ repeatedly for many times. 
Thus, the probability distribution of the states at each time the information debt hits $\zeta$ are distinct and should be characterize carefully.

Denote $\pi_{\zeta}^{(l)},l\ge 1$ as the conditional probability distribution of the states, given that currently $I_d(t) = \zeta$ and this is the $i$-th time that $I_d(t)$ hits $\zeta$ before it hits back to zero. 
Recall that $T_{0\rightarrow 0}$ denotes the transition matrix of the probability distribution of the states between any two adjacent times that $I_d(t)$ hits zero.
Similarly, let $T_{0\rightarrow \zeta}$ denote the transition matrix of the probability distribution of the states between any two timeslots $t_{i_0}$ and $\tau_j^{\dag}$, which satisfies that $\tau_j^{\dag} = \inf \{t':t_{i_0}<t'<t_{i_0+1}, I_d(t')=\zeta\}$.
Literally, $\tau_j^{\dag}$ is the first time after $t_{i_0}$ that $I_d(t)$ hits $\zeta$ before it hits back to zero.
Moreover, let $T_{\zeta\rightarrow \zeta}$ denote the transition matrix of the probability distribution of the states between any two adjacent times that $I_d(t)$ hits $\zeta$ and within which $I_d(t)$ doesn't hit 0.
In the same method of Proposition \ref{proposition:2}, we can derive
\begin{align}
    T_{0\rightarrow 0} = \big[\mathbf{\Gamma}_{0,0} + \mathbf{\Gamma}_{s}(\mathbf{I}_{2\zeta}-\mathbf{T}_{\zeta}\mathbf{Q})^{-1}\mathbf{T}_{\zeta}\mathbf{\Gamma}_{e}\big]\mathbf{T}_1.
\end{align}
\begin{align}
       T_{0\rightarrow \zeta} \!=\! \big[\mathbf{\Gamma}_{0,\zeta} 
        \!+\! \mathbf{\Gamma}_{0,\phi} (\mathbf{I}_{2\zeta-2} \!-\! \mathbf{T}_{\zeta-1}\mathbf{\Gamma}_{\phi,\phi})^{-1}\mathbf{T}_{\zeta-1}\mathbf{\Gamma}_{\phi,\zeta} \big]
        \mathbf{T}_{1},
\end{align}
\begin{align}
       T_{\zeta\rightarrow \zeta} \!=\! \big[\mathbf{\Gamma}_{\zeta,\zeta} 
        \!+\! \mathbf{\Gamma}_{\zeta,\phi} (\mathbf{I}_{2\zeta-2} \!-\! \mathbf{T}_{\zeta-1}\mathbf{\Gamma}_{\phi,\phi})^{-1}\mathbf{T}_{\zeta-1}\mathbf{\Gamma}_{\phi,\zeta}\big]
        \mathbf{T}_{1}.
\end{align}
Then we define event $A_l$. The event $A_l$ represents that currently $I_d(t) = \zeta$ and this has been the $l$-th time that the information debt hits $\zeta$ after it starts from zero and before it hits back to zero.
Further define the probability $\text{Pr}(0\rightarrow \zeta)$ and the conditional probabilities $\text{Pr}(\zeta\rightarrow \zeta|A_l), l\ge 1$.  
$\text{Pr}(0\rightarrow \zeta)$ is the probability that after $I_d(t)$ starts from zero, it hits $\zeta$ before hitting back to zero.
$\text{Pr}(\zeta\rightarrow \zeta|A_l)$ is the conditional probability that after $I_d(t)$ starts from $\zeta$, it hits back to $\zeta$ again before hitting zero, given that $I_d(t)$ has already hits $\zeta$ for $l$ times after it starts from zero and before it hits back to zero.
Recall the denotation $(\mathbf{I}_{2\zeta-2} - \mathbf{T}_{\zeta-1}\mathbf{\Gamma}_{\phi,\phi})^{-1} = \mathbf{M}$. By the definitions stated above, $\text{Pr}(0\rightarrow \zeta)$ and $\text{Pr}(\zeta\rightarrow \zeta|A_l), l\ge 1$ can be given as follows:
\begin{align}
    \text{Pr}(0\rightarrow \zeta) = \pi^{(0)}\!\Bigg\{\!\!\begin{bmatrix}
            \Gamma_{0,\zeta}^G \\
            \Gamma_{0,\zeta}^B\end{bmatrix} 
        + \mathbf{\Gamma}_{0,\phi} 
  \mathbf{M} \mathbf{T}_{\zeta-1} \begin{bmatrix}
            \Gamma_{\phi,\zeta}^G \\
            \Gamma_{\phi,\zeta}^B
        \end{bmatrix}\!\!\Bigg\},
\end{align}
\begin{align}
    \text{Pr}(\zeta\rightarrow \zeta|A_l) = \pi_{\zeta}^{(l)}\!\Bigg\{\!\!\begin{bmatrix}
            \Gamma_{\zeta,\zeta}^G \\
            \Gamma_{\zeta,\zeta}^B\end{bmatrix} 
        + \mathbf{\Gamma}_{\zeta,\phi} 
  \mathbf{M} \mathbf{T}_{\zeta-1} \begin{bmatrix}
            \Gamma_{\phi,\zeta}^G \\
            \Gamma_{\phi,\zeta}^B
        \end{bmatrix}\!\!\Bigg\}.
\end{align}
For ease of presentation, we denote $\vec{c} = \begin{bmatrix}
            \Gamma_{\zeta,\zeta}^G \\
            \Gamma_{\zeta,\zeta}^B\end{bmatrix} 
        + \mathbf{\Gamma}_{\zeta,\phi} 
  \mathbf{M} \mathbf{T}_{\zeta-1} \begin{bmatrix}
            \Gamma_{\phi,\zeta}^G \\
            \Gamma_{\phi,\zeta}^B
        \end{bmatrix}$ and thus $\text{Pr}(\zeta\rightarrow \zeta|A_l) = \pi_{\zeta}^{(l)}\cdot\vec{c}$.
With the definitions above, $\pi_{\zeta}^{(l)},l\ge 1$ can be derived in a iterative form as follows:
\begin{align}
    \pi_{\zeta}^{(1)} = \frac{\pi^{(0)}\cdot T_{0\rightarrow\zeta}}{\text{Pr}(0\rightarrow \zeta)},
\end{align}
\begin{align}\label{equation:recursive pi(l)}
    \pi_{\zeta}^{(l)} = \frac{\pi_{\zeta}^{(l-1)}\cdot T_{\zeta\rightarrow\zeta}}{\text{Pr}(\zeta\rightarrow \zeta|A_{l-1})}, l\ge 2.
\end{align}
By substituting (\ref{equation:recursive pi(l)}) into itself iteratively, we can derive 
\begin{align}
    \pi_{\zeta}^{(l)} &= \frac{\pi_{\zeta}^{(l-1)}\cdot T_{\zeta\rightarrow\zeta}}{\text{Pr}(\zeta\rightarrow \zeta|A_{l-1})}\\
    &= \frac{\pi_{\zeta}^{(l-1)}\cdot T_{\zeta\rightarrow\zeta}}{\pi_{\zeta}^{(l-1)}\cdot\vec{c}}\\
    &= \frac{\frac{\pi_{\zeta}^{(l-2)}\cdot T_{\zeta\rightarrow\zeta}}{\pi_{\zeta}^{(l-2)}\cdot\vec{c}}\cdot T_{\zeta\rightarrow\zeta}}{\frac{\pi_{\zeta}^{(l-2)}\cdot T_{\zeta\rightarrow\zeta}}{\pi_{\zeta}^{(l-2)}\cdot\vec{c}}\cdot\vec{c}}\\
    &\overset{(e)}{=} \frac{\pi_{\zeta}^{(l-2)}\cdot T_{\zeta\rightarrow\zeta}\cdot T_{\zeta\rightarrow\zeta}}{\pi_{\zeta}^{(l-2)}\cdot T_{\zeta\rightarrow\zeta}\cdot\vec{c}}\\
    &= \cdots\\
    &= \frac{\pi_{\zeta}^{(1)}\cdot T_{\zeta\rightarrow\zeta}^{l-1}}{\pi_{\zeta}^{(1)}\cdot T_{\zeta\rightarrow\zeta}^{l-2}\cdot\vec{c}}, \forall l\ge 2,
\end{align}
where equality (e) is due to fact that $\pi_{\zeta}^{(l-2)}\cdot\vec{c}$ is a scalar.

To proceed forward, as in \cite{RLSCs}, for any $t$, we denote the time interval between $t$ and the first time after $t$ that the information debt hits $x$ as 
\begin{equation}
    H_t(x) \triangleq \inf\{\tau>0:I_d(t+\tau)=x\}.
\end{equation}
We then denote the time interval between $t$ and the last time after $t$ that the information debt hits $\zeta$ before hitting zero as 
\begin{equation}\label{equation:definition of Lambda_t}
    \Lambda_t \triangleq \sup\{\tau\ge 0:I_d(t+\tau)=\zeta \ \text{and} \ \tau \le H_t(0)\}.
\end{equation}
Notice that in \cite{RLSCs}, for any $t$, they were interested in the value of $\mathbb{E}\{\Lambda_t|I_d(t)=\zeta\}$, the average number of timeslots it takes for the information debt to start from $\zeta$ and hit the last $\zeta$ before hitting 0.
Due to the Markov property and the i.i.d. SEC, $\mathbb{E}\{\Lambda_t|I_d(t)=\zeta\}$ is not a function of $t$, thus $\mathbb{E}\{\Lambda_t|I_d(t)=\zeta\}$ is independent of how many times that the information debt has hit $\zeta$ before hitting 0.

On the contrary, in G-ESEC, $\mathbb{E}\{\Lambda_t|I_d(t)=\zeta\}$ depends on the probability distribution of the channel states at current timeslot. For example, when at timeslot $t$, if $I_d(t)=\zeta$ and the probability of $G$ is much larger than the probability of $B$, then the information debt may decease quickly to zero.
Therefore, $t$ is more likely to be the last time that information debt hits $\zeta$ before hitting zero, and thus lead to a smaller value of $\mathbb{E}\{\Lambda_t|I_d(t)=\zeta\}$.
To account for this argument, different from that in \cite{RLSCs}, we further consider another conditional expectation $\mathbb{E}\{\Lambda_t|A_l\}$, where the number of times that the information debt has hit $\zeta$ is also included in the condition term.

In \cite{RLSCs}, $\mathbb{E}\{\Lambda_t|I_d(t)=\zeta\}$ is derived in i.i.d. SEC with a recursive equation, which no more holds in the G-ESEC.
This is because the probability distributions of the states at each time the information debt hits $\zeta$ are distinct.
Thus, we modify the recursive equation into a iterative equation as follows:
\begin{align}\label{equation:iterative equation}
    \mathbb{E}&\{\Lambda_t|A_l\} = \pi_{\zeta}^{(l)} \begin{bmatrix}
        \Gamma_{\zeta,\zeta}^G\\
        \Gamma_{\zeta,\zeta}^B
        \end{bmatrix}(1 + \mathbb{E}\{\Lambda_t|A_{l+1}\}) + \sum_{k=2}^\infty \mathbf{\Gamma}_{\zeta,\phi} (\mathbf{T}_{\zeta-1}\mathbf{\Gamma}_{\phi,\phi})^{k-2} \mathbf{T}_{\zeta-1} \!\!
        \begin{bmatrix}
        \Gamma_{\phi,\zeta}^G\\
        \Gamma_{\phi,\zeta}^B
        \end{bmatrix} \!\!(k+\mathbb{E}\{\Lambda_t|A_{l+1}\}).
\end{align}
Recall the denotations of $\vec{m}$ and $\vec{c}$, (\ref{equation:iterative equation}) can be simplified into 
\begin{align}\label{equation:iterative equation_simplified}
    \mathbb{E}\{\Lambda_t|A_l\} = \pi_{\zeta}^{(l)}\cdot\vec{m} + \pi_{\zeta}^{(l)}\cdot\vec{c}\cdot\mathbb{E}\{\Lambda_t|A_{l+1}\}.
\end{align}
To derive $\mathbb{E}\{L_{B_1}\}$, we are interested in the value of $\mathbb{E}\{\Lambda_t|A_1\}$, which can be obtain by iteratively substitute (\ref{equation:iterative equation_simplified}) into itself.
\begin{align}
    &\mathbb{E}\{\Lambda_t|A_1\} =\pi_{\zeta}^{(1)}\vec{m} + \pi_{\zeta}^{(1)}\vec{c}\cdot\mathbb{E}\{\Lambda_t|A_{2}\}\\
    &=\pi_{\zeta}^{(1)}\vec{m} + \pi_{\zeta}^{(1)}\vec{c}\cdot\pi_{\zeta}^{(2)}\vec{m} + \pi_{\zeta}^{(1)}\vec{c}\cdot\pi_{\zeta}^{(2)}\vec{c}\cdot\mathbb{E}\{\Lambda_t|A_{3}\}\\
    &= \cdots \\
    &= \pi_{\zeta}^{(1)}\vec{m} + \cdots \!+\!\left(\prod_{i=1}^{k-1} \pi_{\zeta}^{(i)}\vec{c}\right)\!\cdot\!\pi_{\zeta}^{(k)}\vec{m} +\! \left(\prod_{i=1}^{k} \pi_{\zeta}^{(i)}\vec{c}\right)\!\cdot\!\mathbb{E}\{\Lambda_t|A_{k+1}\}.\label{equation:iterative equation_infinity}
\end{align}
When $k$ approaches infinity, (\ref{equation:iterative equation_infinity}) can be given by 
\begin{align}\label{equation:Lamda_1}
    &\mathbb{E}\{\Lambda_t|A_1\} = \sum_{k=1}^\infty\left(\prod_{i=1}^{k-1} \pi_{\zeta}^{(i)}\vec{c}\right)\cdot \pi_{\zeta}^{(k)}\vec{m} + \lim_{k\rightarrow\infty} \left(\prod_{i=1}^{k} \pi_{\zeta}^{(i)}\vec{c}\right)\cdot \mathbb{E}\{\Lambda_t|A_{k+1}\}.
\end{align}
First consider the remainder term in (\ref{equation:Lamda_1}).
\begin{align}\label{equation:remainder=0}
    &\lim_{k\rightarrow\infty} \left(\prod_{i=1}^{k} \pi_{\zeta}^{(i)}\vec{c}\right)\cdot \mathbb{E}\{\Lambda_t|A_{k+1}\} \\
    &=\lim_{k\rightarrow\infty} \pi_{\zeta}^{(1)}\vec{c}\cdot\frac{\pi_{\zeta}^{(1)}T_{\zeta\rightarrow\zeta}\vec{c}}{\pi_{\zeta}^{(1)}\vec{c}}\cdot\frac{\pi_{\zeta}^{(1)}T_{\zeta\rightarrow\zeta}^2\vec{c}}{\pi_{\zeta}^{(1)}T_{\zeta\rightarrow\zeta}\vec{c}}\cdots \frac{\pi_{\zeta}^{(1)}T_{\zeta\rightarrow\zeta}^{k-1}\vec{c}}{\pi_{\zeta}^{(1)}T_{\zeta\rightarrow\zeta}^{k-2}\vec{c}}\cdot\mathbb{E}\{\Lambda_t|A_{k+1}\}\\
    &= \lim_{k\rightarrow\infty} \pi_{\zeta}^{(1)}T_{\zeta\rightarrow\zeta}^{k-1}\vec{c}\cdot\mathbb{E}\{\Lambda_t|A_{k+1}\}\\
    &= \pi_{\zeta}^{(1)} \left(\lim_{k\rightarrow\infty}T_{\zeta\rightarrow\zeta}^{k-1}\right)\vec{c}\cdot \left(\lim_{k\rightarrow\infty}\mathbb{E}\{\Lambda_t|A_{k+1}\}\right).
\end{align}
Note that $T_{\zeta\rightarrow\zeta}$ is a non-stochastic matrix. Thus, $\rho(T_{\zeta\rightarrow\zeta})<1$ and $\lim_{k\rightarrow\infty}T_{\zeta\rightarrow\zeta}^{k-1}=\mathbf{0}$. Also note that $\lim_{k\rightarrow\infty}\mathbb{E}\{\Lambda_t|A_{k+1}\}<\infty$. 
Therefore, the remainder term in (\ref{equation:Lamda_1}) approaches zero when $k$ goes infinity. Then (\ref{equation:Lamda_1}) can be further given by
\begin{align}\label{equation:E(Lambda|1)}
       \mathbb{E}\{\Lambda_t|A_1\} &=\sum_{k=1}^\infty\left(\prod_{i=1}^{k-1} \pi_{\zeta}^{(i)}\vec{c}\right)\cdot \pi_{\zeta}^{(k)}\vec{m}\\
        &=\pi_{\zeta}^{(1)}\left(\sum_{k=1}^\infty T_{\zeta\rightarrow\zeta}^{k-1}\right)\vec{m}\\
       &= \frac{\pi^{(0)}\cdot T_{0\rightarrow \zeta}}{\text{Pr}(0\rightarrow \zeta)}(\mathbf{I}_2-T_{\zeta\rightarrow\zeta})^{-1}\Vec{m}.
\end{align}

Now we are ready to derive $\mathbb{E}\{L_{B_1}\}$.
\begin{align}
    &\mathbb{E}\{L_{B_1}\} = \pi^{(0)}\begin{bmatrix}
        \Gamma_{0,\zeta}^G\\
        \Gamma_{0,\zeta}^B
    \end{bmatrix}\cdot(1+\mathbb{E}\{\Lambda_t|A_1\}) 
    + \sum_{k=2}^\infty \pi^{(0)} \mathbf{\Gamma}_{0,\phi} (\mathbf{T}_{\zeta-1}\mathbf{\Gamma}_{\phi,\phi})^{k-2} \mathbf{T}_{\zeta-1}\begin{bmatrix}
        \Gamma_{\phi,\zeta}^G\\
        \Gamma_{\phi,\zeta}^B
    \end{bmatrix}\cdot (k+\mathbb{E}\{\Lambda_t|A_1\})\\
    &= \text{Pr}(0\rightarrow \zeta)\cdot(1+\mathbb{E}\{\Lambda_t|A_1\})+\pi^{(0)}\mathbf{\Gamma}_{0,\phi}\mathbf{M}^{2}\mathbf{T}_{\zeta-1}\begin{bmatrix}
        \Gamma_{\phi,\zeta}^G\\
        \Gamma_{\phi,\zeta}^B
    \end{bmatrix}\\
    &= \pi^{(0)}\cdot T_{0\rightarrow \zeta}(\mathbf{I}_2-T_{\zeta\rightarrow\zeta})^{-1}\Vec{m} + \text{Pr}(0\rightarrow \zeta) + \pi^{(0)}\mathbf{\Gamma}_{0,\phi}\mathbf{M}^{2}\mathbf{T}_{\zeta-1}\begin{bmatrix}
        \Gamma_{\phi,\zeta}^G\\
        \Gamma_{\phi,\zeta}^B
    \end{bmatrix}\\
    &=\pi^{(0)} \cdot \left[T_{0\rightarrow \zeta} \cdot (\mathbf{I}_2-T_{\zeta\rightarrow\zeta})^{-1}\cdot \Vec{m}+\Vec{n}\right].
\end{align}

\subsection{$\mathbb{E}\{L_{B_2}\}$}

Similar to the definition of $\Lambda_t$ in (\ref{equation:definition of Lambda_t}), for any $t$, we denote the time interval between $t$ and the first time after $t$ that the information debt hits zero before hitting $\zeta$ as 
\begin{equation}\label{equation:definition of V_t}
    V_t \triangleq \inf\{\tau\ge 0:I_d(t+\tau)=0 \ \text{and} \ \tau \le H_t(\zeta)\}.
\end{equation}
Recall that $L_{B_2} \triangleq \mathds{1}\{\tau_j \in (t_{i_0},t_{i_0+1})\}\cdot\max(-\alpha,t_{i_0+1} - \Delta - 1 - \tau_{j^*})$. 
Thus, $\mathbb{E}\{L_{B_2}\} = \mathbb{E}\{\max(-\alpha,V_t - \Delta - 1)|I_d(t)=\zeta\}$.
Then we denote $H_l \triangleq \mathbb{E}\{\max(-\alpha, V_t- \Delta - 1)|A_l\}$.
Similar to (\ref{equation:iterative equation}), iterative equation can be also derived for $H_l$ as follows:
\begin{align}\label{equation:iterative H_l}
    H_l 
    &= \pi_\zeta^{(l)} \max(-\alpha,-\Delta)+\sum_{k=2}^\infty \pi_\zeta^{(l)} \mathbf{\Gamma}_{\zeta,\phi} (\mathbf{T}_{\zeta-1}\mathbf{\Gamma}_{\phi,\phi})^{k-2}\mathbf{T}_{\zeta-1} \cdot\begin{bmatrix}
        \Gamma_{\phi,0}^G\\
        \Gamma_{\phi,0}^B
    \end{bmatrix} \max(-\alpha,k-\Delta-1) + \text{Pr}(\zeta\rightarrow \zeta|A_l)\cdot H_{l+1}\\
    &= \pi_\zeta^{(l)}\cdot \vec{b} + \text{Pr}(\zeta\rightarrow \zeta|A_l)\cdot H_{l+1},
\end{align}
where $\vec{b}$ was defined in (\ref{equation:b}). Iteratively substitute (\ref{equation:iterative H_l}) into itself, we obtain 
\begin{align}
    H_1 &= \pi_\zeta^{(1)}\vec{b} + \text{Pr}(\zeta\rightarrow \zeta|A_l)H_{2}\\
    &= \pi_\zeta^{(1)}\vec{b} + \text{Pr}(\zeta\rightarrow \zeta|A_{1})\pi_\zeta^{(2)} \vec{b} + \text{Pr}(\zeta\rightarrow \zeta|A_{1})\text{Pr}(\zeta\rightarrow \zeta|A_{2})H_3\\
    &= \pi_\zeta^{(1)}\vec{b} + \cdots + \prod_{i=0}^{k-1}\text{Pr}(\zeta\rightarrow \zeta|A_{i})\pi_\zeta^{(k)}\vec{b} + \prod_{i=0}^{k}\text{Pr}(\zeta\rightarrow \zeta|A_{i})H_{k+1}\\
    &= \sum_{k=1}^l \left(\prod_{i=0}^{k-1}\text{Pr}(\zeta\rightarrow \zeta|A_{i})\right)\pi_\zeta^{(k)}\vec{b} + \prod_{i=0}^{k}\text{Pr}(\zeta\rightarrow \zeta|A_{i})H_{k+1}.\label{equation:iterative equation_infinity_2}
\end{align}
Similar to (\ref{equation:remainder=0}), it is easy to verify that the remainder term of (\ref{equation:iterative equation_infinity_2}) approaches zero when $k\rightarrow \infty$. Thus, we have
\begin{align}
    H_1 &= \sum_{k=1}^\infty \left(\prod_{i=0}^{k-1}\text{Pr}(\zeta\rightarrow \zeta|A_{i})\right)\pi_\zeta^{(k)}\vec{b}\\
    &= \sum_{k=1}^\infty \left(\prod_{i=0}^{k-1}\pi_{\zeta}^{(l)}\vec{c}\right)\pi_\zeta^{(k)}\vec{b}\\
    &= \sum_{k=1}^\infty \left(\pi_{\zeta}^{(1)}\vec{c}\cdots \pi_{\zeta}^{(k-1)}\vec{c}\right)\pi_\zeta^{(k)}\vec{b}\\
    &= \sum_{k=1}^\infty \pi_{\zeta}^{(1)}\vec{c}\cdot\frac{\pi_{\zeta}^{(1)}T_{\zeta\rightarrow\zeta}\vec{c}}{\pi_{\zeta}^{(1)}\vec{c}}\cdot\frac{\pi_{\zeta}^{(1)}T_{\zeta\rightarrow\zeta}^2\vec{c}}{\pi_{\zeta}^{(1)}T_{\zeta\rightarrow\zeta}\vec{c}}\cdots \frac{\pi_{\zeta}^{(1)}T_{\zeta\rightarrow\zeta}^{k-1}\vec{b}}{\pi_{\zeta}^{(1)}T_{\zeta\rightarrow\zeta}^{k-2}\vec{c}}\cdot \\
    &= \pi_{\zeta}^{(1)}\left(\sum_{k=1}^\infty T_{\zeta\rightarrow\zeta}^{k-1}\right)\vec{b}\\
    &= \pi_{\zeta}^{(1)}\left(\mathbf{I}_2-T_{\zeta\rightarrow\zeta}\right)^{-1}\vec{b}.
\end{align}

Now we are ready to derive $\mathbb{E}\{L_{B_2}\}$.
\begin{align}
    \mathbb{E}\{L_{B_2}\} &= \left[1-\text{Pr}(0\rightarrow \zeta)\right]\cdot 0 + 
    \text{Pr}(0\rightarrow \zeta)\cdot H_1\\
    &= \text{Pr}(0\rightarrow \zeta)\pi_{\zeta}^{(1)}\left(\mathbf{I}_2-T_{\zeta\rightarrow\zeta}\right)^{-1}\vec{b}\\
    &= \pi^{(0)}\cdot T_{0\rightarrow\zeta}\left(\mathbf{I}_2-T_{\zeta\rightarrow\zeta}\right)^{-1}\vec{b}.
\end{align}
Therefore, Lemma \ref{lemma:2} is proved.

\section{Proof of Lemma \ref{lemma:E(ti0+1-ti0) when r=1/2}}\label{appendix:lemma E(ti0+1-ti0) when r=1/2}
Recall that we assume $\alpha\rightarrow\infty$ and $K=N-K$.
Note that a similar case with random source arrival and $K = 1$ has been investigated in \cite{asymptotics2}.
Similar to \cite{asymptotics2}, for any fixed finite integers $n\in [0,\infty)$, we first define the event 
\begin{equation}
    \mathcal{A}_n=\left\{\max\{I_d(\tau):\tau\in( t_{i_0},t_{i_0+1}]\}<n\right\}.
\end{equation}
Specifically, $\mathcal{A}_n$ is the event that the entire trajectory of $I_d(t)$ in the interval $( t_{i_0},t_{i_0+1}]$ is strictly below a ceiling value $n$. Considering this event restricts the original infinite-state-space problem into a finite-state Markov chain, which is much more tractable than the original problem. 
By the monotone convergence theorem, we have 
\begin{align}\label{equation:monotone convergence theorem}
    \lim_{n\rightarrow \infty} \text{Pr}(\mathcal{A}_n)\cdot\mathbb{E}\big\{t_{i_0+1} - t_{i_0}\big|\mathcal{A}_n\big\}
    = \lim_{n\rightarrow \infty} \mathbb{E}\big\{\mathds{1}\{\mathcal{A}_n\}\cdot (t_{i_0+1} - t_{i_0})\big\}
    = \mathbb{E}\big\{t_{i_0+1} - t_{i_0}\big\}.
\end{align}

\begin{figure}
    \centering
    \includegraphics[width=0.8\linewidth]{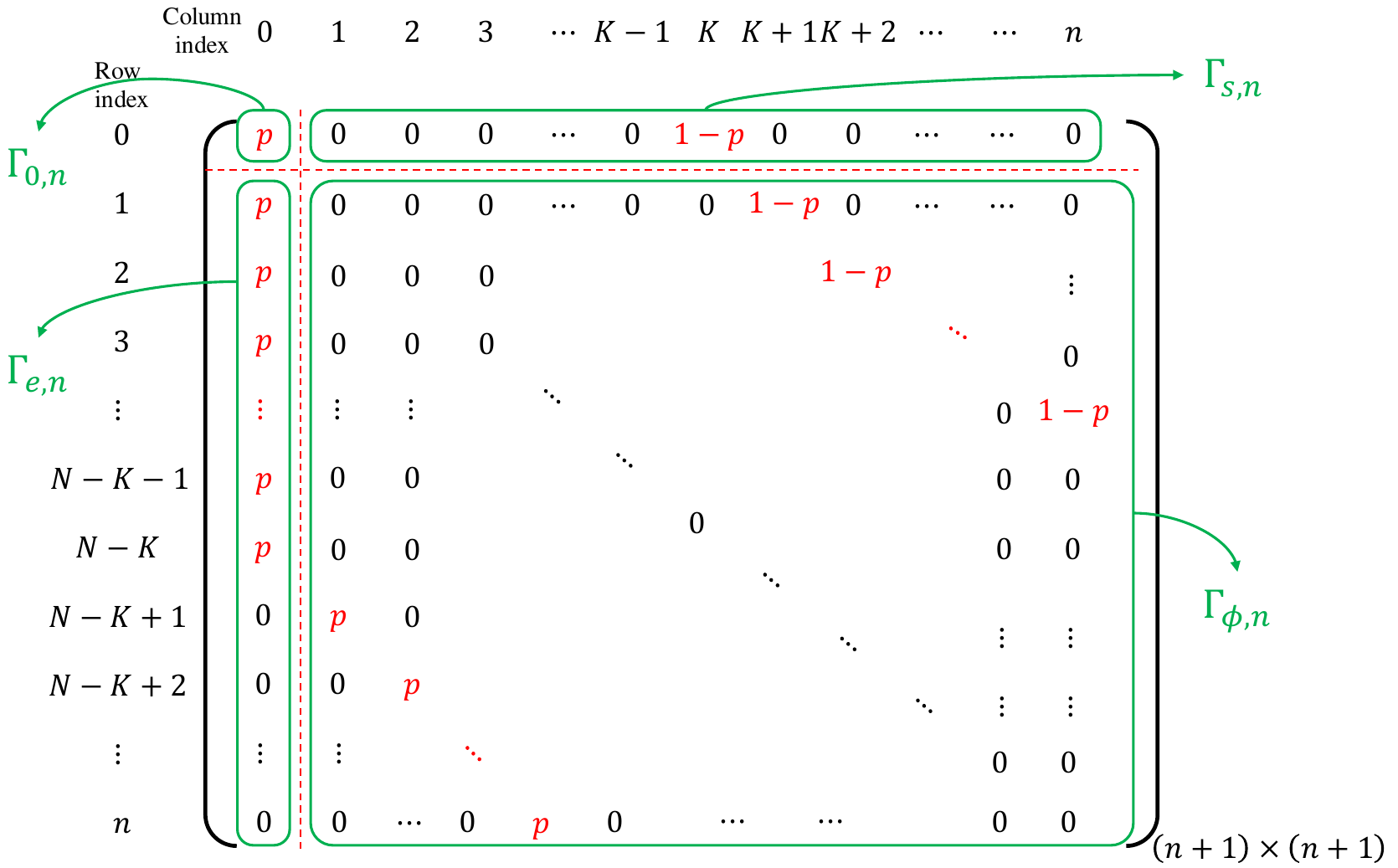}
    \caption{Illustration of $\Gamma$, the transition matrix of information debt in i.i.d. PEC.}
    \label{fig:matrix_division}
\end{figure}

Since that when $n\rightarrow\infty$, $\text{Pr}(\mathcal{A}_n)\rightarrow 1$ almost sure, in the following, we first derive the analytical expression of $\mathbb{E}\big\{t_{i_0+1} - t_{i_0}\big|\mathcal{A}_n\big\}$ and then let $n$ approach infinity. Denote the transition matrix of the information debt in the i.i.d. PEC as $\Gamma$. With Definition \ref{definition:informtion debt new}, when $\mathcal{A}_n$ happens, $\Gamma$ can be written as shown in Fig. \ref{fig:matrix_division}. Note that $\Gamma$ is with the size of $(n+1)\times(n+1)$ and can be divided into four non-overlapping parts $\Gamma_{0,n},\Gamma_{s,n},\Gamma_{e,n}$ and $\Gamma_{\phi,n}$. $\Gamma_{0,n},\Gamma_{s,n},\Gamma_{e,n}$ are a scalar and two vectors in very simple form, respectively. Specially, $\Gamma_{\phi,n}$ is a banded Toeplitz matrix with two symmetric off-diagonals. Recall that we assume $K=N-K$ in this case. The starting points of the two off-diagonals of $\Gamma_{\phi,n}$ are at column index $K+1$ and row index $N-K+1$, respectively, which are symmetrical to the main diagonal. It is also worthy to point out that $\Gamma$ is not presented as a standard stochastic matrix, such that all entries of each row will sum up to 1. Actually, the entries in the last column with row index $N-K-1$ to $n$ are hardwired to zero. This is because $\Gamma$ will be essentially considered as a transition matrix with infinite size, when we let $n$ approach infinity. Thus, it can be imagined that the right and bottom sides of the matrix $\Gamma$ are extending indefinitely. Therefore, the entries in column $n$ with row index $N-K-1$ to $n$ will not be constrained to $1-p$. 

According to the characteristic of Toeplitz matrix with two symmetric off-diagonals\cite{Toeplitz}, first define $\beta$ as the remainder of the Euclidian division of $n$ by $K$, that is
\begin{align}
    \beta = n - K n_K
\end{align}
or, in other words, $\beta = n|K$, and $n_K$ is the quotient.
Then, $\Gamma_{\phi,n}$ has the following eigendecomposition.
\begin{lemma}\label{lemma:matrix decomposition}
    When $\Gamma_{\phi,n}$ follows the structure in Fig. \ref{fig:matrix_division}, it can be decomposed as follows:
    \begin{align}
        \Gamma_{\phi,n} = U_n \cdot \Lambda_n \cdot U_n^{-1},
    \end{align}
where the matrix of eigenvalues $\Lambda_n$ and the matrices of eigenvectors $U_n, U_n^{-1}$ explicitly satisfies
    \begin{align}\label{equation:Lamda_n}
        \Lambda_n = 2\sqrt{p(1-p)}\cdot diag\Big[\overbrace{\cos{\frac{\pi}{n_K+1}},\cdots,\cos{\frac{\pi}{n_K+1}}}^{K-\beta},\overbrace{\cos{\frac{2\pi}{n_K+1}},\cdots,\cos{\frac{2\pi}{n_K+1}}}^{K-\beta},\cdots,\overbrace{\cos{\frac{n_K\pi}{n_K+1}},\cdots,\cos{\frac{n_K\pi}{n_K+1}}}^{K-\beta}, \nonumber\\
        \overbrace{\cos{\frac{\pi}{n_K+2}},\cdots,\cos{\frac{\pi}{n_K+2}}}^{\beta},\overbrace{\cos{\frac{2\pi}{n_K+2}},\cdots,\cos{\frac{2\pi}{n_K+2}}}^{\beta},\cdots,\overbrace{\cos{\frac{(n_K+1)\pi}{n_K+2}},\cdots,\cos{\frac{(n_K+1)\pi}{n_K+2}}}^{\beta}\Big],
    \end{align}
    \begin{align}\label{equation:U_n}
        \Gamma_{s,n} \cdot U_n = \sqrt{p(1-p)} \cdot \Big[\overbrace{0,\cdots,0,\sin{\frac{\pi}{n_K+1}}}^{K-\beta},\overbrace{0,\cdots,0,\sin{\frac{2\pi}{n_K+1}}}^{K-\beta},\cdots,\overbrace{0,\cdots,0,\sin{\frac{n_K\pi}{n_K+1}}}^{K-\beta},\overbrace{0,\cdots,0}^{\beta\cdot(n_K+1)}\Big],
    \end{align}
    \begin{align}\label{equation:U_n-1}
       U_n^{-1} \cdot \Gamma_{e,n} =& \begin{bmatrix}
           \frac{2}{n_K+1}\mathbf{I}_{n_K(K-\beta)} & \\
            & \frac{2}{n_K+2}\mathbf{I}_{(n_K+1)\beta}
       \end{bmatrix}\cdot
       \sqrt{p(1-p)} \cdot \nonumber\\
       &\Big[\overbrace{\sin{\frac{\pi}{n_K+1}},\cdots,\sin{\frac{\pi}{n_K+1}}}^{K-\beta},
                                \overbrace{\sin{\frac{2\pi}{n_K+1}},\cdots,\sin{\frac{2\pi}{n_K+1}}}^{K-\beta},
                                \cdots,
                                \overbrace{\sin{\frac{n_K\pi}{n_K+1}},\cdots,\sin{\frac{n_K\pi}{n_K+1}}}^{K-\beta},\nonumber\\
                                &\quad\quad \overbrace{\sin{\frac{\pi}{n_K+2}},\cdots,\sin{\frac{\pi}{n_K+2}}}^{\beta},
                                \overbrace{\sin{\frac{2\pi}{n_K+2}},\cdots,\sin{\frac{2\pi}{n_K+2}}}^{\beta},
                                \cdots,
                                \overbrace{\sin{\frac{(n_K+1)\pi}{n_K+2}},\cdots,\sin{\frac{(n_K+1)\pi}{n_K+2}}}^{\beta}\Big].
    \end{align}
    \end{lemma}
    Derivations of equations (\ref{equation:Lamda_n}), (\ref{equation:U_n}) and (\ref{equation:U_n-1}) in Lemma \ref{lemma:matrix decomposition} can directly obtained from\cite{Toeplitz} and is thus omitted. 
    
    By definition, $\mathbb{E}\big\{t_{i_0+1} - t_{i_0}\big|\mathcal{A}_n\big\}$ can be given by
    \begin{align}
        \mathbb{E}\big\{t_{i_0+1} - t_{i_0}\big|\mathcal{A}_n\big\} &=  1 \cdot \Gamma_{0,n} + \sum_{k=2}^\infty k \cdot \Gamma_{s,n} \cdot\Gamma_{\phi,n}^{k-2} \cdot\Gamma_{e,n}\\
        &= \Gamma_{0,n} + \sum_{k=2}^\infty k \cdot \Gamma_{s,n} \cdot U_n \cdot\Lambda_n^{k-2} \cdot U_n^{-1} \cdot\Gamma_{e,n}\\
        &= \Gamma_{0,n} + \Gamma_{s,n} \cdot U_n  \cdot[\mathbf{I}_n + (\mathbf{I}_n -\Lambda_n)^{-1}]\cdot (\mathbf{I}_n -\Lambda_n)^{-1}  \cdot U_n^{-1}\cdot\Gamma_{e,n}. \label{equation:165}
    \end{align}

    With (\ref{equation:Lamda_n}), the terms $(\mathbf{I}_n -\Lambda_n)^{-1}$ and $[\mathbf{I}_n + (\mathbf{I}_n -\Lambda_n)^{-1}]$ can be directly derived as (\ref{equation:(In-Ln)-1}) and (\ref{equation:In+(In-Ln)-1}) below, where we only explicitly present the first term of the diagonal vector and the other terms are omitted for ease of presentation. 
    \begin{align}\label{equation:(In-Ln)-1}
        (\mathbf{I}_n -\Lambda_n)^{-1} = diag\Big[\frac{1}{1-2\sqrt{p(1-p)}\cos{\frac{\pi}{n_K+1}}},\cdots\Big],
    \end{align}
    \begin{align}\label{equation:In+(In-Ln)-1}
         \big[\mathbf{I}_n + (\mathbf{I}_n -\Lambda_n)^{-1}\big]\cdot(\mathbf{I}_n -\Lambda_n)^{-1} = diag\Big[\big(1+\frac{1}{1-2\sqrt{p(1-p)}\cos{\frac{\pi}{n_K+1}}}\big)\cdot\frac{1}{1-2\sqrt{p(1-p)}\cos{\frac{\pi}{n_K+1}}},\cdots\Big].
    \end{align}

    Therefore, equation (\ref{equation:165}) can be further written as 
    \begin{align}
    \mathbb{E}\big\{t_{i_0+1} - t_{i_0}\big|\mathcal{A}_n\big\} &= p + p(1-p)\cdot \sum_{j=1}^{n_K} \sin{\frac{j\pi}{n_K+1}}\cdot \big(1+\frac{1}{1-2\sqrt{p(1-p)}\cos{\frac{j\pi}{n_K+1}}}\big)\cdot\frac{\sin{\frac{j\pi}{n_K+1}} \cdot \frac{2}{n_K+1}}{1-2\sqrt{p(1-p)}\cos{\frac{j\pi}{n_K+1}}}\\
    &= p + \frac{2p(1-p)}{n_K+1}\cdot \sum_{j=1}^{n_K} \big(1+\frac{1}{1-2\sqrt{p(1-p)}\cos{\frac{j\pi}{n_K+1}}}\big)\cdot\frac{\sin^2{\frac{j\pi}{n_K+1}} }{1-2\sqrt{p(1-p)}\cos{\frac{j\pi}{n_K+1}}}\label{equation:168}. 
    \end{align}
    Note that when $n$ approaches infinity, the quotient $n_K$ also approaches infinity. When $n_K\rightarrow \infty$, let $\frac{1}{n_K+1} = dx$ be the integration variable. Subsequently, the summation from $j=1$ to $n_K$ can be transformed to a integral from $x=0$ to 1. 
    Therefore, equation (\ref{equation:168}) can be further written as
    \begin{align}
        \mathbb{E}\{t_{i_0+1} - t_{i_0}\} 
            =p + 2p(1-p)\cdot \int_0^1 \left(1+\frac{1}{1-2\sqrt{p(1-p)}\cos{\pi x}}\right)\frac{\sin^2{\pi x}}{1-2\sqrt{p(1-p)}\cos{\pi x}} dx.
    \end{align}
    
    Thus, Lemma \ref{lemma:E(ti0+1-ti0) when r=1/2} is proved.

\section{Proof of Lemma \ref{lemma:E(N_e) when r=1/2}}\label{appendix:E(N_e) when r=1/2}
By definition, the expectation $\mathbb{E}\{N_e\}$ should be averaged over all possible paths. First classify all possible paths by their lengths. The length can be 1 or all even numbers that are larger than 2. 
Let $L$ denote the length of path. $L=1$ represents that a perfect delivery appears in the first timeslot of information debt transition, such that $t_{i_0+1} - t_{i_0} = 1$. When $L\ge 2$, recall that we assume $K=N-K$, which means the increase of information debt when erasure appears equals to the decrease of information debt when the perfect delivery appears.
Therefore, in a round $t\in(t_{i_0},t_{i_0+1}]$, the number of erasures must equal to the number of perfect delivery, which indicates that $L$ should be an even number. 

\begin{figure}
    \centering
    \includegraphics[width=0.7\linewidth]{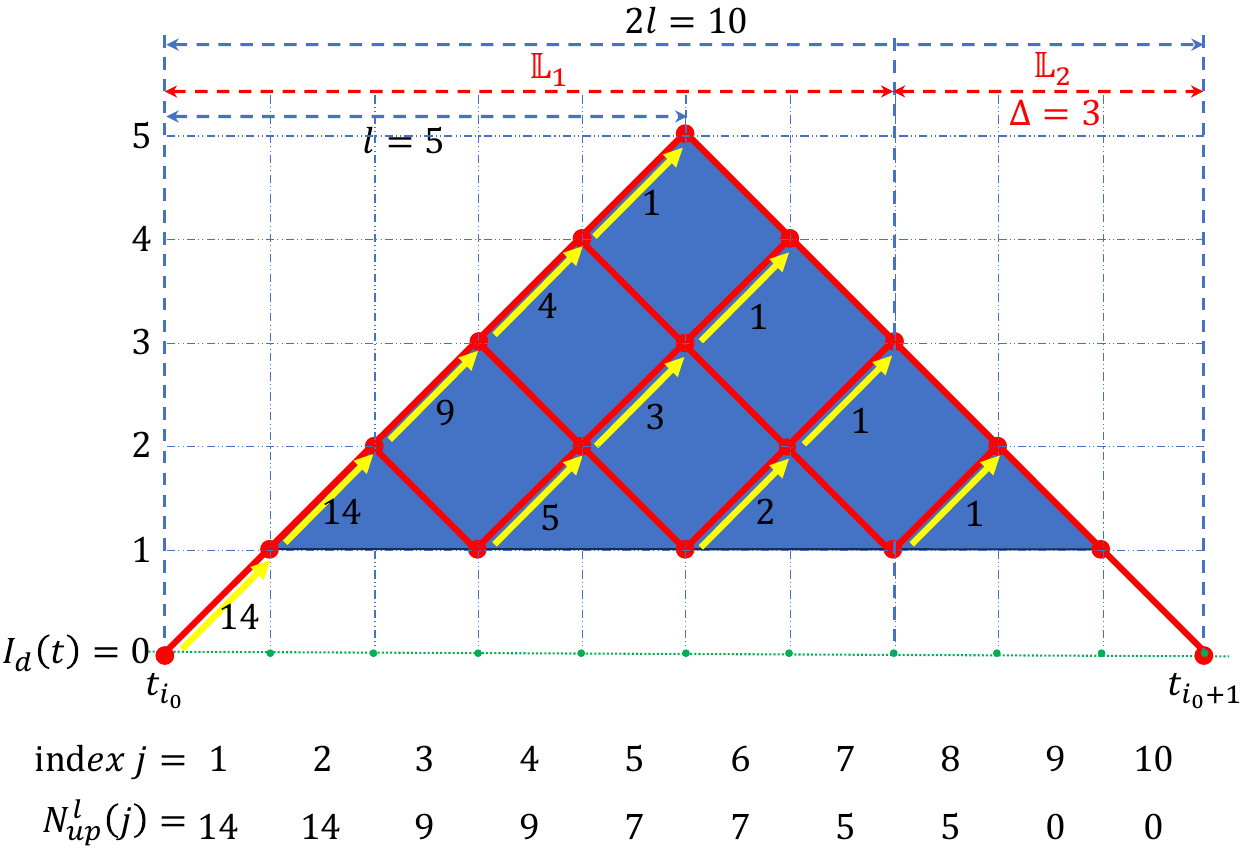}
    \caption{An example of the problem conversion with $l=5$ and $\Delta=3$.}
    \label{fig:Catalan}
\end{figure}

Let's focus on the paths with a certain length $L$, which are referred to as $L$-paths thereafter. Denote $p_L$ as the probability that a $L$-path appears.
Recall we assume that the probability of perfect delivery is $p$, and the corresponding erasure probability equals to $1-p$ in i.i.d. PEC.
When $L=1$, we have $p_1 = p$ and $t_{i_0+1} - t_{i_0} = 1$.
When $L\ge 2$, let $L=2l$, where $l\ge 1$ is an integer. In a $2l$-path, there must be $l$ erasures and $l$ perfect deliveries. 
Therefore, we have $p_{2l} = p^l(1-p)^l$. Note that a valid $2l$-path must satisfy that the $l$ erasures and $l$ perfect deliveries are distributed in the path such that $I_d(t)>0,\forall t\in(t_{i_0},t_{i_0+1})$ and $I_d(t_{i_0})=0,I_d(t_{i_0+1})=0$.
The number of such valid $2l$-paths is very similar to the Catalan Numbers \cite{Catalan}, which is a concept in the combinatorial mathematics. 
For any integer $n\ge 0$, Catalan Numbers $C_n = \frac{1}{n+1} \binom{2n}{n}$ can represent the number of monotonic lattice paths along the edges of a grid with $n \times n$ square cells, which do not pass the diagonal. 
By the definition of Catalan Numbers, it is easy to verify that the number of valid $2l$-paths equals to $C_{l-1}$.
An illustration of $l=5$ can be found in Fig. \ref{fig:Catalan}. The red lattice in Fig. \ref{fig:Catalan} represents the union of the routes of all valid $2l$-paths. Note that the number of monotonic paths of the red lattice in the blue triangle area is exactly in line with the standard definition of Catalan Numbers. Thus, when $l=5$, the number of valid $2l$-paths equals to $C_{l-1}=C_4 = 14$.

Denote the number of erasures in $\mathbb{L}_1=\left(t_{i_0},\max( t_{i_0+1}-\Delta,t_{i_0})\right)$ of the $i$-th $2l$-paths as $N_e(l,i)$, for any $l\ge 1$ and $i\in [C_{l-1}]$.
By definition, $\mathbb{E}\{N_e\}$ can be derived as
\begin{align}
    \mathbb{E}\{N_e\} &= p_1 \cdot 0 + \sum_{l=1}^\infty \sum_{i=1}^{C_{l-1}} p_{2l} \cdot N_e(l,i) \\
    &= \sum_{l = \lceil\frac{\Delta}{2}\rceil + 1}^\infty p^l(1-p)^l  \sum_{i=1}^{C_{l-1}}  N_e(l,i), \label{equation:169}
\end{align}
where the last equality is due to the fact that when $2l \le \Delta + 1$, we have $\mathbb{L}_1 = \Phi$ and thus $N_e(l,i) = 0, \forall i\in [C_{l-1}]$.
                                                                                                                                    
Then, we derive the exact expression of $\sum_{i=1}^{C_{l-1}} N_e(l,i)$. For any given $l$, the physical meaning of $\sum_{i=1}^{C_{l-1}} N_e(l,i)$ is the sum of erasures in $\mathbb{L}_1$ among all $2l$-paths. 
Let $Era(l,i,j)$ denote the indicator function of the erasure for the $j$-th step of the $i$-th $2l$-paths. If an erasure appears in that step, then $Era(l,i,j)=1$.
Thus, we have 
\begin{align}
    \sum_{i=1}^{C_{l-1}} N_e(l,i)  &= \sum_{i=1}^{C_{l-1}} \sum_{j=1}^{2l-\Delta-1} Era(l,i,j) \\
    &=  \sum_{j=1}^{2l-\Delta-1} \sum_{i=1}^{C_{l-1}} Era(l,i,j) \\
    &=  \sum_{j=1}^{2l-\Delta-1}  N_{up}^l(j),\label{equation:172}
\end{align}
where $N_{up}^l(j) \triangleq \sum_{i=1}^{C_{l-1}} Era(l,i,j)$ represents the sum of erasures of the $j$-th step among all $2l$-paths. In the following, we first convert the problem of deriving $N_{up}^l(j)$ into a counting problem of monotonic lattice paths.
Then we leverage the feature of Catalan Numbers to derive $N_{up}^l(j)$. 

\begin{prop}\label{proposition:problem convert to Catalan}
    (\textit{Problem conversion})  For any fixed index $l\ge 1$, construct the monotonic lattice of all $2l$-paths as shown in Fig. \ref{fig:Catalan}. $N_{up}^l(j)$ can be derived by counting the number of \textit{upward path} in the $j$-th step of all $2l$-paths. The upward paths are displayed by yellow arrows in Fig. \ref{fig:Catalan}.
\end{prop}

    Proposition \ref{proposition:problem convert to Catalan} directly holds. An example of the problem conversion can be found in Fig. \ref{fig:Catalan}. In Fig. \ref{fig:Catalan}, the black number below the yellow arrow counts how many paths will go through this upward path among all $2l$-paths. By counting the upward paths, one can notice that for index $j=1$ and $j=2$, all paths will go through the same upward path. Thus $N_{up}^l(1)=N_{up}^l(2)=14$. For $j=3$ and $j=4$, 9 out of 14 paths will go through the upward path, thus $N_{up}^l(3)=N_{up}^l(4)=9$. For $j=5,6$, $N_{up}^l(3)=N_{up}^l(4)=7$. And similarly, $N_{up}^l(7)=N_{up}^l(8)=5,N_{up}^l(9)=N_{up}^l(10)=0$. 

\begin{prop}\label{proposition:Catalan}
    (\textit{The feature of Catalan Numbers}) 
    For any fixed index $l\ge 1$, $N_{up}^l(j)$ satisfies that 
    \begin{itemize}
        \item When $j=1$, $N_{up}^l(j)=C_{l-1}$.
        \item When $2\le j \le 2l$ and $j|2=0$, $N_{up}^l(j)=N_{up}^l(j-1)$.
        \item When $2\le j \le 2l$ and $j|2=1$, $N_{up}^l(j)=N_{up}^l(j-1) - C_{\frac{j-1}{2}-1}\cdot C_{\frac{2l+1-j}{2}-1}$.
    \end{itemize}
\end{prop}

Proposition \ref{proposition:Catalan} can be directly proved by analyzing the structure of the iterative relationship of the numbers in the monotonic lattice paths. The proof of Proposition \ref{proposition:Catalan} is tedious and thus omitted. 

With Proposition \ref{proposition:problem convert to Catalan} and Proposition \ref{proposition:Catalan}, let $J = \lfloor \frac{2l-\Delta-1}{2}\rfloor$, it can be derived that 
    \begin{equation}\label{equation:Nuplj}
        \sum_{j=1}^{2l-\Delta-1}  N_{up}^l(j) = \left\{
        \begin{aligned}
            &2\left[J\cdot C_{l-1}-\sum_{j=0}^{J-2}C_j\cdot C_{l-2-j}\cdot\big(J-1-j\big)\right] \qquad\qquad\quad \text{if} \  \Delta|2=1, \\
            &\big(2J+1\big)C_{l-1} - \sum_{j=0}^{J-2} C_j\cdot C_{l-2-j}\cdot\big(2J-1-2j\big)  \qquad\qquad  \text{if} \ \Delta|2=0.
        \end{aligned}
        \right
        .
    \end{equation}
Plug $J = \lfloor \frac{2l-\Delta-1}{2}\rfloor$ in, the two lines of equation (\ref{equation:Nuplj}) can be written in a unified form as follows
\begin{align}
    \sum_{j=1}^{2l-\Delta-1}  N_{up}^l(j) = \Big(2l-\Delta-1\Big)\cdot C_{l-1} - \sum_{i=0}^{l-\lfloor\frac{\Delta}{2}\rfloor - 3} C_i \cdot C_{l-2-i} \cdot \Big(2l - \Delta - 3 - 2i\Big).\label{equation:174}
\end{align}

With equations (\ref{equation:169}), (\ref{equation:172}) and (\ref{equation:174}), we directly obtain 
\begin{align}
    \mathbb{E}\{N_e\} = \sum_{l = \lceil\frac{\Delta}{2}\rceil + 1}^\infty p^l(1-p)^l  \left[\Big(2l-\Delta-1\Big)\cdot C_{l-1} - \sum_{i=0}^{l-\lfloor\frac{\Delta}{2}\rfloor - 3} C_i \cdot C_{l-2-i} \cdot \Big(2l - \Delta - 3 - 2i\Big)\right].
\end{align}
Thus, Lemma \ref{lemma:E(N_e) when r=1/2} is proved.

\end{appendices}

\bibliographystyle{IEEEtran}
\begin{spacing}{1.0}
		
\end{spacing}

\end{document}